%% file: offset_dipole.tex
\documentclass[useAMS,usenatbib,referee]{mn2e}
\usepackage{amsmath, amssymb, aas_macros}
\usepackage[dvips]{graphicx}
\usepackage{lscape}
\usepackage{tikz}   
\usepackage{gensymb}

\usepackage{tikz-3dplot} 

\newcommand{\grad}{\mathbf{\nabla}}
\newcommand{\rot}{\mathbf{\nabla} \times}
\newcommand{\divg}{\mathbf{\nabla}\cdot}

\newcommand{\ex}{\vec{e}_{\rm x}}
\newcommand{\ey}{\vec{e}_{\rm y}}
\newcommand{\ez}{\vec{e}_{\rm z}}
\newcommand{\rlight}{r_{\rm L}}

\title[Radiation of an off-centred dipole]{Radiation from an off-centred rotating dipole in vacuum}

\author[J. P\'etri]{J.  P\'etri$^{1}$
\thanks{E-mail: jerome.petri@astro.unistra.fr} \\
  $^{1}$Observatoire astronomique de Strasbourg, Universit\'e de Strasbourg, CNRS, UMR 7550, \\11 rue de l'universit\'e, F-67000 Strasbourg, France.}

\begin{document}

\date{Accepted . Received ; in original form }

\pagerange{\pageref{firstpage}--\pageref{lastpage}} 

\maketitle

\label{firstpage}

\begin{abstract}
When a neutron star forms, after the collapse of its progenitor, a strong magnetic field survives in its interior. This magnetic topology is usually assumed to be well approximated by a dipole located right at the centre of the star. However, there is no particular reason why this dipole should be attached to this very special point. A slight shift from the stellar centre could have strong implications for the surrounding electromagnetic field configuration leading to clear observational signatures. We study the effect of the most general off-centred dipole anchored in the neutron star interior. Exact analytical solutions are given in vacuum outside the star to any order of accuracy in the small parameter $\epsilon = d/R$, where $d$ is the displacement of the dipole from the stellar centre and $R$ the neutron star radius. As a simple diagnostic of this decentred dipole, the spin-down luminosity and the torque exerted on its crust are computed to the lowest leading order in~$\epsilon$. Results are compared to earlier works and a discussion on repercussions on pulsar braking index and multi-wavelength light curves is proposed.
\end{abstract}


\section{Introduction}

Final stage of stellar evolution, neutron stars are expected to harbour huge magnetic fields in their interior. To simplify their study, it is often assumed that the magnetic field is a dipole located exactly at the centre of the star. Such assumption could hardly be supported by a physical explanation. Actually the situation should be quite opposite. The probability to find such a peculiar topology is rather weak. So it is highly probable that during the collapse of the magnetized progenitor, the induced field is not centred and maybe even not strictly dipolar already at the beginning of its life. An asymmetry in the magnetic field topology with respect to the spherical neutron star was already mentioned by \cite{1975ApJ...201..447H} in order to explain the high kick velocity above 100~km/s given to the newly born star. Polarization effects and thus contributions from an additional electric field component have been added to this picture by \cite{1976ApJ...209..245T}.

In the case of a centred dipole, the torque applied on a neutron star in vacuum always predicts an alignment as explain by \cite{1970ApJ...159L..81D} and \cite{1970ApJ...160L..11G}. This fact is also discussed by \cite{2014MNRAS.441.1879P}. Moreover, this electromagnetic torque has been computed following several expressions, either using the electromagnetic stress-energy tensor or the Lorentz force acting inside and on the neutron star surface. Both approaches agree on the regular torque but for the anomalous torque for which a single analytical formula has been found, the constant proportionality factor depends on the expression used, stress-energy tensor or Lorentz force. Also the electric field contribution, sometimes omitted, should be included, see for instance \cite{2014PhyU...57..799B} for a critical review of these discrepancies. These complications did not prevent \cite{1985ApJ...299..706G} from considering contributions from a quadrupole component to the total torque. Deviation from a pure and centred rotating dipole is an attractive assumption to get more insight into pulsar magnetospheres. Indeed, a simple prescription for a distorted dipolar field was proposed by \cite{2011ApJ...726L..10H} to enhance the pair production rate above polar caps. Non dipolar fields have also been suggested to explain the anomalous braking index of pulsars as discussed by \cite{2010MNRAS.409.1077B}.

Off centred dipoles are not a privilege of neutron stars. They have already been investigated in main sequence stars. \cite{1974MNRAS.169..471S} looked at decentred dipoles in stars in the special case of a displacement along the magnetic axis. Off centred dipoles are invoked in AP Stars to solve the asymmetry problem between the north and south hemisphere \citep{1970ApJ...159.1001L}. An off-centred dipole is also the preferred way to interpret Zeeman line profiles as explained in \cite{1974ApJ...187..271B}. A dominant dipolar fossil field may not exclude higher order multipoles such as quadrupoles and hexapoles in magnetic Ap-Bp stars \citep{1981A&A...103..244M}.
In planets, \cite{1976PASAu...3...51K} also noted that a decentred pure dipole can better fit the Pioneer data about Jupiter than a centred dipole. It was used to explain the asymmetries in the radio emission by minimizing the quadrupolar term. \cite{1980AJ.....85..611L} presented a review about magnetic field in non degenerated star and concluded that a decentred oblique dipole can reasonably fit the magnetic topology. Hints for off-centred dipole or dipole plus quadrupole fields are given by \cite{1995ApJ...449..863P} from polarization observations of white dwarfs. Distinguishing an off-centred dipole from a combination of multipoles seems difficult on the base only of observations \citep{1984MNRAS.206..407M}. For a review on magnetic stars, see for instance \cite{1986PhR...140....1M}.

Several authors have also focused attention on the magnetic field structure near the surface of a neutron star. Hall effects inside the star can generate quadrupole toroidal fields from the differential rotation of a poloidal dipole field as shown in recent simulations by \cite{2014MNRAS.438.1618G}. In the same vain, \cite{2014MNRAS.444.3198G} investigated the formation of magnetic spots, that is, strong and highly curved magnetic field lines from the Hall drift of crustal toroidal fields in order to justify the partially screened gap model presented by~\cite{2003A&A...407..315G}. However they found that quadrupolar components seem to produce too weak field intensities. Multipolar field contributions to the axisymmetric force-free magnetosphere starts to appear also in the literature like for instance in \cite{2016arXiv160404625G}.

Corrections to a centred dipole did not get full attention for neutron stars. However, multipoles certainly exist in all stars. \cite{2015A&A...573A..51B} and \cite{2015MNRAS.450..714P} gave exact analytical solutions  for any rotating multipole in vacuum. These expressions are very useful to get more physical insight into the off-centred dipole that can be expanded in a series of centred multipoles. Recently \cite{2016ApJ...819L..16A} discovered a pulsar with a braking index of $n=3.15$, thus larger than the fiducial value of~$n=3$ obtained by pure magnetodipole losses. This is a hint for a quadrupolar component in the magnetic field or maybe for quadrupolar gravitational radiation although other explanations such as magnetic field decay and obliquity decay are not excluded. Following the reasoning of \cite{2000A&A...354..163P} assuming quadrupolar gravitational radiation, it is possible to constrain the respective weights of dipole and quadrupole contribution according to this braking index. 

With the wealth of new and very accurate multi-wavelength observations of pulsed emission in pulsars including polarization, time is ripe to go further than the almost exclusively used centred dipole to sharpen our view on pulsar magnetospheres. Rotating an off-centred dipole anchored in a perfectly conducting sphere with finite radius leads very naturally to multipolar components to any order with weighting coefficients solely related to the geometry of the dipole. We use this assumption as a starting point for the justification of multipolar components. In this paper we propose to compute exact analytical expressions for the electromagnetic field in vacuum outside an off-centred rotating dipole to any order in the small parameter $\epsilon = d/R$, where $d$ is the distance from the centre of the star and $R$ the neutron star radius. In section \ref{sec:DipoleDecentre}, we show how to compute this expansion to any order by use of the properties of spherical harmonics. The exact solution is then presented in section~\ref{sec:RadiatingDipole}. Next, in section~\ref{sec:Luminosite} we compute the spin-down luminosity expected from such a system and compare previous works assuming a point dipole to our finite size dipole. The same comparison is done for the braking index. The associated torque is also estimated in section~\ref{sec:Couple} followed by a discussion in section~\ref{sec:Discussion}. Conclusions and implications are given in section~\ref{sec:Conclusion}.

\section{Off-centred dipole: static limit}
\label{sec:DipoleDecentre}

To study the effect of an off-centred dipole on various physical mechanisms occurring in the magnetosphere, finding exact analytical expressions for the electromagnetic field induced in vacuum by such rotating magnet is a good starting point. We explain in this section how to get these exact formulas.

\subsection{Radial magnetic field decomposition}

In almost all studies about pulsar magnetospheres and related emission properties, the magnetic moment~$\pmb{\mu} = \mu \, \bmath{m}$ is located at the centre of the star, $\bmath{m}$ being an unit vector. Although this is the simplest assumption, it is not necessarily the most reliable. In this paper, we assume that the magnetic moment is off-centred with respect to the stellar centre (supposed to correspond also to the origin of the spherical coordinate system) by an amount given by the displacement vector~$\mathbf{d}$. Modifying the classical vectorial formula for a dipole in vacuum \citep{2001elcl.book.....J} according to the shift of its centre, the off-centred magnetic dipole is given by
\begin{equation}
\label{eq:OffcentredDipole}
\mathbf{B} = \frac{B\,R^3}{\|\mathbf{r} - \mathbf{d}\|^3} \, \left[ \frac{3\,\bmath{m} \cdot (\mathbf{r} - \mathbf{d})}{\|\mathbf{r} - \mathbf{d}\|^2} \, (\mathbf{r} - \mathbf{d}) - \bmath{m} \right] \ .
\end{equation}
where $B$ is the magnetic field intensity at the equator. See also \cite{2014MNRAS.440.2519B} who started their study on the same ground. Such magnetic field is easily connected to the multipolar expansion. Indeed \cite{2015MNRAS.450..714P} has shown that in order to compute the exact radiating solution in vacuum, only the radial component of the magnetic field~$B^r = \bmath B \cdot \bmath{n}$ at the stellar surface is required to fully and explicitly get all components of the electric and magnetic field. $B^r$ is expanded into spherical harmonics~$Y_{\ell,m}$ and requires an expansion of expressions like $\|\mathbf{r} - \mathbf{d}\|^{-3}$ and $\|\mathbf{r} - \mathbf{d}\|^{-5}$ in terms of these spherical harmonics~$Y_{\ell,m}$. Let us outline the procedure.

Starting from the generating function for the Legendre polynomials \citep{2005mmp..book.....A} defined as
\begin{equation}
 \frac{1}{\sqrt{1-2\,x\,t+t^2}} = \sum_{\ell=0}^{+\infty} P_\ell(x) \, t^\ell
\end{equation}
and differentiating twice this expression with respect to the variable~$x$ we get
\begin{subequations}
\begin{align}
 \frac{1}{(1-2\,x\,t+t^2)^{3/2}} & = \sum_{\ell=1}^{+\infty} P'_\ell(x) \, t^{\ell-1} \\
 \frac{3}{(1-2\,x\,t+t^2)^{5/2}} & = \sum_{\ell=1}^{+\infty} P''_\ell(x) \, t^{\ell-2} .
\end{align}
\end{subequations}
To be more specific, we introduce a Cartesian coordinate system in which the position vector is represented by $\bmath{r} = r\,\bmath n$ with $\bmath n = (\sin\vartheta \, \cos\varphi, \sin\vartheta\,\sin\varphi, \cos\vartheta)$ the unit vector along the radial direction in spherical coordinates $(r, \vartheta, \varphi)$. Without loss of generality, it is always possible to bring back the dipole into the $(xOz)$~plane. Thus by assumption, the magnetic moment is located at $\bmath{d}=d\,(\sin\delta, 0, \cos \delta)$ and directed along the unit vector $\bmath m = (\sin\alpha\, \cos\beta, \sin\alpha\,\sin\beta, \cos\alpha)$. The angle between the position vector~$\bmath r$ and the displacement~$\bmath d$ is denoted by $\gamma$ and its cosinus is given by
\begin{equation}
 \cos \gamma = \cos \delta \, \cos \vartheta + \sin \delta \, \sin\vartheta \, \cos\varphi.
\end{equation}
To summarize the characteristics of the off-centred dipole, instead of specifying only the obliquity of the pulsar, that is one parameter denoted by $\chi$ for a centred dipole, we now have to determine three additional parameters so in total four parameters which are, see fig.~\ref{fig:Dipole}
\begin{itemize}
\item the obliquity~$\alpha$.
\item the out of meridional plane angle~$\beta$.
\item the distance~$d$ from the centre of the star.
\item the angle~$\delta$ between the rotation axis and the line joining the dipole to the centre.
\end{itemize}
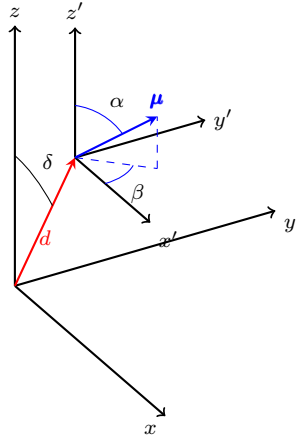
\begin{figure}
\centering


\tdplotsetmaincoords{60}{60}

\pgfmathsetmacro{\rvec}{.8}
\pgfmathsetmacro{\thetavec}{30}
\pgfmathsetmacro{\phivec}{60}

\begin{tikzpicture}[scale=4,tdplot_main_coords]

\coordinate (O) at (0,0,0);

\tdplotsetcoord{P}{\rvec}{\thetavec}{0}


\draw[thick,->] (0,0,0) -- (1,0,0) node[anchor=north east]{$x$};
\draw[thick,->] (0,0,0) -- (0,1,0) node[anchor=north west]{$y$};
\draw[thick,->] (0,0,0) -- (0,0,1) node[anchor=south]{$z$};

\draw[-stealth,thick,color=red] (O) -- (P) node [midway, below] {$d$} ;



\tdplotsetthetaplanecoords{0}

\tdplotdrawarc[tdplot_rotated_coords]{(0,0,0)}{0.5}{0}{\thetavec}{anchor=south west}{$\delta$}


\tdplotsetrotatedcoords{0}{0}{0}

\tdplotsetrotatedcoordsorigin{(P)}

\draw[thick,tdplot_rotated_coords,->] (0,0,0) -- (.5,0,0) node[anchor=north west]{$x'$};
\draw[thick,tdplot_rotated_coords,->] (0,0,0) -- (0,.5,0) node[anchor=west]{$y'$};
\draw[thick,tdplot_rotated_coords,->] (0,0,0) -- (0,0,.5) node[anchor=south]{$z'$};


\draw[-stealth,thick,color=blue,tdplot_rotated_coords] (0,0,0) -- (.2,.2,.2) node [above] {$\pmb{\mu}$} ;
\draw[dashed,color=blue,tdplot_rotated_coords] (0,0,0) -- (.2,.2,0);
\draw[dashed,color=blue,tdplot_rotated_coords] (.2,.2,0) -- (.2,.2,.2);

\tdplotdrawarc[tdplot_rotated_coords,color=blue]{(0,0,0)}{0.2}{0}{45}{anchor=north west,color=black}{$\beta$}

\tdplotsetrotatedthetaplanecoords{45}

\tdplotdrawarc[tdplot_rotated_coords,color=blue]{(0,0,0)}{0.2}{0}{55}{anchor=south west,color=black}{$\alpha$}

\end{tikzpicture}
\caption{Geometry of the decentred dipole showing the three important angles $\{\alpha, \beta, \delta\}$ and the distance $d$.}
\label{fig:Dipole}
\end{figure}
The centre of the magnetic dipole should obviously remain inside the neutron star surface i.e. the condition $d<R$ holds so that we can appropriately write it as $d=\epsilon\,R$ with $\epsilon\in[0,1[$. According to \cite{1964AJ.....69..567G} the $m$-th derivative of the Legendre polynomials, written as an $(m)$ exponent, can be developed onto Legendre polynomials themselves as
\begin{subequations}
\begin{align}
P_n^{(m)} & = \sum_{i=0}^{E[(n-m)/2]} c_i^{n,m} \, P_{n-m-2\,i} \\
c_i^{n,m} & = \frac{(2\,n - 1 - 2\,i)!!}{(2\,n - 2\,m + 1 - 2\,i)!!} \, (2\,n - 2\,m + 1 - 4\,i) \, \binom{m-1+i}{i}
\end{align}
\end{subequations}
where $E[n]$ represents the integer part of~$n$ and $\binom{m-1+i}{i}$ the binomial coefficients. Actually we will only require expressions for the first and second derivatives. For such cases, we can use the well known differential equations
\begin{subequations}
\begin{align}
 (1-x^2) \, P_n'(x) & = n \, P_{n-1}(x) - n \, x \, P_n(x) \\
 (1-x^2) \, P_n''(x) & = 2 \, x \, P_n'(x) - n \, (n+1) \, P_n(x) \ .
\end{align}
\end{subequations}
Moreover the addition theorem for spherical harmonics \citep{2005mmp..book.....A} stipulates that
\begin{equation}
P_\ell(\cos\gamma) = \frac{4\,\upi}{2\,\ell+1} \, \sum_{m=-\ell}^\ell Y_{\ell,m}(\vartheta,\varphi) \, Y_{\ell,m}(\delta,0)^* .
\end{equation}
The symbol $^*$ denotes the complex conjugate operation. Summarizing all the properties needed for the Legendre polynomials, we get the following results
\begin{subequations}
\begin{align}
 \frac{1}{\|\mathbf{r} - \mathbf{d}\|^3} & = \frac{1}{r^3} \, \sum_{\ell=1}^{+\infty} \left( \frac{d}{r} \right)^{\ell-1} \, P'_\ell(\cos \gamma) \\
 \frac{1}{\|\mathbf{r} - \mathbf{d}\|^5} & = \frac{1}{3\,r^5} \, \sum_{\ell=2}^{+\infty} \left( \frac{d}{r} \right)^{\ell-2} \, P''_\ell(\cos \gamma) \ .
\end{align}
\end{subequations}
To get the expansion of the magnetic field component~$B^r$ we develop its numerator into spherical harmonics such that
\begin{equation}
 \|\mathbf{r} - \mathbf{d}\|^5 \, B^r = \sum_{\ell} \sum_{m=-\ell}^{m=\ell} \mathcal{N}^r_{\ell,m} \, Y_{\ell,m}(\vartheta,\varphi)
\end{equation}
with the expansion coefficients explicitly given for the off-centred dipole field as
\begin{subequations}
\begin{align}
 \mathcal{N}^r_{0,0} & = - \frac{20\sqrt{\upi}}{3} \, r \, d \, ( \cos \alpha \, \cos \delta + \sin \alpha \, \sin \delta \, \cos \beta) \\
 \mathcal{N}^r_{1,0} & = \frac{\sqrt{\upi}}{3} \, [ \cos \alpha \, ( d^2 + 4 \, r^2 + 3 \, d^2 \, \cos 2\,\delta ) + 3 \, d^2 \, \sin \alpha \, \sin 2\,\delta \, \cos \beta) \\
 \mathcal{N}^r_{1,1} & = \frac{\sqrt{\upi}}{6} \, [ \cos \beta \, \sin \alpha \, ( - d^2 - 4 \, r^2 + 3 \, d^2 \, \cos 2\,\delta ) - 2 \, i \, ( d^2 - 2 \, r^2 ) \, \sin \alpha \, \sin \beta - 6 \, d^2 \, \cos \alpha \, \sin \delta \, \cos \delta ) \\
 \mathcal{N}^r_{2,0} & = - \frac{2}{3} \, \sqrt{\frac{\upi}{5}} \, r \, d \, ( 2 \, \cos \alpha \, \cos \delta - \sin \alpha \, \sin \delta \, \cos \beta ) \\
 \mathcal{N}^r_{2,1} & = - \frac{1}{2} \, \sqrt{\frac{\upi}{30}} \, i \, r \, d \, e^{-i\,(\alpha+\beta+\delta)} \, (-1 + e^{2\,i\,\alpha} - e^{i\,\beta} - e^{i\,(2\,\alpha+\beta)} - e^{2\,i\,\delta} + e^{2\,i\,(\alpha+\delta)} + e^{i\,(\beta+2\,\delta)} + e^{i\,(2\,\alpha+\beta+2\,\delta)} ) \\
 \mathcal{N}^r_{2,2} & = - \sqrt{\frac{2\,\upi}{15}} \, r \, d \, e^{-i\,\beta} \, \sin \alpha \, \sin \delta \ .
\end{align}
\end{subequations}
From the symmetry properties of the spherical harmonics with $m>0$ we have $Y_{\ell,-m} = (-1)^m \, Y^*_{\ell,m}$ thus the coefficients with negative mode numbers are deduced by the expression $\mathcal{N}^r_{\ell,-m}(r) = (-1)^m \, \mathcal{N}^r_{\ell,m}(r)^*$.

According to the well established theory of angular momentum in quantum mechanics, a product of spherical harmonics can be decomposed into single spherical harmonics by introducing the Clebsch-Gordan coefficients $\begin{bmatrix}
 \ell_1 & \ell_2 & \ell \\
 m_1 & m_2 & m
 \end{bmatrix}$
\citep{devanathan2006angular} such that
\begin{equation}
 Y_{\ell_1,m_1}(\vartheta,\varphi) \, Y_{\ell_2,m_2}(\vartheta,\varphi) = \sum_\ell \sqrt{\frac{(2\,\ell_1+1)\,(2\,\ell_2+1)}{4\,\upi\,(2\,\ell+1)}} \,
 \begin{bmatrix}
 \ell_1 & \ell_2 & \ell \\
 m_1 & m_2 & m
 \end{bmatrix}
 \,
 \begin{bmatrix}
 \ell_1 & \ell_2 & \ell \\
 0 & 0 & 0
 \end{bmatrix}
 \,
Y_{\ell,m}(\vartheta,\varphi)
\end{equation}
where the only non-vanishing Clebsch-Gordan coefficients are given by $m=m_1+m_2$ and $\ell = \ell_1 + \ell_2, \ell_1 + \ell_2 - 2, \ell_1 + \ell_2 - 4, \hdots ,|\ell_1 - \ell_2|$ \citep{cohen1973mecanique2}. Thus the expansion of the magnetic field component~$B^r$ onto spherical harmonics becomes
\begin{equation}
B^r = \sum_{\ell,m} b_{\ell,m}(\alpha, \beta, \delta, d) \, Y_{\ell, m}(\vartheta,\varphi)
\end{equation}
where $b_{\ell,m}(\alpha, \beta, \delta, d)$ are constant coefficients determined by the above procedure. Each term of the summation does only contain one spherical harmonic~$Y_{\ell, m}(\vartheta,\varphi)$ which depends on the spherical coordinates $\vartheta$ and $\varphi$. The coefficients associated to each $Y_{\ell, m}(\vartheta,\varphi)$ depend only on the particular geometry under study $(\alpha, \beta, \delta, d)$ and can be expanded in a power series of the small parameter~$\epsilon=d/R$. The number of multipole components used to approximate an off-centred dipole is linked to the order of expansion in~$\epsilon$. By identification with the divergencelessness expansion of the same field component
\begin{equation}
\label{eq:BrYlm}
 B^r = \sum_{\ell,m} - \frac{\sqrt{\ell\,(\ell+1)}}{r} \, f^{\rm B}_{\ell,m}(r) \, Y_{\ell,m}(\vartheta,\varphi)
\end{equation}
we get the ``potential functions''~$f^{\rm B}_{\ell,m}$ that fully and analytically solve the problem of the radiating off-centred dipole in vacuum. Unfortunately, this identification reveals not very handy to deal with to get the multipolar expansion of the off-centred dipole. Nevertheless it has the merit to show that the boundary condition imposed by $B^r$ on the surface induces an infinite series of spherical harmonics coefficients containing any positive power of $\epsilon$. As a practical application to this development, in the next paragraph we straightforwardly compute the expansion coefficients~$ f^{\rm B}_{\ell,m}(r)$ up to the third order in $\epsilon$. Physically, it means that we add corrections from the quadrupole~($\ell=2$), the hexapole~($\ell=3$) up to the octupole~($\ell=4$).

\subsection{Magnetic potential expansion}

A more elegant way to find the expansion coefficients of the radial magnetic field uses the magnetic potential $\mathbf{B}  = - \grad \Psi_M$ with the condition $\divg \mathbf{B} = 0$. $\mathbf{B}$ is then by definition curl-free. Thus the magnetic potential satisfies Laplace equation $\Delta \Psi_M = 0$. In the coordinate system attached to the dipole, setting the $z'$ axis along the magnetic moment~$\pmb{\mu}$, the potential $\Psi_M$ is simply
\begin{equation}
 \Psi_M = \sqrt{\frac{4\,\upi}{3}} \, B \, R^3 \, \frac{Y_{1,0}(\vartheta',\varphi')}{{r'}^2} = B \, R^3 \, \frac{\cos \vartheta'}{r'^2} \ .
\end{equation}
If the centre of the coordinate system is at $(0,0,z=a)$ then according to \cite{morse1953methodsvol2}
\begin{equation}
 \frac{Y_{\ell,m}(\vartheta', \varphi)}{{r'}^{\ell+1}} = \frac{1}{a^{\ell+1}} \, \sum_{\ell'=\ell}^{+\infty} \frac{1}{(\ell'-\ell)!} \, \sqrt{\frac{2\,\ell+1}{2\,\ell'+1} \, \frac{(\ell'-m)!\,(\ell'+m)!}{(\ell+m)! \, (\ell-m)!}} \, \left( \frac{a}{r} \right)^{\ell'+1} \, Y_{\ell',m}(\vartheta, \varphi) \ .
\end{equation}
The distance between the two origins is $OO'=d$ and the angles are related by 
\begin{equation}
r \, \cos \vartheta = d + r' \, \cos \vartheta'
\end{equation}
from which we deduce that 
\begin{equation}
 \cos \vartheta' = \frac{r \, \cos \vartheta - d}{r'} \ .
\end{equation}
\begin{figure}
\centering
\begin{tikzpicture}[]
\draw[->] (0,0) node [below] {$O$} -- (0,5) node [above] {$z'=z$};
\draw[->] (0,0) -- (3,0) node [right] {$x'=x$};
\draw[->] (0,0) -- (2,4) node [above] {$M$};
\draw[->] (0,2) node [left] {$O'$}-- (2,4) ;
\draw (0,1) arc (90:63:1) node [above] {$\vartheta$} ;
\draw (0,3) arc (90:73:3) node [above] {$\vartheta'$}  ;
\end{tikzpicture}
\caption{Translation of the coordinate system in spherical geometry.}
\label{fig:Translation}
\end{figure}
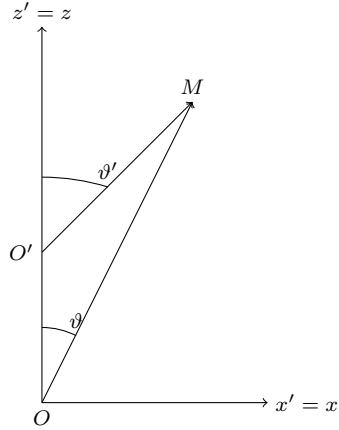
To bring the general dipole into the form above, that is oriented along the $z'$-axis, we perform rotations with appropriate Euler angles such that 
\begin{equation}
Y_{\ell,m}(\vartheta, \varphi) = \sum_{m'=-\ell}^{\ell} \mathcal{D}^\ell_{m',m}(\psi, \theta, \phi) \, Y_{\ell,m'}(\vartheta', \varphi').
\end{equation}
The $\mathcal{D}^\ell_{m',m}$ are known as Wigner rotation matrices and are described in many textbooks on quantum mechanics \citep{macrobert1967spherical, rose1995elementary}. The transformation combining a rotation, followed by a translation and finally by a inverse rotation was proposed by \cite{1979ApJS...41...75R}. This gives another mean to compute the coefficients $b_{\ell,m}(\alpha, \beta, \delta, d)$.

\subsection{Radial magnetic field series expansion}

The last and more direct approach to find solutions employs a series expansion of the same radial component of the magnetic field~$B^r$ in a power series of $d/r$ given by
\begin{equation}
 B^r(r,\vartheta,\varphi) = \sum_{n=0}^{+\infty} B_n^r(\vartheta,\varphi) \, \left(\frac{d}{r}\right)^n \ .
\end{equation}
Inverting relation eq.~(\ref{eq:BrYlm}) using the orthonormality of the $Y_{\ell,m}$, we get
\begin{equation}
f^{\rm B}_{\ell,m}(r) = - \frac{r}{\sqrt{\ell\,(\ell+1)}} \, \sum_{n=0}^{+\infty} \left(\frac{d}{r}\right)^n  \, \iint_{4\,\upi} B_n^r(\vartheta,\varphi) \, Y_{\ell,m}(\vartheta,\varphi)^* \, \sin\vartheta \, d\vartheta \, d\varphi
\end{equation}
Performing the series expansion and integrating in the full solid angle of $4\,\upi$~sr to third order in $d/r$, we get the potentials inside the star as
\begin{subequations}
\label{eq:fBlm}
\begin{align}
 f^{\rm B}_{1,0} & = - \sqrt{\frac{8\,\upi}{3}} \, \frac{B\,R^3}{r^2} \, \cos \alpha \\
 f^{\rm B}_{1,1} & =  \sqrt{\frac{16\,\upi}{3}} \, \frac{B\,R^3}{r^2} \, \sin \alpha \, e^{-i\,\beta} \\
 f^{\rm B}_{2,0} & = - \sqrt{\frac{6\,\upi}{5}} \, \frac{B\,R^3}{r^2} \, \frac{d}{r} \, ( 2 \, \cos \alpha \, \cos \delta - \sin \alpha \, \sin \delta \, \cos \beta ) \\
 f^{\rm B}_{2,1} & = 6 \, \sqrt{\frac{\upi}{5}} \, \frac{B\,R^3}{r^2} \, \frac{d}{r} \, ( \cos \alpha \, \sin \delta + \sin \alpha \, \cos \delta \, e^{-i\,\beta} ) \\
 f^{\rm B}_{2,2} & = - 6 \, \sqrt{\frac{\upi}{5}} \, \frac{B\,R^3}{r^2} \, \frac{d}{r} \, \sin \alpha \, \sin \delta \, e^{-i\,\beta} \\
 f^{\rm B}_{3,0} & = - \sqrt{\frac{3\,\upi}{7}} \, \frac{B\,R^3}{r^2} \, \left(\frac{d}{r}\right)^2 \, ( \cos \alpha \, ( 1 + 3 \, \cos 2\,\delta ) - 4 \, \sin \alpha \, \sin \delta \, \cos \beta \, \cos \delta ) \\
 f^{\rm B}_{3,1} & = \sqrt{\frac{\upi}{7}} \, \frac{B\,R^3}{r^2} \, \left(\frac{d}{r}\right)^2 \, ( \sin \alpha \, ( ( 2 + 6 \, \cos 2\,\delta ) \, e^{-i\,\beta} - 2 \, \sin^2 \delta \, e^{i\,\beta} )  \nonumber \\
 &  + 8 \, \cos \alpha \, \sin 2\,\delta ) \\
 f^{\rm B}_{3,2} & = - 2\, \sqrt{\frac{10\,\upi}{7}} \, \frac{B\,R^3}{r^2} \, \left(\frac{d}{r}\right)^2 \, \sin \delta \, ( \cos \alpha \, \sin \delta  + 2 \, \sin \alpha \, \cos \delta \, e^{-i\,\beta} ) \\
 f^{\rm B}_{3,3} & = 2\, \sqrt{\frac{15\,\upi}{7}} \, \frac{B\,R^3}{r^2} \, \left(\frac{d}{r}\right)^2 \, \sin \alpha \, \sin ^2 \delta \, e^{-i\,\beta} \\
 f^{\rm B}_{4,0} & = \sqrt{\frac{5\,\upi}{24}} \, \frac{B\,R^3}{r^2} \, \left(\frac{d}{r}\right)^3 \, ( 3 \, \sin \alpha \, \cos \beta ( \sin \delta + 5 \sin 3 \, \delta ) - 4 \, \cos \alpha \, ( 3 \, \cos \delta + 5 \, \cos 3\,\delta) ) \\
 f^{\rm B}_{4,1} & = \frac{5\,\sqrt{\upi}}{24} \, \frac{B\,R^3}{r^2} \, \left(\frac{d}{r}\right)^3 \, ( 3 \, \cos \alpha \, ( \sin \delta + 5 \sin 3 \, \delta ) + 4 \, \cos \delta \, \sin \alpha \, ( 5 \, \cos 2\,\delta - 1 )  \, e^{-i\,\beta} \nonumber \\
 & - 6\, \sin \alpha \, \sin \delta \, \sin 2\delta  \, e^{i\,\beta}  ) \\
 f^{\rm B}_{4,2} & = - \frac{5}{6} \, \sqrt{\frac{\upi}{2}} \, \frac{B\,R^3}{r^2} \, \left(\frac{d}{r}\right)^3 \, \sin \delta \, ( 6 \, \cos \alpha \, \sin 2\,\delta + \sin \alpha \, ( ( 5 + 7 \, \cos2\delta -\sin^2\delta) \, e^{-i\,\beta} - \sin^2\delta \, e^{i\,\beta} ) \\
 f^{\rm B}_{4,3} & = \frac{5}{6} \, \sqrt{7\,\upi} \, \frac{B\,R^3}{r^2} \, \left(\frac{d}{r}\right)^3 \, \sin^2 \delta \, ( \cos \alpha \, \sin \delta + 3 \, \sin \alpha \, \cos \delta \, e^{-i\,\beta} ) \\
 f^{\rm B}_{4,4} & = - \frac{5}{3} \, \sqrt{\frac{7\,\upi}{2}} \, \frac{B\,R^3}{r^2} \, \left(\frac{d}{r}\right)^3 \, \sin \alpha \, \sin^3 \delta \, e^{-i\,\beta} .
\end{align}
\end{subequations}
The functions~$f^{\rm B}_{\ell,m}$ are defined everywhere inside the star $r\leqslant R$. To compute the external solution in vacuum $r\geqslant R$, we only need these functions evaluated at $r=R$. In that case $d/r$ can be replaced by the small parameter $\epsilon=d/R$. From the symmetry properties of the spherical harmonics, the coefficients with negative azimuthal numbers~$m<0$ are retrieved from the relation $f^{\rm B}_{\ell,-m}(r) = (-1)^m \, f^{\rm B}_{\ell,m}(r)^*$. Note that these expressions are valid only inside the neutron star $r\leqslant R$ where we assume that the magnetic dipole is not perturbed by any current, it is frozen into the superconducting plasma. The $\ell=1$ mode corresponds to the centred dipole inclined with respect to the rotation axis with an angle~$\alpha$ but also rotated along the same rotation axis with an angle~$\beta$ i.e. a phase shift according to the substitution $\varphi \rightarrow \varphi-\beta$ depicted by the complex factor $e^{-i\,\beta}$. This is expected from the prescription of the magnetic moment~$\mathbf{m}$ that does not lie in the $\varphi=0$ plane but in this new rotated plane. To this lowest approximation, there is no dependence on the small parameter~$\epsilon$. If no multipolar corrections $\ell>1$ are added, we retrieve the standard centred dipole. The first perturbations come from the quadrupole term~$\ell=2$, corresponding to the coefficients in front of the first power of~$\epsilon$. All three azimuthal modes~$m=0$, $m=1$ and $m=2$ contribute to the corrections with complex coefficients depending on the particular geometry studied. The $m=0$ mode as for the centred dipole part is a real number and thus the correction of $(\ell,m)=(2,0)$ is proportional to the centred axisymmetric quadrupole~$(\ell,m)=(2,0)$ mode. For the $m=1$ mode, it is a linear combination of the pure quadrupole and an azimuthally shifted quadrupole by making the substitution~$\varphi \rightarrow \varphi - \beta$ as expected from the complex factor $e^{-i\,\beta}$. Finally, the $m=2$ mode introduces a correction with another azimuthally shifted quadrupole with the substitution~$2\,\varphi \rightarrow 2\,\varphi - \beta$. The same discussion applies to the hexapole and octupole, we do not detail them here. We next explain how to compute the electromagnetic field outside the star from these boundary conditions.

\section{Radiating off-centred dipole}
\label{sec:RadiatingDipole}

So far we only expanded the static offset dipole into multipole components. Outside the star, radiation is permitted in vacuum and can also be described by an expansion into multipoles.

\subsection{Exact solution}

According to standard textbooks such as \cite{2001elcl.book.....J}, the exact solution for the electromagnetic field in vacuum outside the star is judiciously expanded into vector spherical harmonics. Following our slightly different notation introduced in \cite{2015MNRAS.450..714P} and also in \cite{2013MNRAS.433..986P} for the general-relativistic extension, the exact electromagnetic field solution satisfying the divergencelessness constrain for both the electric (displacement) field~$\mathbf{D}$ and the magnetic field~$\mathbf{B}$ is (we recall that the electric displacement field $\mathbf{D}$ is related to the electric field $\mathbf{E}$ by the linear relation $\mathbf{D} = \varepsilon_0\,\mathbf{E}$)
\begin{subequations}
\label{eq:Constantes}
\begin{align}
  \label{eq:Solution_Generale_div_0_D}
  \mathbf{D}(r,\vartheta,\varphi,t) = & \sum_{\ell=1}^\infty \rot [a^{\rm D}_{\ell,0} \, \frac{\mathbf{\Phi}_{\ell,0}}{r^{\ell+1}}] + \\
  & \sum_{\ell=1}^\infty\sum_{m=-\ell,m\neq0}^\ell  \left( \rot [a^{\rm D}_{\ell,m} \, h_\ell^{(1)}(k_m\,r) \, \mathbf{\Phi}_{\ell,m}]  + i \, \varepsilon_0 \, m \, \Omega \, a^{\rm B}_{\ell,m} \, h_\ell^{(1)}(k_m\,r) \, \mathbf{\Phi}_{\ell,m} \right) \, e^{-i\,m\,\Omega\,t} \nonumber \\
  \label{eq:Solution_Generale_div_0_B}
  \mathbf{B}(r,\vartheta,\varphi,t) = & \sum_{\ell=1}^\infty \rot [a^{\rm B}_{\ell,0} \, \frac{\mathbf{\Phi}_{\ell,0}}{r^{\ell+1}}] + \\
  & \sum_{\ell=1}^\infty\sum_{m=-\ell,m\neq0}^\ell  \left( \rot [a^{\rm B}_{\ell,m} \, h_\ell^{(1)}(k_m\,r) \, \mathbf{\Phi}_{\ell,m}]  - i \, \mu_0 \, m \, \Omega \, a^{\rm D}_{\ell,m} \, h_\ell^{(1)}(k_m\,r) \, \mathbf{\Phi}_{\ell,m} \right) \, e^{-i\,m\,\Omega\,t} \nonumber
\end{align}
\end{subequations}
where $\{a^{\rm D}_{\ell,m}, a^{\rm B}_{\ell,m}\}$ are constants depending on the boundary conditions imposed on the stellar surface. We recall that these constants are given for an asymmetric mode $m>0$ by
\begin{subequations}
\begin{align}
\label{eq:aBlm}
 a^{\rm B}_{\ell,m} & = \frac{f^{\rm B}_{\ell,m}(R)}{h_\ell^{(1)}(k_m\,R)}
\end{align}
and for the axisymmetric case~$m=0$ by
\begin{align}
\label{eq:aBl0}
 a^{\rm B}_{\ell,0} = R^{\ell+1} \, f^{\rm B}_{\ell,0}(R).
\end{align}
Specializing to a unique multipole, the two non-vanishing electric field coefficients $a^{\rm D}_{\ell,m}$ are for an asymmetric mode $m>0$
\label{eq:aDlm}
\begin{align}
 a^{\rm D}_{\ell+1,m} \, \left.\partial_r ( r \, h_{\ell+1}^{(1)}(k_m\,r))\right|_{r=R} & = \varepsilon_0 \, R \, \Omega \,  \sqrt{\ell\,(\ell+2)} \, J_{\ell+1,m} \, f^{\rm B}_{\ell,m}(R) \\
 a^{\rm D}_{\ell-1,m} \, \left.\partial_r ( r \, h_{\ell-1}^{(1)}(k_m\,r))\right|_{r=R} & = - \varepsilon_0 \, R \, \Omega \, \sqrt{(\ell-1)\,(\ell+1)} \, J_{\ell,m} \, f^{\rm B}_{\ell,m}(R)
 \end{align}
where we introduced the numbers~$J_{\ell,m} = \sqrt{\frac{\ell^2-m^2}{4\,\ell^2-1}}$. For $\ell=1$ only one solution exists. For the axisymmetric case $m=0$ we find
\begin{align}
\label{eq:aDl0}
 (\ell+1) \, a^{\rm D}_{\ell+1,0} & = - \varepsilon_0 \, R^{\ell+3} \, \Omega \,  \sqrt{\ell\,(\ell+2)} \, J_{\ell+1,0} \, f^{\rm B}_{\ell,0}(R) \\
 (\ell-1) \, a^{\rm D}_{\ell-1,0} & =   \varepsilon_0 \, R^{\ell+1} \, \Omega \, \sqrt{(\ell-1)\,(\ell+1)} \, J_{\ell,0} \, f^{\rm B}_{\ell,0}(R).
 \end{align}
\end{subequations}
Clearly, the constant of integration in equations~(\ref{eq:Constantes}) are fully determined by the coefficients~$f^{\rm B}_{\ell,m}(R)$, defined on the neutron star surface, as claimed above. To summarize, we are able to compute analytically any solution to Maxwell equations in vacuum once the $f^{\rm B}_{\ell,m}(R)$ are known on the surface. In this paper, the $f^{\rm B}_{\ell,m}(R)$ are constrained by the off-centred dipole and given by eq.~(\ref{eq:fBlm}) for the few first terms. However any kind of magnetic field topology can be prescribed on the surface, multipoles being possibly dominant with respect to the dipole. The potentials~$f^{\rm B}_{\ell,m}$ are always determined according to eq.~(\ref{eq:BrYlm}).

In the far zone, where wave emission dominates the electromagnetic field, the magnetic and electric part are written respectively as
\begin{subequations}
\begin{align}
  \label{eq:Solution_Asymptot_B}
  \mathbf{B}_{\rm w} & = \sum_{\ell\geq1,m\neq0} - (-i)^{\ell} \, \frac{e^{i\,m\,\Omega\,(r/c - t)}}{r} \,
    \left( a^{\rm B}_{\ell,m} \, \mathbf{\Psi}_{\ell,m} + \mu_0 \, c \, a^{\rm D}_{\ell,m} \, \mathbf{\Phi}_{\ell,m} \right) \\
  \label{eq:Solution_Asymptot_D}
  \mathbf{D}_{\rm w} & = \sum_{\ell\geq1,m\neq0} - (-i)^{\ell} \, \frac{e^{i\,m\,\Omega\,(r/c - t)}}{r} \,  \left( a^{\rm D}_{\ell,m} \, \mathbf{\Psi}_{\ell,m} - \varepsilon_0 \, c \, a^{\rm B}_{\ell,m} \, \mathbf{\Phi}_{\ell,m} \right)  \\
  \label{eq:OndePlane}
  = & \varepsilon_0 \, c \, \mathbf{B}_{\rm w} \wedge \mathbf{n}.
\end{align}
\end{subequations}
The last equality shows the plane wave nature at large distances $r\gg\rlight$ with the typical $1/r$ decrease in amplitude. These asymptotic values are very useful to compute the spindown luminosity and the recoil induced by the asymmetric energy flux.

\subsection{The off-centred dipole expansion}

Going through the algebraic manipulations of the electromagnetic field for an off-centred dipole, it can be shown that the magnetic field as well as the electric field are decomposed up to the quadrupole component such that
\begin{multline}
\label{eq:DipoleQuardupole}
 \mathbf{F}^{\rm off} = \mathbf{F}^{\rm dip}(\psi \rightarrow \psi-\beta) + 
 \epsilon \, \left[ ( 2 \, \cos \alpha \, \cos \delta - \sin \alpha \, \sin \delta \, \cos \beta ) \, \mathbf{F}_{m=0}^{\rm quad}(\psi) \right. \\
 + \cos\alpha \, \sin\delta \, \mathbf{F}_{m=1}^{\rm quad}(\psi) + 
 \sin\alpha \, \cos\delta \, \mathbf{F}_{m=1}^{\rm quad}(\psi \rightarrow \psi-\beta)\\
\left. + \sin \alpha \, \sin\delta \, \mathbf{F}_{m=2}^{\rm quad}(2\,\psi \rightarrow 2\,\psi-\beta) \right]
\end{multline}
where the field~$\mathbf{F}$ can be replaced by the electric displacement field~$\mathbf{D}$ or the magnetic field~$\mathbf{B}$. The meaning of the superscripts should be obvious, it refers to the dipole $^{\rm dip}$ or quadrupole $^{\rm quad}$ perturbations and the subscripts indicate the azimuthal mode to consider for the specified coefficient. The normalization of the quadrupole field is arbitrary but we choose it to get the weights indicated in equation~(\ref{eq:DipoleQuardupole}). See appendix~\ref{app:C} for details about exact expressions. 
The change $\psi \rightarrow \psi_\beta =\psi-\beta$ in the dipole corresponds to a simple shift in the azimuthal direction as expected from the orientation of the magnetic moment with respect to the $(xOz)$ plane. It reflects in the electric and magnetic part of the field in a similar manner.

\begin{figure}
\centering
\input{offset_aligned_dipole.tex}
\caption{A sample of magnetic field lines for the offset aligned dipole with $R=0.1\,\rlight$ and $\epsilon=0.2$ are shown in red solid line. The centred aligned dipole is shown in blue dashed line for comparison.}
\label{fig:OffsetAligned}
\end{figure}
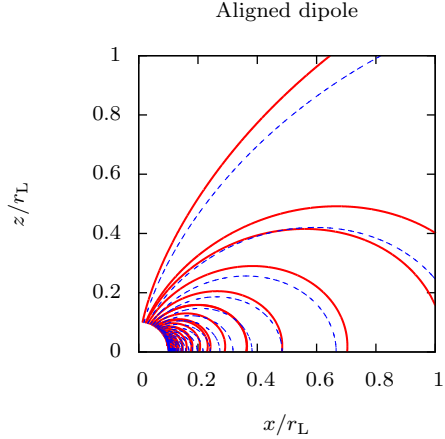
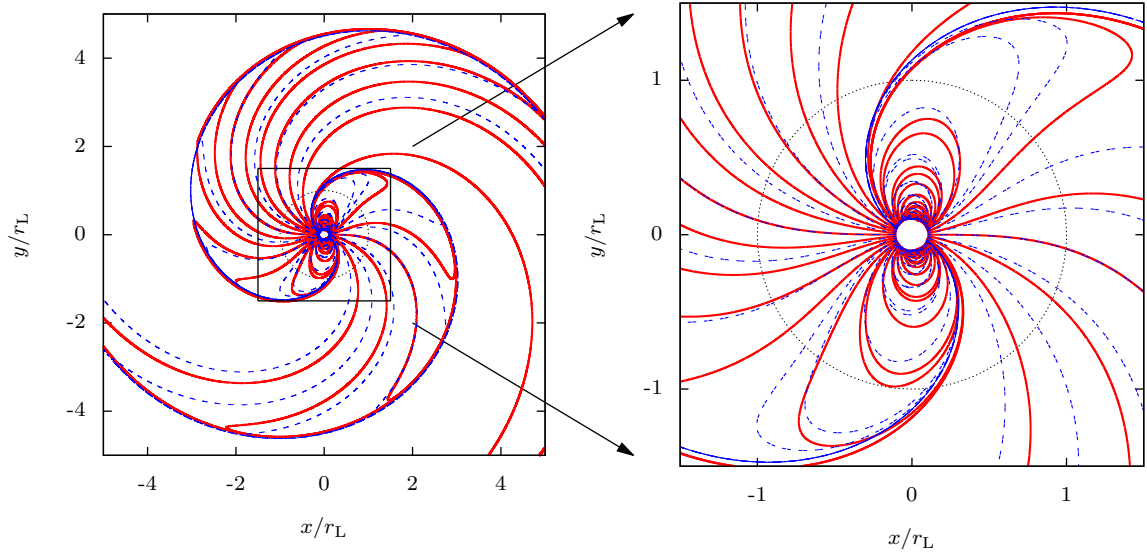
\begin{figure}
\centering
\input{offset_perp_dipole_zoom_b0.tex}
\caption{A sample of magnetic field lines for the offset perpendicular dipole with $R=0.1\,\rlight$ and $\epsilon=0.2$ are shown in red solid line for $\beta=0\degr$. The centred perpendicular dipole is shown in blue dashed line. Zoom into the light-cylinder on the right plot.}
\label{fig:OffsetPerp}
\end{figure}

Some special cases of off-centred dipoles are shown in subsequent figures including these quadrupole corrections. The offset aligned dipole with $R=0.1\,\rlight$ and $\epsilon=0.2$ is shown in fig.~\ref{fig:OffsetAligned} in red solid lines. It is instructive to compare it to the centred aligned dipole as given by the blue dashed line in the same figure. We only shown one quarter of field lines but the reader should convince himself that it corresponds simply to a translation of all field lines in the $z$ direction by a distance $d=\epsilon\,R$ (to the accuracy of the quadrupole perturbations).
The offset perpendicular dipole with $R=0.1\,\rlight$ and $\epsilon=0.2$ is shown in fig.~\ref{fig:OffsetPerp} in red solid lines for $\beta=0\degr$. Comparisons are made with respect to the centred perpendicular dipole as given by the blue dashed line in the same figure. Here also the reader should convince himself that at least in the quasi-static near zone it corresponds to a translation of all field lines in the $x$ direction by a distance $d=\epsilon\,R$ (again to the accuracy of the quadrupole perturbations). Of particular interest is the two arm spiral singular magnetic field line that remains open. For the centred dipole, the two arms perfectly overlap when a rotation of $\upi$ is applied to one of it. However, the two spiral arms are no more symmetric with respect to this phase shift when $\epsilon>0$. Such asymmetries should also reflect into the geometry of the striped wind outside the light-cylinder. The two pulses emanating from this wind would not look symmetric any more. The current sheet north-south symmetry is broken and allows more flexibility in adjusting the high-energy light curves as seem by Fermi/LAT \citep{2013ApJS..208...17A}. Implications of an offset dipole remain perceptible at large distance because multipolar components propagate as plan waves with amplitude decreasing like $1/r$ exactly like the dipole part. The respective weights between all the components are easily deduced from the expansion coefficients $f^{\rm B}_{\ell,m}(R)$, eq.~(\ref{eq:fBlm}). Forthcoming force-free simulations of the pulsar magnetosphere should quantitatively demonstrate this fact. However we stress that multipolar effects at the light-cylinder are only discernible for the fastest pulsars, those with millisecond period. Indeed, assuming that all multipole fields are of similar strength at the stellar surface, knowing that static multipoles decrease like $B\,(R/r)^{\ell+2}$ the ratio between a multipole~$B_\ell$ and the dipole~$B_{\rm dip}$ at the light-cylinder is 
\begin{equation}
 \frac{B_{\ell}}{B_{\rm dip}} \approx \left(\frac{R}{\rlight}\right)^{\ell-1} \ .
\end{equation}
Already for the quadrupole the relative strength drops to $R/\rlight$ which becomes very small for normal pulsars. Because the wave zone starts in the vicinity of the light-cylinder and the amplitude of a multipolar wave is directly related to the field at the light-cylinder, the impact of multipoles outside the light-cylinder is drastically reduced except for the fastest millisecond pulsars.

The right panel of fig.~\ref{fig:OffsetPerp} zooms into the deep magnetosphere where it can be seen that the singular open field line is shifted to the right at both poles.
The translation in the positive~$x$ direction is emphasized in this zoom of fig.~\ref{fig:OffsetPerp}. Compared to the two blue spiral arms, it is clear that the red arms are both approximately shifted to the right by an amount~$d=\epsilon\,R$.

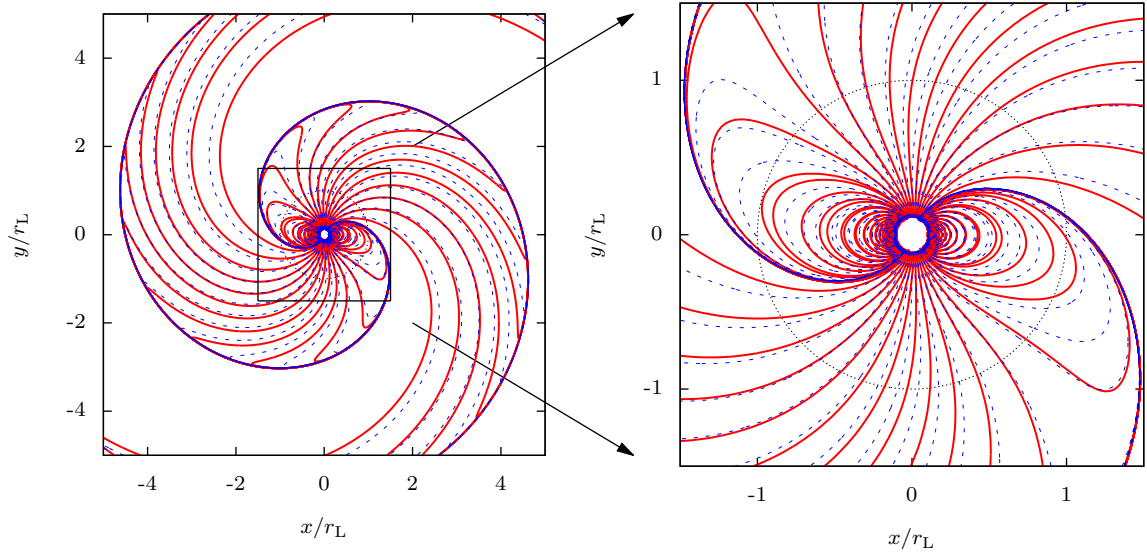
\begin{figure}
\centering
\input{offset_perp_dipole_zoom_b1.tex}
\caption{A sample of magnetic field lines for the offset perpendicular dipole with $R=0.1\,\rlight$ and $\epsilon=0.2$ are shown in red solid line for $\beta=90\degr$. The centred perpendicular dipole is shown in blue dashed line. Zoom into the light-cylinder on the right plot.}
\label{fig:OffsetPerpB1}
\end{figure}

A second offset perpendicular dipole with $R=0.1\,\rlight$ and $\epsilon=0.2$ is shown in fig.~\ref{fig:OffsetPerpB1} in red solid lines for $\beta=90\degr$. In that case at large distances, the two spiral arms always overlap. The right panel of fig.~\ref{fig:OffsetPerpB1} zooms into the magnetosphere where we see that the singular open field line almost joins the centred case already at the light cylinder at both poles.
To pin point the straightforward and illuminating consequences of these quadrupolar corrections, we compute the spin-down luminosity, the electromagnetic recoil force, the braking index and the torque in the next section.

\section{Spin-down luminosity}
\label{sec:Luminosite}

An interesting diagnostic to quantify the implications of an off-centred dipole is to compute the spin-down luminosity of the magnet depending on the geometrical input parameters. First we perform the calculation for a point dipole and compare our estimate with previous results given in the literature and then take into account the finite size of the star.

\subsection{General comments}

We recall that for the orthogonal centred point dipole in vacuum, the spindown luminosity is
\begin{equation}
\label{eq:SpindownDipole}
L_{\rm dip} = \frac{8\,\upi\,\Omega^4\,B^2\,R^6}{3\,\mu_0\,c^3} \ .
\end{equation} 
In the wave zone the radial component of the Poynting vector simplifies into
\begin{subequations}
\begin{align}
 S^r & = \frac{c}{2\,\mu_0} \, \mathbf{B} \cdot \mathbf{B}^* = \frac{c}{2\,\mu_0} \, \sum_{\ell\geq1,m\neq0,\ell'} i^{\ell'-\ell} \, \frac{e^{i\,(m-m')\,\Omega\,(r/c - t)}}{r^2} \\
 & ( a^{\rm B}_{\ell,m} \, \mathbf{\Psi}_{\ell,m} + \mu_0 \, c \, a^{\rm D}_{\ell,m} \, \mathbf{\Phi}_{\ell,m} ) \, ( a^{\rm{B}^*}_{\ell',m'} \, \mathbf{\Psi}_{\ell',m'}^* + \mu_0 \, c \, a^{\rm{D}^*}_{\ell',m'} \, \mathbf{\Phi}_{\ell',m'}^* ) \ .
\end{align}
\end{subequations}
Using the orthonormality condition of the vector spherical harmonics, we get the spindown luminosity as 
\begin{equation}
 L = \iint S^r \, r^2 \, d\Omega = \frac{c}{2\,\mu_0} \, \sum_{\ell\geq1,m\neq0}  \left( |a^{\rm B}_{\ell,m}|^2 + \mu_0^2 \, c^2 \, |a^{\rm D}_{\ell,m}|^2 \right) \ .
\end{equation}
We use this expression to compute the Poynting flux for the off-centred dipole in the most general case. In the same vein, the force along the rotational axis is deduced by projecting along the $z$-axis to get 
\begin{subequations}
\begin{align}
 F & = \frac{1}{c} \, \iint S^r \, r^2 \, \cos\vartheta \, d\Omega \\
 & = \frac{1}{2\,\mu_0} \, \sum_{\ell,\ell'\geq1,m\neq0} i^{\ell'-\ell} \, \left( ( a^{\rm B}_{\ell,m} \, a^{\rm{B}^*}_{\ell',m} + \mu_0^2 \, c^2 \, a^{\rm D}_{\ell,m} \, a^{\rm{D}^*}_{\ell',m} ) \, I_1 + \mu_0 \, c \, ( a^{\rm B}_{\ell,m} \, a^{\rm{D}^*}_{\ell',m} - a^{\rm{B}^*}_{\ell',m} \, a^{\rm D}_{\ell,m} ) \, I_3 \right)
\end{align}
\end{subequations}
where the integrals~$I_1,I_3$ are defined in the appendix~\ref{app:B}. All the interesting features of the decentred dipole or of any magnetic field topology are enclosed in the constant coefficients $a^{\rm B}_{\ell,m}$ and $a^{\rm D}_{\ell,m}$.

\subsection{Point dipole}

The Poynting flux for a point dipole is given in \cite{2001elcl.book.....J}. But for a periodic signal it is more convenient to represent it in terms of a Fourier series rather than a Fourier transform such that the radiated power at the harmonic frequency $\omega = m\,\Omega$ is given by
\begin{equation}
\label{eq:dIsdOmega}
\left.\frac{dI}{d\Omega}\right|_{\omega = m\,\Omega} = \frac{\mu_0\,\mu}{4\,\upi} \, \frac{\omega^4}{2\,\upi\,c^3} \left| \frac{\sqrt{2\,\upi}}{T} \, \int_{0}^{T} \bmath{n} \wedge \bmath{m}(t) \, e^{-i\,m\,\Omega\,( t - \bmath{n} \cdot \bmath{d}(t)/c)} \, dt \right|^2
\end{equation}
where $T=2\,\upi/\Omega$ is the rotation period of the neutron star and $\mu$ the magnetic moment. See the appendix~\ref{app:A} for the convention used to compute the Fourier transform. Because the dipole motion is periodic, we expect Dirac distributions at harmonic frequencies of the fundamental frequency~$\Omega$. Thus it is equivalent to compute the Fourier series instead of the Fourier transform. The correspondence between both transforms is discussed in appendix~\ref{app:A}. 

Let us briefly go through the outline of the computation. The cross product between the position vector and the magnetic moment gives
\begin{subequations}
\begin{align}
\bmath{n} \wedge \bmath{m}(t) & = ( \cos \alpha \, \sin \vartheta \, \sin \varphi - \sin \alpha \, \cos \vartheta \, \sin (\Omega\,t+\beta) ) \, \ex \nonumber \\
& + ( \sin \alpha \, \cos \vartheta \, \cos (\Omega\,t+\beta) - \cos \alpha \, \sin \vartheta \, \cos \varphi ) \, \ey \nonumber \\
& + \sin \alpha \, \sin \vartheta \, \sin(\Omega\,t+\beta-\varphi) \, \ez 
\end{align}
\end{subequations}
and the argument in the retarded time of the complex exponential in eq.~(\ref{eq:dIsdOmega}) is
\begin{equation}
\bmath{n} \cdot \bmath{d}(t) = d \, ( \cos \delta \, \cos \vartheta + \sin \delta \, \sin \vartheta \, \cos (\Omega\,t - \varphi) ) \ .
\end{equation}
Because of rotational symmetry along the $\ez$ axis, without loss of generality, we can set $\varphi=\upi/2$. Let us compute the integrals of each component in eq.~(\ref{eq:dIsdOmega}). Starting with the $\ex$ component, making the change of variable $x=\Omega\,t$ and introducing two more auxiliary variables (that have nothing to do with Cartesian coordinates)
\begin{subequations}
\begin{align}
 y & = \frac{\Omega \, d \, \cos \delta \, \cos \vartheta}{c} \\
 z & = \frac{\Omega \, d \, \sin \delta \, \sin \vartheta}{c}
\end{align}
\end{subequations}
the projection of eq.~(\ref{eq:dIsdOmega}) along $\ex$ gives
\begin{subequations}
\begin{align}
 & \frac{1}{T} \, \int_{0}^{T} \bmath{n} \wedge \bmath{m}(t) \, e^{-i\,m\,\Omega\,( t - \bmath{n} \cdot \bmath{d}(t)/c)} \, dt \cdot \ex \nonumber \\
 & = \frac{1}{2\,\upi} \, \int_{-\upi}^{+\upi} ( \cos \alpha \, \sin \vartheta - \sin \alpha \, \cos \vartheta \, \sin (x+\beta) ) \, e^{-i\,m\,( x - y - z \, \sin x)} \, dx \nonumber \\
 & = e^{i\,m\,y} \, \left[ ( \cos \alpha \, \sin \vartheta - \frac{\sin \alpha \, \sin \beta \, \cos \vartheta}{z}) \, J_m( m \, z) + i \, \sin \alpha \, \cos \beta \, \cos \vartheta \, J_m'( m \, z) \right] \ .
\end{align}
The same kind of projection applies to the other components along $\ey$
\begin{align}
 \frac{1}{T} \, \int_{0}^{T} \bmath{n} \wedge \bmath{m}(t) \, e^{-i\,m\,\Omega\,( t - \bmath{n} \cdot \bmath{d}(t)/c)} \, dt \cdot \ey & = e^{i\,m\,y} \, \sin \alpha \, \cos \vartheta \, \left[ \cos \beta \, \frac{J_m \left( m \, z \right)}{z} + i \, \sin \beta \, J_m' (m \, z) \right]
\end{align}
and along $\ez$
\begin{align}
 \frac{1}{T} \, \int_{0}^{T} \bmath{n} \wedge \bmath{m}(t) \, e^{-i\,m\,\Omega\,( t - \bmath{n} \cdot \bmath{d}(t)/c)} \, dt \cdot \ez & = - e^{i\,m\,y} \, \sin \alpha \, \sin \vartheta \, \left[ \cos \beta \, \frac{J_m \left( m \, z \right)}{z} + i \, \sin \beta \, J_m' (m \, z) \right] \ .
\end{align}
\end{subequations}
The spindown luminosity associated to the $m$-th harmonic ($\omega = m\,\Omega)$ is therefore
\begin{subequations}
\begin{align}
 \left.\frac{dI}{d\Omega}\right|_{\omega = m\,\Omega} & = \frac{\mu_0\,\mu^2}{4\,\upi} \, \frac{m^4\,\Omega^4}{2\,\upi\,c^3} \, \left[ \left( \cos \alpha \, \sin \vartheta - \frac{\sin \alpha \, \sin \beta \, \cos \vartheta}{z} \right)^2 \, J_m^2( m \, z) \right. \nonumber \\
 & \left. + \sin^2 \alpha \, \cos^2 \beta \, \cos^2 \vartheta \, J_m'^2( m \, z) + \sin^2 \alpha \, \left( \cos^2 \beta \, \frac{J_m^2 ( m \, z )}{z^2} + \sin^2 \beta \, J_m'^2 (m \, z) \right) \right] \ .
\end{align}
\end{subequations}
The total luminosity is obtained by integration over the solid angle where contributions from all harmonics are included
\begin{equation}
\label{eq:LuminositePonctuelle}
 L = \sum_{m=1}^{+\infty} \iint \left.\frac{dI}{d\Omega}\right|_{\omega = m\,\Omega} \, d\Omega \ .
\end{equation}
As in \cite{1975ApJ...201..447H} we can also compute the net force as another integration like
\begin{equation}
\label{eq:ForcePonctuelle}
 F = \frac{1}{c} \, \sum_{m=1}^{+\infty} \iint \left.\frac{dI}{d\Omega}\right|_{\omega = m\,\Omega} \, \cos \vartheta\, d\Omega \ .
\end{equation}
Integrations are done according to well known integral representations of Bessel functions as reminded in appendix~\ref{app:A}. Contrary to the centred dipole, the star now radiates energy in multiple integer of its rotation frequency~$\Omega$. To the lowest order in $\epsilon$, to compare with \cite{1975ApJ...201..447H}, we set $\delta=\upi/2$ such that $d$ represents simply the distance from the rotation axis. For the spindown luminosity, we find for the fundamental frequency and its first harmonic
\begin{subequations}
\begin{align}
 L_{m=1} & = L_{\rm dip} \, \left[ \left( 1 + \frac{1}{10} \, \epsilon^2 \, a^2 \, ( \cos2\,\beta - 3 ) \right) \, \sin^2\alpha + \frac{2}{5} \, \epsilon^2 \, a^2 \, \cos^2\alpha \right] \\
 L_{m=2} & = \frac{48}{5} \, L_{\rm dip} \, \epsilon^2 \, a^2 \, \sin^2\alpha \ .
\end{align}
\end{subequations}
For the electromagnetic recoil, the fundamental frequency and its first harmonic contribution gives
\begin{subequations}
\begin{align}
 F_{m=1} & = \frac{1}{5} \, \frac{L_{\rm dip}}{c} \, \epsilon \, a \, \cos\alpha \, \sin\alpha \, \sin\beta \\
 F_{m=2} & = \frac{64}{35} \, \frac{L_{\rm dip}}{c} \, \epsilon^3 \, a^3 \, \cos\alpha \, \sin\alpha \, \sin\beta \ .
\end{align}
\end{subequations}
We recast the term $\Omega\,d/c$ as a product of the two fundamental parameters $\epsilon\ll1$ and $a=R/\rlight\ll1$ such that $\Omega\,d/c = \epsilon \, a$ and therefore remains very small. To compare our results with those of \cite{1975ApJ...201..447H}, we need to make the following identifications
\begin{subequations}
\begin{align}
 \mu_\rho & = \mu \, \sin \alpha \, \cos \beta \\
 \mu_\phi & = \mu \, \sin \alpha \, \sin \beta \\
 \mu_z & = \mu \, \cos \alpha \ .
\end{align}
\end{subequations}
Our expressions for $m=1$ coincide with those of \cite{1975ApJ...201..447H} except for a correction in the spindown luminosity where we found an additional term of second order in $\epsilon^2\,a^2$ in front of $\sin^2\alpha$. Indeed, according to their estimates, we have in our notation
\begin{subequations}
\begin{align}
 L_{m=1} & = L_{\rm dip} \, \left[ \sin^2\alpha + \frac{2}{5} \, \epsilon^2 \, a^2 \, \cos^2\alpha \right] \\
 F_{m=1} & = \frac{1}{5} \, \frac{L_{\rm dip}}{c} \, \epsilon \, a \, \cos\alpha \, \sin\alpha \, \sin\beta \ .
\end{align}
\end{subequations}
Corrections to the spindown luminosity by inspection of the quadrupolar formula~$L_{m=2}$ leads to irrelevant increase for realistic parameters $a$ and $\epsilon$. Nevertheless, as pointed out by \cite{1975ApJ...201..447H}, the asymmetric pattern of the Poynting flux leads to a meaningful force kicking the star at its birth. The relevant component of the force acting is dipolar $F_{m=1}$ and the quadrupolar pattern $F_{m=2}$ is already reduced by a factor $a^2\,\epsilon^2$. 

What about the effect of the neutron star finite size on these estimates? Thanks to our analytical approximation of an off-centred dipole it is possible to take into account the finite size of the star through the inner boundary conditions at $r=R$. This is done in the next paragraph.

\subsection{Finite radius dipole}

For an off-centred dipole anchored in a star, the point dipole approximation can not be used because it would mean that the star rotates along an axis which is not passing through the star itself. Moreover, the boundary conditions on the stellar surface lead to discrepancy between both approximations. This reflects into the retardation time for deriving the fields. For a point dipole, retardation starts from the location of the dipole itself but for the neutron star, it should start from the surface not from the centre. Therefore both situations are not strictly equivalent and deviation in the respective results of both approaches are observed although only within reasonable bounds.

Performing the same integrals as in eq.~(\ref{eq:LuminositePonctuelle}) for spindown and in eq.~(\ref{eq:ForcePonctuelle}) for the recoil force, we get to the same order of accuracy and setting $\delta=90\degr$ to allow direct comparison with previous works
\begin{subequations}
\label{eq:Lm1e2}
\begin{align}
 L_{m=1} & = L_{\rm dip} \, \left[ \left( 1 - a^2 \right) \, \sin^2\alpha + \frac{24}{25} \, a^2 \, \epsilon^2 \, \cos^2\alpha \right] \\
 L_{m=2} & = \frac{48}{5} \, L_{\rm dip} \, a^2\,\epsilon^2 \, \sin^2\alpha \ .
\end{align}
\end{subequations}
We recognize the typical $( 1 - a^2) \, \sin^2\alpha$ dependence of the luminosity for a centred finite size dipole when $\epsilon=0$. As for the point dipole, corrections are weak and of second order in both $a$ and $\epsilon$, expressed as a product $a^2\,\epsilon^2$. For the point dipole however there are no corrections including powers of~$a$ terms solely. The electromagnetic recoil is given by
\begin{subequations}
\begin{align}
 F_{m=1} & = \frac{6}{5} \, \frac{L_{\rm dip}}{c} \, a\,\epsilon \, \cos\alpha \, \sin\alpha \, \sin\beta \\
 F_{m=2} & = \frac{256}{105} \, \frac{L_{\rm dip}}{c} \, a^3\,\epsilon^3 \, \cos\alpha \, \sin\alpha \, \sin\beta \ .
\end{align}
\end{subequations}
These values of the spindown luminosity and the recoil force are larger than those for the point dipole even up to a factor~6 for $m=1$ recoil mode and less for $m=2$ recoil mode. The increase can be attributed to the contribution from the electric part of the wave which is absent in case of zero radius star where neither electric current nor electric charge can brake the star.

Finally we discuss the impact on the Poynting flux with respect to the azimuth~$\vartheta$. Two examples are particularly relevant in the case of a perpendicular rotator with $\alpha=90\degr$. The first with $\delta=90\degr$ and $\beta=0\degr$
\begin{equation}
 \frac{L(\vartheta)}{L_{\rm dip}} = \frac{3}{8} \, (1 + \cos^2 \vartheta ) + a^2 \, \left( 6 \, \epsilon^2 \, \sin^2 \vartheta \, (1 + \cos^2 \vartheta ) - \frac{1}{2} \right)
\end{equation}
giving a total spindown of
\begin{equation}
 \frac{L}{L_{\rm dip}} = 1 - a^2 \, \left( 1 - \frac{48}{5} \, \epsilon^2 \right) 
\end{equation}
consistent with eq.~(\ref{eq:Lm1e2}). The same expressions hold for the second case with $\delta=90\degr$ and $\beta=90\degr$. Corrections to the centred dipole are only of the order $a^2\,\epsilon^2$ so we do not expect significant deviation from a simple centred dipole. If spindown is affected by the shift, so does the braking index we now compute explicitly.

\subsection{Braking index}

Knowing the total spindown luminosity we can estimate the braking index following the derivative of the luminosity with respect to the spin as shown in \cite{2015MNRAS.450..714P}. We remind that the braking index is computed according to
\begin{equation}
\label{eq:IndiceFreinage}
 n = \frac{a}{L} \, \frac{dL}{da} - 1 = \frac{d\ln L}{d\ln a} - 1 \ .
\end{equation}
As for the spindown and recoil, to keep expressions in a readable format, we only show results for special geometries of the off-centred dipole to get insight into the behaviour of the braking index with respect to the geometrical parameters~$\{\alpha, \beta, \delta, d\}$.

For the point dipole approximation, the braking index is given to lowest order with a lengthy expression. Here we give some interesting limiting case for an aligned rotator~$\alpha=0\degr$ and a perpendicular rotator~$\alpha=90\degr$ to second order in $a$ and $\epsilon$. These read
\begin{subequations}
\begin{align}
 n(\alpha=0\degr) & = 5 + \frac{1}{7} \, a^2 \, \epsilon^2 \, ( 97 - 95 \, \cos2\,\delta) \\
 n(\alpha=90\degr) & = 3 + \frac{1}{5} \, a^2 \, \epsilon^2 \, ( 93 + \cos2\,\beta ) \, \sin^2\delta \ .
\end{align}
\end{subequations}
For the finite size off-centred dipole we find
\begin{subequations}
\begin{align}
 n(\alpha=0\degr) & = 5 + \frac{19}{12} \, a^2 \\
 n(\alpha=90\degr) & = 3 - \frac{2}{25} \, a^2 \, ( 25 + \epsilon^2 \, ( 27 \, \cos2\,\delta - 33 ) ) \ .
\end{align}
\end{subequations}
For an off-centred but aligned dipole, the braking index is close to the pure quadrupole case of $n=5$ within correction of the order $a^2$. As the obliquity increases, the braking index falls down to the pure dipole case $n=3$ within correction of the order $a^2$. The transition occurs more or less sharply depending on the peculiar geometry and could for instance explain the high braking index pulsar observed by \cite{2016ApJ...819L..16A}. Such value can in principle be induced by an off-centred dipole because the braking index~$n$ varies between 3 and 5 depending on the particular geometry. We will come back to this example in more detail in the discussion of  sec~\ref{sec:Discussion}.

\section{Torque exerted on the star}
\label{sec:Couple}

In addition to the spindown luminosity, the torque exerted by the electromagnetic field on the star via its interaction on the crust and interior is also of interest to understand the slow variation of the neutron star geometry. Although the torque responsible for the spindown is well established, there seems to exist a controversy about the anomalous torque liable for the secular evolution of the magnetic moment orientation with respect to the rotation axis. Note also that such torque cannot be estimated in the point dipole approximation because of lack of an interacting volume. So \cite{2000MNRAS.313..217M} computed the exact torque on a finite size dipole derived from the electromagnetic stress-energy tensor. However, there is a debate on the exact value of this anomalous torque, its strength depending on the way to compute it. One way, often quoted in papers such as in \cite{1970ApL.....5...21M} and still recently by \cite{2015MNRAS.453.3540A} starts indeed from the electromagnetic stress-energy tensor, giving the torque as
\begin{equation}
 \frac{dJ_i}{dt} = - \iint \varepsilon_{ikl} \, X^k \, T_{\rm em}^{lm} \, dS_m
\end{equation}
where $X^k$ is the position vector, $dS_m$ the surface element vector, $\varepsilon_{ikl}$ the Levi-Civita symbol and $T_{\rm em}^{lm}$ the electromagnetic stress-energy tensor. Written explicitly for the electromagnetic field it becomes
\begin{equation}
\label{eq:Couple_Tem}
 \mathbf K_1 = \frac{d \mathbf J}{dt} = \varepsilon_0 \, r^3 \, \iint \mathbf n \wedge ( (\mathbf E \cdot \mathbf n ) \, \mathbf E + c^2 \, (\mathbf B \cdot \mathbf n ) \, \mathbf B ) \, d\Omega \ .
\end{equation}
This relation is derived under restrictive assumptions not necessarily met in the context of neutron star electrodynamics. Another but we think more reliable way to compute the torque starts from the Laplace and electric forces applied on the neutron star surface only. Indeed, in the perfect conductor limit, the total volume force vanishes as explained by \cite{2014PhyU...57..799B}. So this torque reads
\begin{equation}
\label{eq:Couple_Laplace}
 \mathbf K_2 = \frac{d \mathbf J}{dt} = \iint \mathbf r \wedge ( \sigma_{\rm s} \, \mathbf E + \mathbf i_{\rm s} \wedge \mathbf B ) \, dS = r^3 \, \iint \mathbf ( \sigma_{\rm s} \, \mathbf n \wedge \mathbf E + ( \mathbf B \cdot \mathbf n ) \, \mathbf i_{\rm s} ) \, d\Omega 
\end{equation}
where $\sigma_{\rm s} = \varepsilon_0 \, [\mathbf{E}] \cdot \mathbf{n}$ represents the surface charge density and $\mathbf i_{\rm s} = [\mathbf{B}] \wedge \mathbf{n} / \mu_0$ the surface current density. The notation $[\mathbf{F}]$ means the jump of the vector field $\mathbf{F}$ across the layer. Both expressions eq.~(\ref{eq:Couple_Tem}) and eq.~(\ref{eq:Couple_Laplace}) are not equivalent, see \cite{2014PhyU...57..799B} for a discussion on different results and their discrepancy. Their calculation of the torque applied on the neutron star differs from that computed through the stress-energy tensor because the flux of stress-energy tensor through the neutron star surface describes the total angular momentum losses, which is given by $dL_{\rm matter}/dt + dL_{\rm field}/dt$. \cite{2014PhyU...57..799B} emphasized that $dL_{\rm field}/dt$ does not vanish inside the star. Adopting another point of view, the total angular momentum losses could be considered, without separating the matter and electromagnetic field contributions. In other words, the magnetic field contribution could be added to the effective inertia tensor of the neutron star. Such approach introducing a effective neutron star inertia tensor has been used by \cite{2015MNRAS.451.2564G} to look for its precession motion. 
See also \cite{2015MNRAS.451..695Z} who introduced magnetic dipole and quadrupole inertia tensors to compute again some electromagnetic torques and precession of the neutron star.

\subsection{Centred dipole}

In order to quantitatively point out the difference in the results, we compute the torques deduced in both manner starting with the centred dipole. We separate the magnetic force contribution from the electric force contribution, that is, terms containing only $\mathbf{E}$ and terms containing only $\mathbf{B}$. With the first method we get, irrespective of the inner magnetisation because the solution outside is insensitive to this magnetisation, the Cartesian components of the magnetic $\mathbf{K}_1^{\rm B}$ and electric $\mathbf{K}_1^{\rm E}$ torque as
\begin{subequations}
\begin{align}
 K_{1,x}^{\rm B} & = \frac{L_{\rm dip}}{\Omega} \, \frac{\left(a^6 - 9 \, a^4 + 90 \right) \, \sin 2\,\alpha }{5 \, \left(a^2+1\right) \, \left(a^6 - 3 \, a^4 + 36 \right)} \\
 K_{1,y}^{\rm B} & = \frac{L_{\rm dip}}{\Omega} \, \frac{\left(5 \, a^6 - 24 \, a^4 - 63 \, a^2 + 126 \right) \, \sin \alpha \, \cos \alpha}{5 \, a \, \left(a^2+1\right) \, \left(a^6 - 3 \, a^4 + 36 \right)} \\
 K_{1,z}^{\rm B} & = - \frac{L_{\rm dip}}{\Omega} \, \frac{4 \, \left(2 \, a^6 - 3 \, a^4 + 45 \right) \, \sin^2 \alpha }{5 \, \left(a^2+1\right) \, \left(a^6 - 3 \, a^4 + 36 \right)} \\
 K_{1,x}^{\rm E} & = \frac{L_{\rm dip}}{\Omega} \, \frac{3 \, a^4 \, \sin 2\,\alpha }{5 \, \left(a^6-3 \, a^4+36\right)} \\
 K_{1,y}^{\rm E} & = - \frac{L_{\rm dip}}{\Omega} \, \frac{\left(a^6-3 \, a^4-9 \, a^2-18\right) \, \sin 2\,\alpha }{5 \, a\,  \left(a^6-3\,  a^4+36\right)} \ .
\end{align}
\end{subequations}
In case of a perfectly centred dipole, we retrieve the exact expressions derived from the Hankel functions. These results agree with those given in the appendix of \cite{2000MNRAS.313..217M}. 
In the second method, we have to distinguish between a uniform and a dipolar interior magnetisation. In the former case, the torques are given for the uniform magnetization for the Cartesian components of the magnetic $\mathbf{K}_2^{\rm B}$ and electric $\mathbf{K}_2^{\rm E}$ torque as
\begin{subequations}
\begin{align}
 K_{2,x}^{\rm B(uni)} & = K_{1,x}^{\rm B} \\
 K_{2,y}^{\rm B(uni)} & = K_{1,y}^{\rm B} \\ 
 K_{2,z}^{\rm B(uni)} & = K_{1,z}^{\rm B} \\
 K_{2,x}^{\rm E(uni)} & = K_{1,x}^{\rm E} \\
 K_{2,y}^{\rm E(uni)} & = - \frac{L_{\rm dip}}{\Omega} \,\frac{\left(4 a^6-12 a^4-9 a^2+90\right) \sin 2\,\alpha }{5 a \left(a^6-3 a^4+36\right)}
\end{align}
\end{subequations}
and for the dipolar magnetization
\begin{subequations}
\begin{align}
 K_{2,x}^{\rm B(dip)} & = K_{1,x}^{\rm B} \\
 K_{2,y}^{\rm B(dip)} & = K_{1,y}^{\rm B} \\
 K_{2,z}^{\rm B(dip)} & = K_{1,z}^{\rm B} \\
 K_{2,x}^{\rm E(dip)} & = K_{1,x}^{\rm E} \\
 K_{2,y}^{\rm E(dip)} & = \frac{L_{\rm dip}}{\Omega} \frac{\left(a^6-3 a^4+18 a^2+144\right) \sin \alpha 
   \cos \alpha }{5 a \left(a^6-3 a^4+36\right)} \ .
\end{align}
\end{subequations}
Consequently the magnetic part of the torque is the same in all three cases $\mathbf{K}_{1}^{\rm B} = \mathbf{K}_{2}^{\rm B(uni)} = \mathbf{K}_{2}^{\rm B(dip)}$. It is useful to get the lowest order corrections in~$a$ to the torque in these cases. We find
\begin{subequations}
\begin{align}
 K_{1,x}^{\rm B} & \approx \frac{L_{\rm dip}}{\Omega} \, (1-a^2) \, \cos\alpha \, \sin\alpha \\
 K_{1,y}^{\rm B} & \approx \frac{L_{\rm dip}}{\Omega} \, \left( \frac{7}{10\,a} - \frac{21}{20} \, a \right) \, \cos\alpha \, \sin\alpha \\
 K_{1,z}^{\rm B} & \approx - \frac{L_{\rm dip}}{\Omega} \, (1-a^2) \, \sin^2\alpha \\
 K_{1,y}^{\rm E} & \approx \frac{L_{\rm dip}}{\Omega} \, \left( \frac{a}{10} + \frac{1}{5\,a} \right) \, \cos\alpha \, \sin\alpha \\
 K_{2,y}^{\rm E(uni)} & \approx \frac{L_{\rm dip}}{\Omega} \, \left( \frac{a}{10} -\frac{1}{a} \right) \, \cos\alpha \, \sin\alpha \\
 K_{2,y}^{\rm E(dip)} & \approx \frac{L_{\rm dip}}{\Omega} \, \left( \frac{a}{10} + \frac{4}{5\,a} \right) \, \cos\alpha \, \sin\alpha \ .
\end{align}
\end{subequations}
Note that we neglected the $K_{2,x}^{\rm E}$ component because it is of leading order~$a^4$ and the $z$~component of the electric field torque always vanishes $K_{1,z}^{\rm E} = K_{2,z}^{\rm E(uni)} = K_{2,z}^{\rm E(dip)} = 0$.
Both ways to compute the full torque agree on several components related to the spindown but not on the anomalous torque as claimed previously. The torque exerted by the magnetic part is independent of the method, stress-energy tensor or Lorentz force. Actually there is also no distinction to be made between uniform and dipolar magnetization. The same remark applies to the $x$ component of the electrically induced torque. Discrepancies are only observed for the $y$ component of the electric torque~$K_y^{\rm E}$ which is just the anomalous torque.

A summary of various results about the anomalous torque is presented in table~\ref{tab:AnomalousTorque} showing the constant value~$\xi$ defined to lowest order in~$a$ by
\begin{equation}
 K_y = \frac{3}{2} \, \xi \, \frac{L_{\rm dip}}{a\,\Omega} \, \cos\alpha \, \sin\alpha \ .
\end{equation}
Results depend on the stress-energy tensor or Lorentz force approach used and on neglecting or not the electric part in the latter case.
\begin{table}
\centering
\begin{tabular}{cc}
\hline
$\xi$ & Reference \\
\hline
$1$ & \cite{1970ApJ...160L..11G, 1970ApJ...159L..81D} \\
$3/5$ & \cite{2000MNRAS.313..217M} \\
$1/5$ & \cite{1985ApJ...299..706G} \\
$2/3$ & \cite{1969ApJ...157.1395O} \\
$0$ & \cite{2005pnsr.conf...27I} \\
\hline
\end{tabular}
\caption{Summary of anomalous torques constant~$\xi$.}
\label{tab:AnomalousTorque}
\end{table}

\subsection{Off-centred dipole}

Now we consider the lowest approximation including quadrupolar terms. For the off-centred dipole we only consider a dipolar magnetization such that to lowest order the torque from the stress-energy tensor method is 
\begin{subequations}
\begin{align}
 K_{1,x}^{\rm B} & = \frac{L_{\rm dip}}{\Omega} \, \frac{\sin 2\,\alpha  \left(140 \, a \, \left(a^2 \, \left(93 \, \epsilon^2-25\right)+25\right) \cos \beta -\left(147 \, a^2 \left(66 \, \epsilon^2 - 25 \right)+3240 \, \epsilon^2 + 2450\right) \, \sin \beta \right)}{7000 \, a} \\
 K_{1,y}^{\rm B} & = \frac{L_{\rm dip}}{\Omega} \, \frac{\sin \alpha \, \cos \alpha}{3500 \, a} \\
 & \left(70 \, a \left(a^2 \, \left(207 \, \epsilon^2-50\right)+50\right) \, \sin \beta + \left(a^2\,  \left(8328 \, \epsilon^2-3675\right)+2160 \, \epsilon^2+2450\right) \, \cos \beta \right) \nonumber \\
 K_{1,z}^{\rm B} & = \frac{L_{\rm dip}}{\Omega} \, \frac{1}{50} \, ( a^2 \, \left(\left(216 \, \epsilon^2-25\right) \, \cos 2\,\alpha -264 \, \epsilon ^2+25\right)-\sin^2\alpha ) \\
 K_{1,x}^{\rm E} & = \frac{L_{\rm dip}}{\Omega} \, \frac{\sin 2\,\alpha \, \left(1890 \, a^3 \, \epsilon^2 \, \cos \beta +\left(a^2 \, \left(2187 \, \epsilon^2-175\right)+5 \, \left(243\,  \epsilon ^2-70\right)\right) \, \sin \beta \right)}{3500 \, a} \\
 K_{1,y}^{\rm E} & = \frac{L_{\rm dip}}{\Omega} \, \frac{\sin 2\,\alpha \,  \left(2835 \, a^3 \, \epsilon^2 \, \sin \beta +\left(-7 \, a^2 \, \left(468 \, \epsilon^2-25\right)-810 \, \epsilon^2 + 350\right)\, \cos \beta \right)}{3500 \, a} \ .
\end{align}
\end{subequations}
If instead we use the Lorentz force we get
\begin{subequations}
\begin{align}
 K_{2,x}^{\rm B} & = K_{1,x}^{\rm B} \\
 K_{2,y}^{\rm B} & = K_{1,y}^{\rm B} \\
 K_{2,z}^{\rm B} & = K_{1,z}^{\rm B} \\
 K_{2,x}^{\rm E} & = \frac{L_{\rm dip}}{\Omega} \, \frac{\sin 2\,\alpha  \left(1890 a^3 \epsilon ^2 \cos \beta +\left(a^2 \left(2187 \epsilon ^2-175\right)-10 \left(81 \epsilon
   ^2+140\right)\right) \sin \beta \right)}{3500 a} \\
 K_{2,y}^{\rm E} & = \frac{L_{\rm dip}}{\Omega} \, \frac{\sin 2\,\alpha  \left(2835 a^3 \epsilon ^2 \sin \beta +\left(20 \left(27 \epsilon ^2+70\right)-7 a^2 \left(468 \epsilon
   ^2-25\right)\right) \cos \beta \right)}{3500 a} \ .
\end{align}
\end{subequations}
In these calculations, we did not take into account the electromagnetic perturbations induced by the corotation of the stellar volume charge as done by \cite{2014PhyU...57..799B}. In the limit of $\epsilon\rightarrow0$ and $\beta\rightarrow0$ we retrieve the expressions of the previous section for the centred dipole as it should.

\section{Discussion}
\label{sec:Discussion}

\begin{figure}
\centering
\input{indice_freinage.tex}
\caption{Variation in the braking index for $a=10^{-3}$, $\epsilon=\{0.01,0.02,0.05,0.1,0.2\}$ as shown in the legend, $\delta=90\degr$ and $\beta$ is irrelevant.}
\label{fig:IndiceFreinage}
\end{figure}
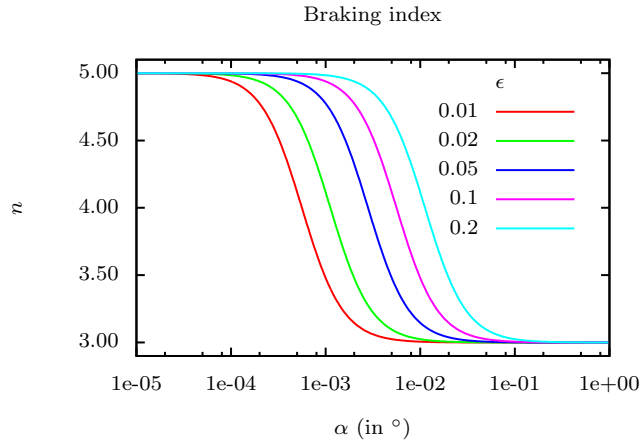

Recently, a pulsar with high braking index of $n=3.15$ and period 206~ms has been discovered by \cite{2016ApJ...819L..16A} which represents the first ever pulsar known to have a braking index larger than~3. Indeed, all other pulsars with measured braking index possess a value less than this fiducial number~3, see \cite{2016ApJ...823...97C} for a recent summary of all measured braking indices. Without any a priori knowledge of the secular evolution of all pulsar parameters such as magnetic field, electric equivalent radius (the parameter denoted by $x_\nu$ in \cite{1997MNRAS.288.1049M}), moment of inertia, inclination angle, the braking index according to the vacuum point magnetodipole losses is
\begin{equation}
 n = 3 + \frac{\Omega}{\dot\Omega} \, \left[ 2 \, \frac{\dot B}{B} + 2 \, \dot\chi \, \cot \chi + 6 \, \frac{\dot R}{R} - \frac{\dot I}{I} \right] .
\end{equation}
Thus formally, braking indices above or below the fiducial~$n=3$ value can be found depending on the physical justification of the change in the parameters entering the dipole losses. The minimal assumptions claims that there is no magnetic field evolution $\dot B = 0$, no change in the equivalent radius $ \dot R = 0$ and no moment of inertia variation $\dot I = 0$. Then the vacuum electromagnetic torque implies $\Omega\,\cos\chi=\textrm{cste}$ leading to
\begin{equation}
n = 3 + 2 \, \cot^2 \chi > 3 \ .
\end{equation}
Thus the braking index of PSR~J1640-4631 would require $\chi \approx 90\degr$ that is an almost orthogonal rotator that contradicts pulse profile observations as reminded by \cite{2016ApJ...823...34E} who proposed a plasma filled magnetosphere model to explain this high braking index. In our off-centred dipole, to the contrary we need an almost aligned pulsar to explain such braking index. Indeed, assuming again for simplicity that $\delta=90\degr$ and a moderate offset of~ $\epsilon=0.2$ we get for $a=10^{-3}$ corresponding to a 206~ms period an angle of  $\alpha=0.04\degr$ ridiculously small. An example is shown in fig.~\ref{fig:IndiceFreinage} where the transition from 5 to 3 is evident when the obliquity~$\alpha$ is increased. This happens already for very small inclinations.
Alternative origins for pulsar braking indices have been proposed in the literature. However, as long as the magnetosphere electrodynamics remains fuzzy it seems pretentious to get firm answers. 

\begin{figure}
\centering
\input{spirale_in_zoom_b0.tex}
\caption{Comparison of the two arm spirals of the centred dipole (blue dashed line) and the off-centred dipole (red solid line) for $a=0.1$, $\beta=0\degr$ and $\epsilon=0.2$.}
\label{fig:SpiraleB0}
\end{figure}
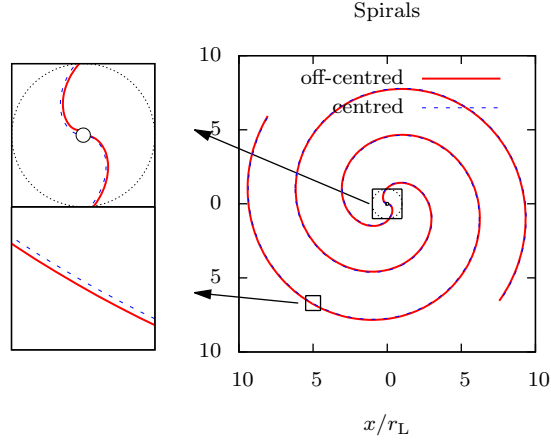

\begin{figure}
\centering
\input{spirale_in_zoom_b1.tex}
\caption{Comparison of the two arm spirals of the centred dipole (blue dashed line) and the off-centred dipole (red solid line) for $a=0.1$, $\beta=90\degr$ and $\epsilon=0.2$.}
\label{fig:SpiraleB1}
\end{figure}
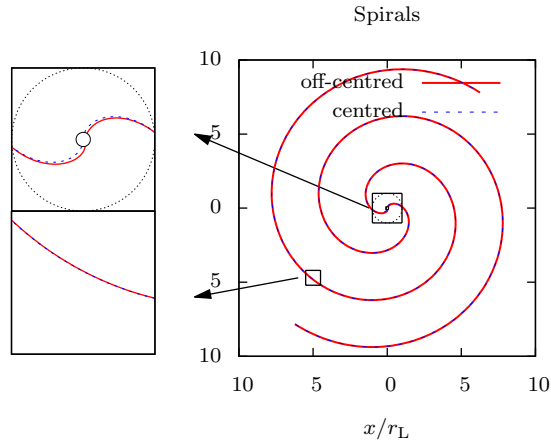

More promising predictions or indirect evidence for off-centred dipole are given by multi-wavelength light-curves fitting from radio up to Mev/GeV. It is often claimed in the literature that the multipolar components decay faster than the dipole and that at large distances, only the dipole magnetic field remains perceptible. Such assertions, whereas right for a {\it static field}, are wrong for a time dependent electromagnetic field. The physical reason for such a behaviour is extremely simple. When multipoles are present on the surface and the star is rotating, it emits electromagnetic waves containing the fundamental mode of a dipole but also harmonic frequencies. The amplitude of these multipole waves decay in the far zone only as~$1/r$ as it should for spherical waves evolving in a three dimensional space. The dipole field behaves obviously in the same way. This means, at least in principle, that we can probe multipolar fields even at large distances, outside the light cylinder. Figure~\ref{fig:SpiraleB0} shows indeed the offset perpendicular rotator for $a=0.1$, $\alpha=90\degr$, $\beta=0\degr$, $\delta=90\degr$ and $\epsilon=0.2$. We recognize the two arm spiral for the centred dipole in blue dashed lines. The off-centred dipole keeps the same geometric structure except for a small shift to the right (at this particular time of rotation phase) according to the displacement vector $\mathbf d$. Both spirals do not overlap, neither inside the light cylinder nor outside in the far wave zone. This shift is clearly seen in the zoom of the upper left inset close to the neutron star surface. The lower left inset shows the phase delay between both spirals at large distances. Asymptotically, the phase lag converges to a constant value depending on the off centred dipole geometry.
A second example is shown in figure~\ref{fig:SpiraleB1} where the only change is the switching to $\beta=90\degr$. In that case, at large distances, the off-centred structure joins the centred dipole spiral arm. There is no phase lag perceptible for $r\gg\rlight$. Therefore, the imprint of multipolar fields for a distant observer is very geometry dependent.
The asymmetries in the wave zone would also reflect into the structure of plasma filled magnetosphere such as the striped wind when plasma back reaction is taken into account. Asymmetries in the high energy light-curves of the striped wind model are related to this new freedom in the magnetic topology. From careful analysis of MeV/GeV pulse profiles, it is possible to constrain the geometry of the quadrupolar component at large distance $r\gg\rlight$ and then trace back its weight with respect to the dipole inside the magnetosphere.

The geometry of the spiral structure is conveniently investigated in the equatorial plane for an orthogonal rotator by the polar equation for the two arm spiral given by
\begin{equation}
\label{eq:SpiraleTheorique}
 \varphi = \Omega \, t - k\,(r-R) + \psi(r,a,\epsilon) + \varphi_0 \ .
\end{equation}
$\varphi_0$ is an arbitrary constant to fix orientation at time $t=0$ and $\psi(r,a,\epsilon)$ depicts the deviation from a Archimedean spiral ($d\psi/dr=0$). At large distance the function $\psi(r,a,\epsilon)$ converges to a constant value denoted by~$\psi_\infty$. Geometrically, this implies that the spiral tends towards a perfect Archimedean spiral. For our discussion on implications about pulsed emission, two other important regions are the light-cylinder with value $\psi_{\rm L}$ and the stellar surface with value $\psi_*$. Mathematically we then define them by
\begin{subequations}
\begin{align}
\psi_* & = \psi(R,a,\epsilon) \\
\psi_{\rm L} & = \psi(\rlight,a,\epsilon) \\
\psi_\infty & = \psi(\infty,a,\epsilon) \ .
\end{align}
\end{subequations}
These constants are shown in table~\ref{tab:Psi} for several parameters $(a,\epsilon)$.
\begin{table}
\centering
\begin{tabular}{c|ccc|ccc}
\hline
$a$ & \multicolumn{3}{|c|}{0.1} & \multicolumn{3}{|c|}{0.2} \\
\hline
$\epsilon$ & $\psi_*$ & $\psi_L$ & $\psi_\infty$ & $\psi_*$ & $\psi_L$ & $\psi_\infty$\\
\hline
\hline
0.0 & 2.94/-0.20 & 2.58/-0.57 & 3.04/-0.10 & 2.75/-0.39 & 2.47/-0.67 & 2.94/-0.20 \\
0.1 & 2.81/-0.16 & 2.55/-0.54 & 3.02/-0.08 & 2.55/-0.32 & 2.42/-0.63 & 2.90/-0.16 \\
0.2 & 2.34/-0.15 & 2.53/-0.52 & 3.00/-0.06 & 2.19/-0.29 & 2.37/-0.58 & 2.87/-0.12 \\
\hline
\end{tabular}
\begin{tabular}{c|ccc|ccc}
\hline
$a$ & \multicolumn{3}{|c|}{0.1} & \multicolumn{3}{|c|}{0.2} \\
\hline
$\epsilon$ & $\psi_*$ & $\psi_L$ & $\psi_\infty$ & $\psi_*$ & $\psi_L$ & $\psi_\infty$\\
\hline
\hline
0.0 & 1.37/4.51 & 1.00/4.15 & 1.47/4.61 & 1.18/4.32 & 0.90/4.04 & 1.37/4.51 \\
0.1 & 1.14/4.80 & 1.00/4.15 & 1.47/4.61 & 0.98/4.62 & 0.89/4.05 & 1.37/4.52 \\
0.2 & 1.02/4.99 & 1.00/4.15 & 1.47/4.61 & 0.89/4.84 & 0.89/4.06 & 1.37/4.52 \\
\hline
\end{tabular}
\caption{The constants~$\psi_*, \psi_L, \psi_\infty$ for $\beta=0\degr$ (upper table) and $\beta=90\degr$ (lower table).}
\label{tab:Psi}
\end{table}
For large distances, the phase shift between the stellar centred and offset dipole becomes constant. In the example shown in the figures these phase delays are $\Delta \psi_1<0$ for one arm and $\Delta \psi_2>0$ for the other arm. Note the change in sign responsible for the asymmetric two arm spiral in red solid line. The constant $\psi_\infty$ to estimate the phase lag between several configurations for the perpendicular rotator is to lowest order assuming $\beta=0\degr$
\begin{equation}
 \psi_\infty = n \, \upi - a \, ( 1 + 2 \, \epsilon ) \textrm{ with } n  \in \mathbb{Z}
\end{equation}
and for $\beta=90\degr$ it becomes independent of $\epsilon$
\begin{equation}
 \psi_\infty = n \, \upi + \frac{\upi}{2} - a \textrm{ with } n  \in \mathbb{Z}\ .
\end{equation}
Therefore in this last case the translation of the dipole centre does not impact on the asymptotic spiral structure compared to the centred dipole.

Let us investigate the phase lag in photon time of arrival when emitted from different parts of the magnetosphere. To generalize the solution to plasma filled magnetospheres, we assume that the radial expansion is not exactly equal to the speed of light~$c$ but slightly less given by~$V=\beta\,c$ to allow for plasma load with finite inertia. The two arm spiral is therefore described by the polar equation
\begin{equation}
 \varphi = \Omega \, \left( t - \frac{r-R}{V} \right) + \psi(r,a,\epsilon) + \varphi_0\
\end{equation}
thus very similar to eq.~(\ref{eq:SpiraleTheorique}) except for the plasma speed~$V$ instead of the speed of light~$c$. We compute the time lag between two photons emitted at locations $(r_1,\varphi_1=0)$ and $(r_2,\varphi_2=0)$ at time $t^e_1$ and $t^e_2$ and received by an observer at time $t^r_1$ and $t^r_2$. The spiral arm forces a difference in emission time given by
\begin{equation}
 \Delta t^e = t^e_2 - t^e_1 = \frac{r_2-r_1}{V} + \frac{\psi_1-\psi_2}{\Omega}\ .
\end{equation}
If photon propagation at finite speed is taken into account, the observer time lag becomes
\begin{equation}
 \Delta t^r = t^r_2 - t^r_1 = \frac{r_1-r_2}{c} + \Delta t^e\ .
\end{equation}
Normalized to the pulsar period we finally have
\begin{equation}
 \frac{\Delta t^r}{P} = \frac{1}{2\,\upi} \, \left[ \left( \frac{1}{\beta} - 1 \right) \, \frac{r_2-r_1}{\rlight} + \psi_1 - \psi_2 \right]\ .
\end{equation}
For ultrarelativistic flows, the relation $\beta-1\ll1$ applies and the time lag is approximated by
\begin{equation}
 \frac{\Delta t^r}{P} \approx \frac{1}{2\,\upi} \, \left[ \frac{r_2-r_1}{2\,\Gamma^2\,\rlight} + \psi_1 - \psi_2 \right]\ .
\end{equation}
If photons are emitted inside the light-cylinder or not faraway outside $(r\ll\Gamma^2\,\rlight)$, the first term is negligible (and exactly zero for a vacuum wave). Consequently, the time delay is only related to the geometry of the spiral arm because it reduces to
\begin{equation}
 \frac{\Delta t^r}{P} \approx \frac{\psi_1 - \psi_2}{2\,\upi} .
\end{equation}
From the values of $(\psi_*, \psi_L, \psi_\infty)$ in table~\ref{tab:Psi}, all interesting time lags between photons can be deduced. For instance assuming radio emission at the surface and high-energy emission at the light-cylinder or in the asymptotic region of the wind, observations of Fermi/LAT can be compared to the simple predictions presented here.

\section{Conclusion}
\label{sec:Conclusion}

The simple idea of an off-centred rotating dipole in vacuum can lead to drastic implications on stellar electrodynamics and related properties. In this paper we showed how to computed analytically the radiating field to any order in the shifted dipole and gave exact analytical expressions for the low order multipole corrections that constitute a useful starting point to study secular evolution of the star, its free precession, spindown luminosity, braking index and electromagnetic torques. We found that corrections are of the order $a\,\epsilon$ or some power of it. Unfortunately such observables are hardly detectable with current instrumentation. Nevertheless, an off-centred dipole could be directly observed through its imprint on pulsed radio and high-energy emission profiles and polarization. Depending on the emission sites, gaps or wind, and radiation mechanisms, synchrotron, inverse Compton or curvature radiation, the signature will be very different. In the near future, we plan to investigate such effects as an extension and realistic application of \cite{2015MNRAS.450..714P} work by setting the weight of multipoles with respect to the dipole according to the geometry introduced here. 

All our discussion was based on multipoles derived from an offset dipole so that $\epsilon$ remains less than unity. If for some reason, a multipolar component should be the dominant part of the magnetic field at the surface, then we should revise all our conclusions drawn in this paper and account for large deviations from the bunch of well-known results about a pure dipole. The clues to several characteristics of rotating neutron stars lie in the topology of its surface magnetic field. Severe improvements in our understanding of such stars are expected from a better study and observational constraints of this multipolar electromagnetic field.

\section*{Acknowledgements}

I am very grateful to the referee for his valuable comments and suggestions. This work has been supported by the French National Research Agency (ANR) through the grant No. ANR-13-JS05-0003-01 (project EMPERE).


\appendix

\section{Fourier transform and series}
\label{app:A}

The spindown luminosity for a arbitrary motion of the dipole is given by \cite{2001elcl.book.....J} as
\begin{equation}
\frac{d^2 I}{d\Omega\,d\omega} = \frac{\mu_0}{4\,\upi} \, \frac{\omega^4}{4\,\upi^2\,c^3} \left| \int_{-\infty}^{+\infty} \bmath{n} \wedge \bmath{\mu}(t) \, e^{i\,\omega\,( t - \bmath{n}\cdot \bmath{r}(t)/c)} \, dt \right|^2
\end{equation}
according to its definition for the Fourier transform. However, they are several ways to define this Fourier transform of a real or complex valued function. We prefer to use the following normalization
\begin{equation}
\label{eq:DefinitionTF}
 \mathcal{F}[f] = \hat f(\omega) = \dfrac{1}{\sqrt{2\,\upi}} \, \int_{-\infty}^{+\infty} f(t) \, e^{-i\,\omega\,t} \, dt
\end{equation}
which corresponds to a change in the sign of the frequency argument $\omega\rightarrow-\omega$ compared to \cite{2001elcl.book.....J}.
In the off-centred rotating dipole, the electromagnetic field as well as the emission pattern are periodic. It is then more judicious to employ Fourier series instead of Fourier transforms. Thus we define the exponential Fourier series as
\begin{equation}
 S_f(t) = \sum_{n=-\infty}^{+\infty} c_n \, e^{i\,n\,\Omega\,t}
\end{equation}
where the coefficients $c_n$ are given by
\begin{equation}
c_n = \dfrac{1}{T} \, \int_0^{T} f(t) \, e^{-i\,n\,\Omega\,t} \, dt \ .
\end{equation} 
The Fourier transform of this periodic signal gives
\begin{equation}
 \mathcal{F}[f] = \sum_{-\infty}^{+\infty} c_n \, \mathcal{F}[e^{i\,n\,\Omega\,t}] = \sum_{-\infty}^{+\infty} \sqrt{2\,\upi} \, c_n \, \delta(\omega - n\,\Omega) \ .
\end{equation}
Therefore, for the periodic signal emanating from the rotating dipole, we can compute the coefficients of the exponential Fourier series and then multiply them by a factor $\sqrt{2\,\upi}$ to get the corresponding Dirac distribution peaks in the Fourier transform.

The spindown luminosity in the $\ell$-th harmonic is therefore
\begin{equation}
\left.\frac{d^2 I}{d\Omega\,d\omega}\right|_{\omega = \ell\,\Omega} = \frac{\mu_0}{4\,\upi} \, \frac{\omega^4}{2\,\upi\,c^3} \left| \frac{\sqrt{2\,\upi}}{T} \, \int_{0}^{T} \bmath{n} \wedge \bmath{\mu}(t) \, e^{-i\,\ell\,\Omega\,( t - \bmath{n}\cdot \bmath{r}(t)/c)} \, dt \right|^2
\end{equation}
where $T=2\,\upi/\Omega$ is the rotation period of the neutron star.
This method avoids the discussion of \cite{1975ApJ...201..447H} and directly leads to the results of the power in a specified harmonic.

In order to compute the integrals involved in the Fourier coefficients we remind the following useful integral representation of the Bessel functions, see for instance \cite{morse1953methodsvol1} or \cite{smirnov1989cours} to get
\begin{subequations}
\begin{align}
 J_n(z) & = \frac{1}{2\,\upi} \, \int_{-\upi}^{+\upi} e^{i\,(n\,\varphi -z\,\sin\varphi)} \, d\varphi = \frac{1}{2\,\upi} \, \int_{-\upi}^{+\upi} e^{-i\,(n\,\varphi -z\,\sin\varphi)} \, d\varphi \\
 \frac{n}{z} \, J_n(z) & = \frac{1}{2\,\upi} \, \int_{-\upi}^{+\upi} e^{i\,(n\,\varphi - z \, \sin\varphi)}  \, \cos\varphi \, d\varphi = \frac{1}{2\,\upi} \, \int_{-\upi}^{+\upi} e^{-i\,(n\,\varphi - z \, \sin\varphi)} \, \cos\varphi \, d\varphi \\
  i \, J_n'(z)  & = \frac{1}{2\,\upi} \, \int_{-\upi}^{+\upi} e^{i\,(n\,\varphi - z \, \sin\varphi)}
 \, \sin\varphi \, d\varphi = - \frac{1}{2\,\upi} \, \int_{-\upi}^{+\upi} e^{-i\,(n\,\varphi - z \, \sin\varphi)}
 \, \sin\varphi \, d\varphi \ .
\end{align}
\end{subequations}

\section{Useful properties of vector spherical harmonics}
\label{app:B}

The analytical computations performed in this paper are greatly eased if one remember some useful properties about vector spherical harmonics such as their orthonormality expressed as
\begin{subequations}
\begin{align}
  \iint \mathbf{\Phi}_{\ell'm'} \cdot \mathbf{\Phi}_{\ell m}^* \, d\Omega & = \delta_{\ell \ell'} \, \delta_{mm'} \\
  \iint \mathbf{\Psi}_{\ell'm'} \cdot \mathbf{\Psi}_{\ell m}^* \, d\Omega & = \delta_{\ell \ell'} \, \delta_{mm'} \\
  \iint \mathbf{\Psi}_{\ell m} \cdot \mathbf{\Phi}_{\ell m}^* \, d\Omega & = 0 \ .
\end{align}
\end{subequations}
Some other important relations useful to estimate the electromagnetic recoil bring in other integrals given by 
\begin{subequations}
\begin{align}
  I_1 = \iint \mathbf{\Phi}_{\ell m} \cdot \mathbf{\Phi}_{\ell' m'}^* \, \cos \vartheta \, d\Omega & = \iint \mathbf{\Psi}_{\ell m} \cdot \mathbf{\Psi}_{\ell' m'}^* \, \cos \vartheta \, d\Omega \\
   = \frac{\sqrt{\ell\,(\ell+2)}}{\ell+1} \, J_{\ell+1,m} \, \delta_{\ell+1,\ell'} \, \delta_{m,m'} & + \frac{\sqrt{(\ell-1)\,(\ell+1)}}{\ell} \, J_{\ell,m} \, \delta_{\ell-1,\ell'} \, \delta_{m,m'} \\
  I_3 = \iint \mathbf{\Psi}_{\ell m} \cdot \mathbf{\Phi}_{\ell' m'}^* \, \cos \vartheta \, d\Omega & = \frac{i\,m}{\ell\,(\ell+1)} \, \delta_{\ell,\ell'} \, \delta_{m,m'} \ .
\end{align}
\end{subequations}
These integrals are found by integration by part and using the definition of scalar spherical harmonics being eigenfunctions of the Laplacian operator. They have already been given in \cite{1979ApJS...41...75R} according to a private communication of Teukolsky.


\section{Exact analytical expressions for multipolar fields}
\label{app:C}

The solution for the offset dipole is completely known once the coefficients $f_{l,m}^{\rm B}(R)$ have been specified on the neutron star surface. It is the boundary condition that completely and uniquely determines the full electromagnetic field outside the star, except for a possible total electric charge of the system that is not necessarily equal to zero. Here we give the exact analytical solution in complex form for any combination of potentials~$f_{l,m}^{\rm B}(R)$.

\subsection{Dipole}

For the dipolar part of the fields, the constant of integrations are
\begin{subequations}
\begin{align}
 a^{\rm B}_{1,0} & = R^2 \, f^{\rm B}_{1,0}(R) \\
 a^{\rm B}_{1,1} & = \frac{f^{\rm B}_{1,1}(R)}{h_1^{(1)}(k\,R)} \\
 a^{\rm D}_{2,0} & = - \frac{1}{\sqrt{5}} \, \varepsilon_0 \, \Omega \, R^4 \, f^{\rm B}_{1,0}(R) \\
 a^{\rm D}_{2,1} & = \sqrt{\frac{3}{5}} \, \varepsilon_0 \, \Omega \, R \, \frac{f^{\rm B}_{1,1}(R)}{\partial_r(r\,h_2^{(1)}(k\,r))|_R} \ .
\end{align}
\end{subequations}
In explicit form, the $m=0$ reads
\begin{subequations}
\begin{align}
 B_r & = - \sqrt{\frac{3}{8\,\upi}} \, a_{1,0}^{\rm B} \, \frac{2\,\cos\vartheta}{r^3} \\
 B_\vartheta & = - \sqrt{\frac{3}{8\,\upi}} \, a_{1,0}^{\rm B} \, \frac{\sin\vartheta}{r^3} \\
 B_\varphi & = 0
\end{align}
\end{subequations}
the $m=1$ reads
\begin{subequations}
\begin{align}
 B_r & = \sqrt{\frac{3}{\upi}} \, a_{1,1}^{\rm B} \, h_1^{(1)}(k\,r) \, \frac{\sin\vartheta}{2\,r} \, e^{i\,\psi} \\
 B_\vartheta & = \frac{e^{i\,\psi}}{4 \,\sqrt{\upi}\,r\,\rlight} \, \cos\vartheta \, \left[ \sqrt{5} \, \mu_0 \, c \, r \, a_{2,1}^{\rm D} \, h_2^{(1)}(k\,r) + \sqrt{3} \, \rlight \, a_{1,1}^{\rm B} \, \partial_r(r\,h_1^{(1)}(k\,r) ) \right] \\
 B_\varphi & = \frac{i \, e^{i\,\psi}}{4 \,\sqrt{\upi}\,r\,\rlight} \, \left[ \sqrt{5} \, \mu_0 \, c \, r \, a_{2,1}^{\rm D} \, h_2^{(1)}(k\,r) \, \cos2\,\vartheta + \sqrt{3} \, \rlight \, a_{1,1}^{\rm B} \, \partial_r(r\,h_1^{(1)}(k\,r) ) \right] \ .
\end{align}
\end{subequations}
For the electric field we have for the $m=0$ 
\begin{subequations}
\begin{align}
 D_r & = - \sqrt{\frac{15}{2\,\upi}} \, a_{2,0}^{\rm D} \, \frac{3\,\cos^2\vartheta-1}{2\,r^4} \\
 D_\vartheta & = - \sqrt{\frac{15}{2\,\upi}} \, a_{2,0}^{\rm D} \, \frac{\sin\vartheta\,\cos\vartheta}{r^4} \\
 D_\varphi & = 0
\end{align}
\end{subequations}
the $m=1$ reads
\begin{subequations}
\begin{align}
 D_r & = 3\,\sqrt{\frac{5}{\upi}} \, a_{2,1}^{\rm D} \, h_2^{(1)}(k\,r) \, \frac{\cos\vartheta\,\sin\vartheta}{2\,r} \, e^{i\,\psi} \\
 D_\vartheta & = \frac{e^{i\,\psi}}{4 \,\sqrt{\upi}\,c\,r\,\rlight\,\mu_0} \, [ - \sqrt{3} \, r \, a_{1,1}^{\rm B} \, h_1^{(1)}(k\,r) + \sqrt{5} \, \mu_0 \, c \, \rlight \, a_{2,1}^{\rm D} \, \partial_r(r\,h_2^{(1)}(k\,r) ) \, \cos 2\,\vartheta ] \\
 D_\varphi & = \frac{i \, e^{i\,\psi} \, \cos \vartheta}{4 \,\sqrt{\upi}\,c\,r\,\rlight\,\mu_0} \, [ - \sqrt{3} \, r \, a_{1,1}^{\rm B} \, h_1^{(1)}(k\,r) + \sqrt{5} \, \mu_0 \, c \, \rlight \, a_{2,1}^{\rm D} \, \partial_r(r\,h_2^{(1)}(k\,r) ) ]
\end{align}
\end{subequations}

\subsection{Quadrupole}

For the quadrupolar part of the fields, the constant of integrations are
\begin{subequations}
\begin{align}
 a^{\rm B}_{2,0} & = R^3 \, f^{\rm B}_{2,0}(R) \\
 a^{\rm B}_{2,1} & = \frac{f^{\rm B}_{2,1}(R)}{h_2^{(1)}(k\,R)} \\
 a^{\rm B}_{2,2} & = \frac{f^{\rm B}_{2,2}(R)}{h_2^{(1)}(2\,k\,R)} \\
 a^{\rm D}_{1,0} & = \frac{2}{\sqrt{5}} \, \varepsilon_0 \, \Omega \, R^3 \, f^{\rm B}_{2,0}(R) \\
 a^{\rm D}_{1,1} & = - \sqrt{\frac{3}{5}} \, \varepsilon_0 \, \Omega \, R \, \frac{f^{\rm B}_{2,1}(R)}{\partial_r(r\,h_1^{(1)}(k\,r))|_R} \\
 a^{\rm D}_{3,0} & = - 2 \, \sqrt{\frac{2}{35}} \, \varepsilon_0 \, \Omega \, R^5 \, f^{\rm B}_{2,0}(R) \\
 a^{\rm D}_{3,1} & = \frac{8}{\sqrt{35}} \, \varepsilon_0 \, \Omega \, R \, \frac{f^{\rm B}_{2,1}(R)}{\partial_r(r\,h_3^{(1)}(k\,r))|_R} \\
 a^{\rm D}_{3,2} & = 2 \, \sqrt{\frac{2}{7}} \, \varepsilon_0 \, \Omega \, R \, \frac{f^{\rm B}_{2,2}(R)}{\partial_r(r\,h_3^{(1)}(2\,k\,r))|_R}  \ .
\end{align}
\end{subequations}
For the quadrupolar moment, the $m=0$ reads
\begin{subequations}
\begin{align}
 B_r & = - \sqrt{\frac{15}{2\,\upi}} \, a_{2,0}^{\rm B} \, \frac{3 \, \cos^2\vartheta - 1}{2\,r^4} \\
 B_\vartheta & = - \sqrt{\frac{15}{2\,\upi}} \, a_{2,0}^{\rm B} \, \frac{\cos\vartheta\, \sin\vartheta }{r^4} \\
 B_\varphi & = 0
\end{align}
\end{subequations}
the $m=1$ reads
\begin{subequations}
\begin{align}
 B_r & = 3 \, \sqrt{\frac{5}{\upi}} \, a_{2,1}^{\rm B} \, h_2^{(1)}(k\,r) \, \frac{\cos\vartheta\,\sin\vartheta}{2\,r} \, e^{i\,\psi} \\
 B_\vartheta & = \frac{e^{i\,\psi}}{32\,\sqrt{\upi}\,r\,\rlight} \, [ \mu_0 \,c \,r \, ( 8\,\sqrt{3} \, a_{1,1}^{\rm D} \, h_1^{(1)}(k\,r) + \sqrt{7} \, ( 3 + 5 \, \cos 2\vartheta) \, a_{3,1}^{\rm D} \, h_3^{(1)}(k\,r) ) + \\
 & 8\,\sqrt{5} \, \rlight \, a_{2,1}^{\rm B} \, \partial_r(r\,h_2^{(1)}(k\,r) \, \cos2\,\vartheta ] \nonumber \\
 B_\varphi & = \frac{i\,e^{i\,\psi}}{32\,\sqrt{\upi}\,r\,\rlight} \, \cos \vartheta \, [ \mu_0 \,c \,r \, ( 8\,\sqrt{3} \, a_{1,1}^{\rm D} \, h_1^{(1)}(k\,r) + \\
 & \sqrt{7} \, ( -7 + 15 \, \cos 2\,\vartheta) \, a_{3,1}^{\rm D} \, h_3^{(1)}(k\,r) ) + 8\,\sqrt{5} \, \rlight \, a_{2,1}^{\rm B} \, \partial_r(r\,h_2^{(1)}(k\,r) ] \nonumber
\end{align}
\end{subequations}
the $m=2$ reads
\begin{subequations}
\begin{align}
 B_r & = - 3 \, \sqrt{\frac{5}{\upi}} \, a_{2,2}^{\rm B} \, h_2^{(1)}(2\,k\,r) \, \frac{\sin^2 \vartheta}{4\,r} \, e^{2\,i\,\psi} \\
 B_\vartheta & = - \sqrt{\frac{5}{\upi}} \, \frac{e^{2\,i\,\psi}}{8\,r\,\rlight} \, \sin 2\,\vartheta \, \left[ \sqrt{14} \, \mu_0 \, c \, r \, a_{3,2}^{\rm D} \, h_3^{(1)}(2\,k\,r) + \rlight \, a_{2,2}^{\rm B} \, \partial_r(r\,h_2^{(1)}(2\,k\,r)) \right] \\
 B_\varphi & = - i \, \sqrt{\frac{5}{\upi}} \, \frac{e^{2\,i\,\psi}}{16\,r\,\rlight} \, \sin \vartheta \, \left[ \sqrt{14} \, \mu_0 \, c \, r \, ( 1 + 3 \,\cos 2\,\vartheta) \, a_{3,2}^{\rm D} \, h_3^{(1)}(2\,k\,r) + 4 \, \rlight \, a_{2,2}^{\rm B} \, \partial_r(r\,h_2^{(1)}(2\,k\,r)) \right] \ .
\end{align}
\end{subequations}
For the electric field we have for the $m=0$ 
\begin{subequations}
\begin{align}
 D_r & = - \sqrt{\frac{3}{\upi}} \, \frac{\cos\vartheta}{4\,r^5} \, ( 2\,\sqrt{2} \, a_{1,0}^{\rm D} \, r^2 + \sqrt{7} \, ( 5 \, \cos2\,\vartheta -1 ) \, a_{3,0}^{\rm D} ) \\
 D_\vartheta & = - \sqrt{\frac{3}{\upi}} \, \frac{\sin\vartheta}{16\,r^5} \, ( 4\,\sqrt{2} \, a_{1,0}^{\rm D} \, r^2 + 3 \, \sqrt{7} \, ( 5 \, \cos2\,\vartheta + 3 ) \, a_{3,0}^{\rm D} ) \\
 D_\varphi & = 0
\end{align}
\end{subequations}
the $m=1$ reads
\begin{subequations}
\begin{align}
 D_r & = \frac{e^{i\,\psi}\,\sin\vartheta}{8\,\sqrt{\upi}\,r} \, ( 4 \, \sqrt{3} \, a_{1,1}^{\rm D} \, h_1^{(1)}(k\,r) + 3 \, \sqrt{7} \, ( 5 \, \cos2\,\vartheta + 3 ) \, a_{3,1}^{\rm D} \, h_3^{(1)}(k\,r) ) \\
 D_\vartheta & = \frac{e^{i\,\psi}\,\cos\vartheta}{32\,\mu_0\,c\,\sqrt{\upi}\,r\,\rlight} \, ( -8\,\sqrt{5} \, a_{2,1}^{\rm B} \, r \, h_2^{(1)}(k\,r) + \\
 & \mu_0 \, c \, \rlight \, ( 8\,\sqrt{3} \, a_{1,1}^{\rm D} \, \partial_r(r\,h_1^{(1)}(k\,r)) + \sqrt{7} \, ( -7 + 15 \, \cos2\,\vartheta ) \, a_{3,1}^{\rm D}  \, \partial_r(r\,h_3^{(1)}(k\,r)) ) ) \nonumber \\
 D_\varphi & = \frac{i\,e^{i\,\psi}}{32\,\mu_0\,c\,\sqrt{\upi}\,r\,\rlight} \, ( -8\,\sqrt{5} \, a_{2,1}^{\rm B} \, r \, h_2^{(1)}(k\,r) \, \cos2\,\vartheta + \\
 & \mu_0 \, c \, \rlight \, ( 8\,\sqrt{3} \, a_{1,1}^{\rm D} \, \partial_r(r\,h_1^{(1)}(k\,r)) + \sqrt{7} \, ( 3 + 5 \, \cos2\,\vartheta ) \, a_{3,1}^{\rm D}  \, \partial_r(r\,h_3^{(1)}(k\,r)) ) ) \nonumber
\end{align}
\end{subequations}
the $m=2$ reads
\begin{subequations}
\begin{align}
 D_r & = - \frac{3\,e^{2\,i\,\psi}\,\cos\vartheta\,\sin^2\vartheta}{2\,r} \, \sqrt{\frac{35}{2\,\upi}} \, a_{3,2}^{\rm D} \, h_3^{(1)}(2\,k\,r) \\
 D_\vartheta & = - \frac{e^{2\,i\,\psi}\,\sin\vartheta}{32\,\mu_0\,c\,r\,\rlight} \, \sqrt{\frac{5}{\upi}} \, [ -16 \, a_{2,2}^{\rm B} \, r \, h_2^{(1)}(2\,k\,r) + \mu_0 \, c \, \rlight \, \sqrt{14} \, a_{3,2}^{\rm D} \, \partial_r(r\,h_3^{(1)}(2\,k\,r))  \, ( 1 + 3 \, \cos2\,\vartheta ) ] \\
 D_\varphi & = - \frac{i\,e^{2\,i\,\psi}\,\sin2\,\vartheta}{16\,\mu_0\,c\,r\,\rlight} \, \sqrt{\frac{5}{\upi}} \, [ -4 \, a_{2,2}^{\rm B} \, r \, h_2^{(1)}(2\,k\,r) + \mu_0 \, c \, \rlight \, \sqrt{14} \, a_{3,2}^{\rm D} \, \partial_r(r\,h_3^{(1)}(2\,k\,r) ) ]  \ .
\end{align}
\end{subequations}

\subsection{Hexapole}

For the hexapolar part of the fields, the constant of integrations are
\begin{subequations}
\begin{align}
 a^{\rm B}_{3,0} & = R^4 \, f^{\rm B}_{3,0}(R) \\
 a^{\rm B}_{3,1} & = \frac{f^{\rm B}_{3,1}(R)}{h_3^{(1)}(k\,R)} \\
 a^{\rm B}_{3,2} & = \frac{f^{\rm B}_{3,2}(R)}{h_3^{(1)}(2\,k\,R)} \\
 a^{\rm B}_{3,3} & = \frac{f^{\rm B}_{3,3}(R)}{h_3^{(1)}(3\,k\,R)} \\
 a^{\rm D}_{2,0} & = 3 \, \sqrt{\frac{2}{35}} \, \varepsilon_0 \, \Omega \, R^4 \, f^{\rm B}_{3,0}(R) \\
 a^{\rm D}_{2,1} & = - \frac{8}{\sqrt{35}} \, \varepsilon_0 \, \Omega \, R \, \frac{f^{\rm B}_{3,1}(R)}{\partial_r(r\,h_2^{(1)}(k\,r))|_R} \\
 a^{\rm D}_{2,2} & = - 2 \, \sqrt{\frac{2}{7}} \, \varepsilon_0 \, \Omega \, R \, \frac{f^{\rm B}_{3,2}(R)}{\partial_r(r\,h_2^{(1)}(2\,k\,r))|_R} \\
 a^{\rm D}_{4,0} & = - \sqrt{\frac{5}{21}} \, \varepsilon_0 \, \Omega \, R^6 \, f^{\rm B}_{3,0}(R) \\
 a^{\rm D}_{4,1} & = \frac{5}{\sqrt{7}} \, \varepsilon_0 \, \Omega \, R \, \frac{f^{\rm B}_{3,1}(R)}{\partial_r(r\,h_4^{(1)}(k\,r))|_R} \\
 a^{\rm D}_{4,2} & = 2 \, \sqrt{\frac{5}{7}} \, \varepsilon_0 \, \Omega \, R \, \frac{f^{\rm B}_{3,2}(R)}{\partial_r(r\,h_4^{(1)}(2\,k\,r))|_R} \\
 a^{\rm D}_{4,3} & = \sqrt{\frac{5}{3}} \, \varepsilon_0 \, \Omega \, R \, \frac{f^{\rm B}_{3,3}(R)}{\partial_r(r\,h_4^{(1)}(3\,k\,r))|_R} \ .
\end{align}
\end{subequations}
For the hexapolar moment, the $m=0$ reads
\begin{subequations}
\begin{align}
 B_r & = - \sqrt{\frac{21}{\upi}}\frac{ (3 \cos \vartheta +5 \cos 3\,\vartheta )\, a^B_{3,0}}{8 r^5} \\
 B_\vartheta & = -3 \sqrt{\frac{21}{\upi }}\frac{ (\sin \vartheta +5 \sin 3\,\vartheta )\, a^B_{3,0}}{32 r^5} \\
 B_\varphi & = 0
\end{align}
\end{subequations}
the $m=1$ reads
\begin{subequations}
\begin{align}
 B_r & = \frac{3 \sqrt{\frac{7}{\upi }} e^{i\,\psi } \sin \vartheta  (5 \cos 2\,\vartheta +3)\, a^B_{3,1} \, h_3^{(1)}(k \, r)}{8 r} \\
 B_\vartheta & = \frac{e^{i\,\psi }}{32\, \sqrt{\upi }\, r \,\rlight} \, \cos \vartheta  (c \mu _0 r \left(3 (7 \cos 2\,\vartheta +1)\, a^D_{4,1} \, h_4^{(1)}(k \, r)+8 \sqrt{5} \, a^D_{2,1} \, h_2^{(1)}(k \, r)\right)+ \nonumber \\
 & \sqrt{7} \,\rlight\, (15 \cos 2\,\vartheta -7)\, a^B_{3,1} \partial_r(r\,h_3^{(1)}(k \, r)) \\
 B_\varphi & = \frac{i e^{i\,\psi }}{32\, \sqrt{\upi }\, r \,\rlight} (c \mu _0 r \left(8 \sqrt{5} \cos 2\,\vartheta  a^D_{2,1} \, h_2^{(1)}(k \, r)+3 (\cos 2\,\vartheta +7 \cos 4\,\vartheta )\, a^D_{4,1} \, h_4^{(1)}(k \, r)\right) + \nonumber \\
 & \sqrt{7} \,\rlight\, (5 \cos 2\,\vartheta +3)\, a^B_{3,1} \partial_r(r\,h_3^{(1)}(k \, r))
\end{align}
\end{subequations}
the $m=2$ reads
\begin{subequations}
\begin{align}
 B_r & = -\frac{3 \sqrt{\frac{35}{2 \upi }} e^{2\, i \,\psi } \sin ^2\vartheta  \cos \vartheta  a^B_{3,2} \, h_3^{(1)}(2 \, k \,  r)}{2 r} \\
 B_\vartheta & = -\frac{e^{2\, i \,\psi }}{16\, \sqrt{\upi }\, r \,\rlight}  \sin \vartheta  \, (c \mu _0 r \left(3 \sqrt{2} (7 \cos 2\,\vartheta +5)\, a^D_{4,2} \, h_4^{(1)}(2 \, k \,  r)+8 \sqrt{5} \, a^D_{2,2} \, h_2^{(1)}(2 \, k \,  r)\right)+\nonumber \\
 & \sqrt{70} \,\rlight\, (3 \cos 2\,\vartheta +1)\, a^B_{3,2} \partial_r(r\,h_3^{(1)}(2 \, k \,  r)) \\
 B_\varphi & = -\frac{i e^{2\, i \,\psi }}{16\, \sqrt{\upi }\, r \,\rlight} \sin 2\,\vartheta \, (c \mu _0 r \left(3 \sqrt{2} (7 \cos 2\,\vartheta -1)\, a^D_{4,2} \, h_4^{(1)}(2 \, k \,  r)+4 \sqrt{5} \, a^D_{2,2} \, h_2^{(1)}(2 \, k \,  r)\right)+\nonumber \\
 & 2 \sqrt{70} \,\rlight\, a^B_{3,2} \partial_r(r\,h_3^{(1)}(2 \, k \,  r))
\end{align}
\end{subequations}
the $m=3$ reads
\begin{subequations}
\begin{align}
 B_r & = \frac{\sqrt{\frac{105}{\upi }} e^{3\, i \,\psi } \sin ^3\vartheta  a^B_{3,3} \, h_3^{(1)}(3 \, k \,  r)}{4 r} \\
 B_\vartheta & = \frac{3 \sqrt{\frac{7}{\upi }} e^{3\, i \,\psi } \sin ^2\vartheta  \cos \vartheta  \left(9 c \mu _0 r a^D_{4,3} \, h_4^{(1)}(3 \, k \,  r)+\sqrt{15} \,\rlight\, a^B_{3,3} \partial_r(r\,h_3^{(1)}(3 \, k \,  r)\right)}{16 r \,\rlight} \\
 B_\varphi & = \frac{3 i \sqrt{\frac{7}{\upi }} e^{3\, i \,\psi } \sin ^2\vartheta  \left(3 c \mu _0 r (2 \cos 2\,\vartheta +1)\, a^D_{4,3} \, h_4^{(1)}(3 \, k \,  r)+\sqrt{15} \,\rlight\, a^B_{3,3} \partial_r(r\,h_3^{(1)}(3 \, k \,  r)\right)}{16 r \,\rlight} \ .
\end{align}
\end{subequations}
For the electric field we have for the $m=0$ 
\begin{subequations}
\begin{align}
 D_r & = -\frac{\sqrt{\frac{5}{\upi }} \left(3 (20 \cos 2\,\vartheta +35 \cos 4\,\vartheta +9)\, a^D_{4,0}+8 \sqrt{6} r^2 (3 \cos 2\,\vartheta +1)\, a^D_{2,0}\right)}{64 r^6} \\
 D_\vartheta & = -\frac{\sqrt{\frac{5}{\upi }} \sin 2\,\vartheta  \left(3 (7 \cos 2\,\vartheta +1)\, a^D_{4,0}+2 \sqrt{6} r^2 a^D_{2,0}\right)}{8 r^6} \\
 D_\varphi & = 0
\end{align}
\end{subequations}
the $m=1$ reads
\begin{subequations}
\begin{align}
 D_r & = \frac{3 e^{i\,\psi } \sin 2\,\vartheta  \left(5 (7 \cos 2\,\vartheta +1)\, a^D_{4,1} \, h_4^{(1)}(k \, r)+4 \sqrt{5} \, a^D_{2,1} \, h_2^{(1)}(k \, r)\right)}{16\, \sqrt{\upi }\, r} \\
 D_\vartheta & = \frac{e^{i\,\psi }}{32 \sqrt{\upi } \, c \, \mu _0 \, r \,\rlight} \, (c \mu _0 \,\rlight\, \left(8 \sqrt{5} \cos 2\,\vartheta  a^D_{2,1} \partial_r(r\,h_2^{(1)}(k \, r)+3 (\cos 2\,\vartheta +7 \cos 4\,\vartheta )\, a^D_{4,1} \partial_r(r\,h_4^{(1)}(k \, r)\right) - \nonumber \\
 & \sqrt{7} r (5 \cos 2\,\vartheta +3)\, a^B_{3,1} \, h_3^{(1)}(k \, r)) \\
 D_\varphi & = \frac{i e^{i\,\psi }}{32 \sqrt{\upi } \, c \, \mu _0 \, r \,\rlight} \cos \vartheta  \,(c \mu _0 \,\rlight\, \left(3 (7 \cos 2\,\vartheta +1)\, a^D_{4,1} \partial_r(r\,h_4^{(1)}(k \, r)+8 \sqrt{5} \, a^D_{2,1} \partial_r(r\,h_2^{(1)}(k \, r)\right)-\nonumber \\
 & \sqrt{7} r (15 \cos 2\,\vartheta -7)\, a^B_{3,1} \, h_3^{(1)}(k \, r))
\end{align}
\end{subequations}
the $m=2$ reads
\begin{subequations}
\begin{align}
 D_r & = -\frac{3 e^{2\, i \,\psi } \sin ^2\vartheta  \left(5 \sqrt{2} (7 \cos 2\,\vartheta +5)\, a^D_{4,2} \, h_4^{(1)}(2 \, k \,  r)+4 \sqrt{5} \, a^D_{2,2} \, h_2^{(1)}(2 \, k \,  r)\right)}{16\, \sqrt{\upi }\, r} \\
D_\vartheta & = \frac{e^{2\, i \,\psi }}{16 \sqrt{\upi } \, c \, \mu _0 \, r \,\rlight} \sin 2\,\vartheta  \, (c \mu _0 \,\rlight\, \left(3 \sqrt{2} (1-7 \cos 2\,\vartheta )\, a^D_{4,2} \partial_r(r\,h_4^{(1)}(2 \, k \,  r)-4 \sqrt{5} \, a^D_{2,2} \partial_r(r\,h_2^{(1)}(2 \, k \,  r)\right)+ \nonumber \\
& 2 \sqrt{70} r a^B_{3,2} \, h_3^{(1)}(2 \, k \,  r)) \\
D_\varphi & = \frac{i e^{2\, i \,\psi }}{16 \sqrt{\upi } \, c \, \mu _0 \, r \,\rlight} \sin \vartheta  \,(\sqrt{70} r (3 \cos 2\,\vartheta +1)\, a^B_{3,2} \, h_3^{(1)}(2 \, k \,  r) - \nonumber \\
& c \mu _0 \,\rlight\, (3 \sqrt{2} (7 \cos 2\,\vartheta +5)\, a^D_{4,2} \partial_r(r\,h_4^{(1)}(2 \, k \,  r) +  8 \sqrt{5} \, a^D_{2,2} \partial_r(r\,h_2^{(1)}(2 \, k \,  r)))
\end{align}
\end{subequations}
the $m=3$ reads
\begin{subequations}
\begin{align}
 D_r & = \frac{15 \sqrt{\frac{7}{\upi }} e^{3\, i \,\psi } \sin ^3\vartheta  \cos \vartheta  a^D_{4,3} \, h_4^{(1)}(3 \, k \,  r)}{4 r} \\
 D_\vartheta & = \frac{3 \sqrt{\frac{7}{\upi }} e^{3\, i \,\psi } \sin ^2\vartheta  \left(3 c \mu _0 \,\rlight\, (2 \cos 2\,\vartheta +1)\, a^D_{4,3} \partial_r(r\,h_4^{(1)}(3 \, k \,  r)-\sqrt{15} r a^B_{3,3} \, h_3^{(1)}(3 \, k \,  r)\right)}{16 c \mu _0 r \,\rlight} \\
 D_\varphi & = \frac{3 i \sqrt{\frac{7}{\upi }} e^{3\, i \,\psi } \sin ^2\vartheta  \cos \vartheta  \left(9 c \mu _0 \,\rlight\, a^D_{4,3} \partial_r(r\,h_4^{(1)}(3 \, k \,  r)-\sqrt{15} r a^B_{3,3} \, h_3^{(1)}(3 \, k \,  r)\right)}{16 c \mu _0 r \,\rlight} \ .
\end{align}
\end{subequations}

\subsection{Octupole}

For the octupolar part of the fields, the constant of integrations are
\begin{subequations}
\begin{align}
 a^{\rm B}_{4,0} & = R^5 \, f^{\rm B}_{4,0}(R) \\
 a^{\rm B}_{4,1} & = \frac{f^{\rm B}_{4,1}(R)}{h_4^{(1)}(k\,R)} \\
 a^{\rm B}_{4,2} & = \frac{f^{\rm B}_{4,2}(R)}{h_4^{(1)}(2\,k\,R)} \\
 a^{\rm B}_{4,3} & = \frac{f^{\rm B}_{4,3}(R)}{h_4^{(1)}(3\,k\,R)} \\
 a^{\rm B}_{4,4} & = \frac{f^{\rm B}_{4,4}(R)}{h_4^{(1)}(4\,k\,R)} \\
 a^{\rm D}_{3,0} & = \frac{4}{3} \, \sqrt{\frac{5}{21}} \, \varepsilon_0 \, \Omega \, R^5 \, f^{\rm B}_{4,0}(R) \\
 a^{\rm D}_{3,1} & = - \frac{5}{\sqrt{7}} \, \varepsilon_0 \, \Omega \, R \, \frac{f^{\rm B}_{4,1}(R)}{\partial_r(r\,h_3^{(1)}(k\,r))|_R} \\
 a^{\rm D}_{3,2} & = - 2 \, \sqrt{\frac{5}{7}} \, \varepsilon_0 \, \Omega \, R \, \frac{f^{\rm B}_{4,2}(R)}{\partial_r(r\,h_3^{(1)}(2\,k\,r))|_R} \\
 a^{\rm D}_{3,3} & = - \sqrt{\frac{5}{3}} \, \varepsilon_0 \, \Omega \, R \, \frac{f^{\rm B}_{4,3}(R)}{\partial_r(r\,h_3^{(1)}(3\,k\,r))|_R} \\
 a^{\rm D}_{5,0} & = - 2 \, \sqrt{\frac{2}{33}} \, \varepsilon_0 \, \Omega \, R^7 \, f^{\rm B}_{4,0}(R) \\
 a^{\rm D}_{5,1} & = \frac{8}{\sqrt{11}} \, \varepsilon_0 \, \Omega \, R \, \frac{f^{\rm B}_{4,1}(R)}{\partial_r(r\,h_5^{(1)}(k\,r))|_R} \\
 a^{\rm D}_{5,2} & = 2 \, \sqrt{\frac{14}{11}} \, \varepsilon_0 \, \Omega \, R \, \frac{f^{\rm B}_{4,2}(R)}{\partial_r(r\,h_5^{(1)}(2\,k\,r))|_R} \\
 a^{\rm D}_{5,3} & = 8 \, \sqrt{\frac{2}{33}} \, \varepsilon_0 \, \Omega \, R \, \frac{f^{\rm B}_{4,3}(R)}{\partial_r(r\,h_5^{(1)}(3\,k\,r))|_R} \\
 a^{\rm D}_{5,4} & = 2 \, \sqrt{\frac{6}{11}} \, \varepsilon_0 \, \Omega \, R \, \frac{f^{\rm B}_{4,4}(R)}{\partial_r(r\,h_5^{(1)}(4\,k\,r))|_R} \ .
\end{align}
\end{subequations}
For the hexapolar moment, the $m=0$ reads
\begin{subequations}
\begin{align}
 B_r & = -\frac{3 \sqrt{\frac{5}{\upi }} (20 \cos 2\,\vartheta +35 \cos 4\,\vartheta +9)\, a^B_{4,0}}{64 r^6} \\
 B_\vartheta & = -\frac{3 \sqrt{\frac{5}{\upi }} (2 \sin 2\,\vartheta +7 \sin 4\,\vartheta )\, a^B_{4,0}}{16 r^6} \\
 B_\varphi & = 0
\end{align}
\end{subequations}
the $m=1$ reads
\begin{subequations}
\begin{align}
 B_r & = \frac{15 e^{i\,\psi } (2 \sin 2\,\vartheta +7 \sin 4\,\vartheta )\, a^B_{4,1} \, h_4^{(1)}(k \, r)}{32\, \sqrt{\upi }\, r} \\
 B_\vartheta & = \frac{e^{i\,\psi }}{256\, \sqrt{\upi }\, r \,\rlight} \, (c \mu _0 r (8 \sqrt{7} (5 \cos 2\,\vartheta +3)\, a^D_{3,1} \, h_3^{(1)}(k \, r)+\sqrt{11} (28 \cos 2\,\vartheta +21 \cos 4\,\vartheta +15)\, a^D_{5,1} h_5(k \, r)) + \nonumber \\
 & 24 \,\rlight\, (\cos 2\,\vartheta +7 \cos 4\,\vartheta )\, a^B_{4,1} \partial_r(r\,h_4^{(1)}(k \, r)) \\
 B_\varphi & = \frac{i e^{i\,\psi }}{256\, \sqrt{\upi }\, r \,\rlight} \cos \vartheta  \, (c \mu _0 r (8 \sqrt{7} (15 \cos 2\,\vartheta -7)\, a^D_{3,1} \, h_3^{(1)}(k \, r) + \nonumber \\
 & \sqrt{11} (-84 \cos 2\,\vartheta +105 \cos 4\,\vartheta +43)\, a^D_{5,1} h_5(k \, r))+24 \,\rlight\, (7 \cos 2\,\vartheta +1)\, a^B_{4,1} \partial_r(r\,h_4^{(1)}(k \, r))
\end{align}
\end{subequations}
the $m=2$ reads
\begin{subequations}
\begin{align}
 B_r & = -\frac{15 e^{2\, i \,\psi } \sin ^2\vartheta  (7 \cos 2\,\vartheta +5)\, a^B_{4,2} \, h_4^{(1)}(2 \, k \,  r)}{8 \sqrt{2 \upi } r} \\
 B_\vartheta & = -\frac{e^{2\, i \,\psi }}{16\, \sqrt{\upi }\, r \,\rlight} \sin 2\,\vartheta  \, (\mu _0 (\sqrt{77} c r (3 \cos 2\,\vartheta +1)\, a^D_{5,2} h_5(2 \, k \,  r)+ \nonumber \\
 & 2 \sqrt{70} c r a^D_{3,2} \, h_3^{(1)}(2 \, k \,  r))+3 \sqrt{2} \,\rlight\, (7 \cos 2\,\vartheta -1)\, a^B_{4,2} \partial_r(r\,h_4^{(1)}(2 \, k \,  r)) \\
 B_\varphi & = -\frac{i e^{2\, i \,\psi }}{64\, \sqrt{\upi }\, r \,\rlight} \sin \vartheta  \, (\sqrt{7} \, c \, \mu _0 \, r (4 \sqrt{10} (3 \cos 2\,\vartheta +1)\, a^D_{3,2} \, h_3^{(1)}(2 \, k \,  r)+\nonumber \\
 &\sqrt{11} (12 \cos 2\,\vartheta +15 \cos 4\,\vartheta +5)\, a^D_{5,2} h_5(2 \, k \,  r))+  12 \sqrt{2} \,\rlight\, (7 \cos 2\,\vartheta +5)\, a^B_{4,2} \partial_r(r\,h_4^{(1)}(2 \, k \,  r))
\end{align}
\end{subequations}
the $m=3$ reads
\begin{subequations}
\begin{align}
 B_r & = \frac{15 \sqrt{\frac{7}{\upi }} e^{3\, i \,\psi } \sin^3\vartheta  \cos \vartheta  a^B_{4,3} \, h_4^{(1)}(3 \, k \,  r)}{4 r} \\
 B_\vartheta & = \frac{3 \sqrt{\frac{7}{\upi }} e^{3\, i \,\psi }}{128 r \,\rlight} \sin ^2\vartheta  (\sqrt{3} \, c \, \mu _0 \, r (\sqrt{22} (9 \cos 2\,\vartheta +7)\, a^D_{5,3} h_5(3 \, k \,  r)+ \nonumber \\
 & 8 \sqrt{5} \, a^D_{3,3} \, h_3^{(1)}(3 \, k \,  r))+24 \,\rlight\, (2 \cos 2\,\vartheta +1)\, a^B_{4,3} \partial_r(r\,h_4^{(1)}(3 \, k \,  r)) \\
 B_\varphi & = \frac{3 i \sqrt{\frac{7}{\upi }}}{128 r \,\rlight} e^{3\, i \,\psi } \sin ^2\vartheta  \cos \vartheta  (\sqrt{3} \, c \, \mu _0 \, r (\sqrt{22} (15 \cos 2\,\vartheta +1)\, a^D_{5,3} h_5(3 \, k \,  r) + \nonumber \\
 & 8 \sqrt{5} \, a^D_{3,3} \, h_3^{(1)}(3 \, k \,  r))+72 \,\rlight\, a^B_{4,3} \partial_r(r\,h_4^{(1)}(3 \, k \,  r))
\end{align}
\end{subequations}
the $m=4$ reads
\begin{subequations}
\begin{align}
 B_r & = -\frac{15 \sqrt{\frac{7}{2 \upi }} e^{4\, i \,\psi } \sin ^4\vartheta  a^B_{4,4} \, h_4^{(1)}(4 \, k \,  r)}{8 r} \\
 B_\vartheta & = -\frac{\sqrt{\frac{7}{\upi }} e^{4\, i \,\psi } \sin ^3\vartheta  \cos \vartheta  \left(2 \sqrt{33} \, c \, \mu _0 \, r a^D_{5,4} h_5(4 \, k \,  r)+3 \sqrt{2} \,\rlight\, a^B_{4,4} \partial_r(r\,h_4^{(1)}(4 \, k \,  r)\right)}{4 r \,\rlight} \\
 B_\varphi & = -\frac{i \sqrt{\frac{7}{\upi }} e^{4\, i \,\psi } \sin ^3\vartheta  \left(\sqrt{33} \, c \, \mu _0 \, r (5 \cos 2\,\vartheta +3)\, a^D_{5,4} h_5(4 \, k \,  r)+12 \sqrt{2} \,\rlight\, a^B_{4,4} \partial_r(r\,h_4^{(1)}(4 \, k \,  r)\right)}{16 r \,\rlight} \ .
\end{align}
\end{subequations}
For the electric field we have for the $m=0$ 
\begin{subequations}
\begin{align}
 D_r & = -\frac{\sqrt{\frac{3}{\upi }} \cos \vartheta  \left(\sqrt{110} (-28 \cos 2\,\vartheta +63 \cos 4\,\vartheta +29)\, a^D_{4,0}+32 \sqrt{7} r^2 (5 \cos 2\,\vartheta -1)\, a^D_{3,0}\right)}{128 r^7}\\
 D_\vartheta & = -\frac{\sqrt{\frac{3}{\upi }} \sin \vartheta  \left(5 \sqrt{110} (28 \cos 2\,\vartheta +21 \cos 4\,\vartheta +15)\, a^D_{4,0}+48 \sqrt{7} r^2 (5 \cos 2\,\vartheta +3)\, a^D_{3,0}\right)}{256 r^7} \\
 D_\varphi & = 0
\end{align}
\end{subequations}
the $m=1$ reads
\begin{subequations}
\begin{align}
 D_r & = \frac{3 e^{i\,\psi } \sin \vartheta  \left(16 \sqrt{7} (5 \cos 2\,\vartheta +3)\, a^D_{3,1} \, h_3^{(1)}(k \, r)+5 \sqrt{11} (28 \cos 2\,\vartheta +21 \cos 4\,\vartheta +15)\, a^D_{5,1} h_5(k \, r)\right)}{128\, \sqrt{\upi }\, r} \\
 D_\vartheta & = \frac{e^{i\,\psi }}{256 \sqrt{\upi } \, c \, \mu _0 \, r \,\rlight} \cos \vartheta  (c \mu _0 \,\rlight\, (8 \sqrt{7} (15 \cos 2\,\vartheta -7)\, a^D_{3,1} \partial_r(r\,h_3^{(1)}(k \, r)+ \nonumber \\
 & \sqrt{11} (-84 \cos 2\,\vartheta +105 \cos 4\,\vartheta +43)\, a^D_{5,1} \partial_r(r\,h_5^{(1)}(k \, r))-24 (7 \cos 2\,\vartheta +1)\, a^B_{4,1} r\,h_4^{(1)}(k \, r)) \\
 D_\varphi & = \frac{i e^{i\,\psi }}{256 \sqrt{\upi } \, c \, \mu _0 \, r \,\rlight} (c \mu _0 \,\rlight\, (8 \sqrt{7} (5 \cos 2\,\vartheta +3)\, a^D_{3,1} \partial_r(r\,h_3^{(1)}(k \, r)+ \nonumber \\
 & \sqrt{11} (28 \cos 2\,\vartheta +21 \cos 4\,\vartheta +15)\, a^D_{5,1} \partial_r(r\,h_5^{(1)}(k \, r))-24 (\cos 2\,\vartheta +7 \cos 4\,\vartheta )\, a^B_{4,1} r\,h_4^{(1)}(k \, r))
\end{align}
\end{subequations}
the $m=2$ reads
\begin{subequations}
\begin{align}
 D_r & = -\frac{3 \sqrt{\frac{7}{\upi }} e^{2\, i \,\psi } \sin ^2\vartheta  \cos \vartheta  \left(5 \sqrt{11} (3 \cos 2\,\vartheta +1)\, a^D_{5,2} h_5(2 \, k \,  r)+4 \sqrt{10} \, a^D_{3,2} \, h_3^{(1)}(2 \, k \,  r)\right)}{16 r} \\
 D_\vartheta & = -\frac{e^{2\, i \,\psi }}{64 \sqrt{\upi } \, c \, \mu _0 \, r \,\rlight} \sin \vartheta  (\sqrt{7} c \mu _0 \,\rlight\, (4 \sqrt{10} (3 \cos 2\,\vartheta +1)\, a^D_{3,2} \partial_r(r\,h_3^{(1)}(2 \, k \,  r)+\nonumber\\
 & \sqrt{11} (12 \cos 2\,\vartheta +15 \cos 4\,\vartheta +5)\, a^D_{5,2} \partial_r(r\,h_5^{(1)}(2 \, k \,  r))-  12 \sqrt{2} (7 \cos 2\,\vartheta +5)\, a^B_{4,2} r\,h_4^{(1)}(2 \, k \,  r)) \\
 D_\varphi & = \frac{i e^{2\, i \,\psi }}{16 \sqrt{\upi } \, c \, \mu _0 \, r \,\rlight} \sin 2\,\vartheta  (3 \sqrt{2} (7 \cos 2\,\vartheta -1)\, a^B_{4,2} r\,h_4^{(1)}(2 \, k \,  r)- \nonumber\\
 & \sqrt{7} c \mu _0 \,\rlight\, (\sqrt{11} (3 \cos 2\,\vartheta +1)\, a^D_{5,2} \partial_r(r\,h_5^{(1)}(2 \, k \,  r)+2 \sqrt{10} \, a^D_{3,2} \partial_r(r\,h_3^{(1)}(2 \, k \,  r)))
\end{align}
\end{subequations}
the $m=3$ reads
\begin{subequations}
\begin{align}
 D_r & = \frac{\sqrt{\frac{21}{\upi }} e^{3\, i \,\psi } \sin ^3\vartheta  \left(5 \sqrt{22} (9 \cos 2\,\vartheta +7)\, a^D_{5,3} h_5(3 \, k \,  r)+16 \sqrt{5} \, a^D_{3,3} \, h_3^{(1)}(3 \, k \,  r)\right)}{64 r} \\
 D_\vartheta & = \frac{3 \sqrt{\frac{7}{\upi }} e^{3\, i \,\psi }}{128 c \mu _0 r \,\rlight} \sin^2\vartheta  \cos \vartheta  (\sqrt{3} c \mu _0 \,\rlight\, (\sqrt{22} (15 \cos 2\,\vartheta +1)\, a^D_{5,3} \partial_r(r\,h_5^{(1)}(3 \, k \,  r)+ \nonumber\\
 & 8 \sqrt{5} \, a^D_{3,3} \partial_r(r\,h_3^{(1)}(3 \, k \,  r))-72 a^B_{4,3} r\,h_4^{(1)}(3 \, k \,  r)) \\
 D_\varphi & = \frac{3 i \sqrt{\frac{7}{\upi }} e^{3\, i \,\psi }}{128 c \mu _0 r \,\rlight} \sin ^2\vartheta  (\sqrt{3} c \mu _0 \,\rlight\, (\sqrt{22} (9 \cos 2\,\vartheta +7)\, a^D_{5,3} \partial_r(r\,h_5^{(1)}(3 \, k \,  r)+\nonumber\\
 &8 \sqrt{5} \, a^D_{3,3} \partial_r(r\,h_3^{(1)}(3 \, k \,  r))-24 (2 \cos 2\,\vartheta +1)\, a^B_{4,3} r\,h_4^{(1)}(3 \, k \,  r))
\end{align}
\end{subequations}
the $m=4$ reads
\begin{subequations}
\begin{align}
 D_r & = -\frac{15 \sqrt{\frac{231}{\upi }} e^{4\, i \,\psi } \sin ^4\vartheta  \cos \vartheta  a^D_{5,4} h_5(4 \, k \,  r)}{16 r} \\
 D_\vartheta & = -\frac{\sqrt{\frac{7}{\upi }} e^{4\, i \,\psi } \sin ^3\vartheta  \left(\sqrt{33} c \mu _0 \,\rlight\, (5 \cos 2\,\vartheta +3)\, a^D_{5,4} \partial_r(r\,h_5^{(1)}(4 \, k \,  r)-12 \sqrt{2} \, a^B_{4,4} r\,h_4^{(1)}(4 \, k \,  r)\right)}{16 c \mu _0 r \,\rlight} \\
 D_\varphi & = \frac{i \sqrt{\frac{7}{\upi }} e^{4\, i \,\psi } \sin ^3\vartheta  \cos \vartheta  \left(3 \sqrt{2} \, a^B_{4,4} r\,h_4^{(1)}(4 \, k \,  r)-2 \sqrt{33} c \mu _0 \,\rlight\, a^D_{5,4} \partial_r(r\,h_5^{(1)}(4 \, k \,  r)\right)}{4 c \mu _0 r \,\rlight} \ .
\end{align}
\end{subequations}

\label{lastpage}

\end{document}

%% file: offset_aligned_dipole.tex
\begingroup
  \makeatletter
  \providecommand\color[2][]{%
    \GenericError{(gnuplot) \space\space\space\@spaces}{%
      Package color not loaded in conjunction with
      terminal option `colourtext'%
    }{See the gnuplot documentation for explanation.%
    }{Either use 'blacktext' in gnuplot or load the package
      color.sty in LaTeX.}%
    \renewcommand\color[2][]{}%
  }%
  \providecommand\includegraphics[2][]{%
    \GenericError{(gnuplot) \space\space\space\@spaces}{%
      Package graphicx or graphics not loaded%
    }{See the gnuplot documentation for explanation.%
    }{The gnuplot epslatex terminal needs graphicx.sty or graphics.sty.}%
    \renewcommand\includegraphics[2][]{}%
  }%
  \providecommand\rotatebox[2]{#2}%
  \@ifundefined{ifGPcolor}{%
    \newif\ifGPcolor
    \GPcolortrue
  }{}%
  \@ifundefined{ifGPblacktext}{%
    \newif\ifGPblacktext
    \GPblacktextfalse
  }{}%
  \let\gplgaddtomacro\g@addto@macro
  \gdef\gplbacktext{}%
  \gdef\gplfronttext{}%
  \makeatother
  \ifGPblacktext
    \def\colorrgb#1{}%
    \def\colorgray#1{}%
  \else
    \ifGPcolor
      \def\colorrgb#1{\color[rgb]{#1}}%
      \def\colorgray#1{\color[gray]{#1}}%
      \expandafter\def\csname LTw\endcsname{\color{white}}%
      \expandafter\def\csname LTb\endcsname{\color{black}}%
      \expandafter\def\csname LTa\endcsname{\color{black}}%
      \expandafter\def\csname LT0\endcsname{\color[rgb]{1,0,0}}%
      \expandafter\def\csname LT1\endcsname{\color[rgb]{0,1,0}}%
      \expandafter\def\csname LT2\endcsname{\color[rgb]{0,0,1}}%
      \expandafter\def\csname LT3\endcsname{\color[rgb]{1,0,1}}%
      \expandafter\def\csname LT4\endcsname{\color[rgb]{0,1,1}}%
      \expandafter\def\csname LT5\endcsname{\color[rgb]{1,1,0}}%
      \expandafter\def\csname LT6\endcsname{\color[rgb]{0,0,0}}%
      \expandafter\def\csname LT7\endcsname{\color[rgb]{1,0.3,0}}%
      \expandafter\def\csname LT8\endcsname{\color[rgb]{0.5,0.5,0.5}}%
    \else
      \def\colorrgb#1{\color{black}}%
      \def\colorgray#1{\color[gray]{#1}}%
      \expandafter\def\csname LTw\endcsname{\color{white}}%
      \expandafter\def\csname LTb\endcsname{\color{black}}%
      \expandafter\def\csname LTa\endcsname{\color{black}}%
      \expandafter\def\csname LT0\endcsname{\color{black}}%
      \expandafter\def\csname LT1\endcsname{\color{black}}%
      \expandafter\def\csname LT2\endcsname{\color{black}}%
      \expandafter\def\csname LT3\endcsname{\color{black}}%
      \expandafter\def\csname LT4\endcsname{\color{black}}%
      \expandafter\def\csname LT5\endcsname{\color{black}}%
      \expandafter\def\csname LT6\endcsname{\color{black}}%
      \expandafter\def\csname LT7\endcsname{\color{black}}%
      \expandafter\def\csname LT8\endcsname{\color{black}}%
    \fi
  \fi
  \setlength{\unitlength}{0.0500bp}%
  \begin{picture}(5040.00,3600.00)%
    \gplgaddtomacro\gplbacktext{%
      \csname LTb\endcsname%
      \put(1611,704){\makebox(0,0)[r]{\strut{} 0}}%
      \put(1611,1151){\makebox(0,0)[r]{\strut{} 0.2}}%
      \put(1611,1598){\makebox(0,0)[r]{\strut{} 0.4}}%
      \put(1611,2045){\makebox(0,0)[r]{\strut{} 0.6}}%
      \put(1611,2492){\makebox(0,0)[r]{\strut{} 0.8}}%
      \put(1611,2939){\makebox(0,0)[r]{\strut{} 1}}%
      \put(1743,484){\makebox(0,0){\strut{} 0}}%
      \put(2190,484){\makebox(0,0){\strut{} 0.2}}%
      \put(2637,484){\makebox(0,0){\strut{} 0.4}}%
      \put(3084,484){\makebox(0,0){\strut{} 0.6}}%
      \put(3531,484){\makebox(0,0){\strut{} 0.8}}%
      \put(3978,484){\makebox(0,0){\strut{} 1}}%
      \put(841,1821){\rotatebox{-270}{\makebox(0,0){\strut{}$z/r_{\rm L}$}}}%
      \put(2860,154){\makebox(0,0){\strut{}$x/r_{\rm L}$}}%
      \put(2860,3269){\makebox(0,0){\strut{}Aligned dipole}}%
    }%
    \gplgaddtomacro\gplfronttext{%
    }%
    \gplbacktext
    \put(0,0){\includegraphics{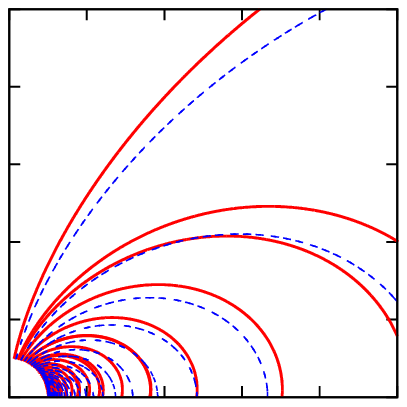}}%
    \gplfronttext
  \end{picture}%
\endgroup

%% file: offset_perp_dipole_zoom_b0.tex
\begingroup
  \makeatletter
  \providecommand\color[2][]{%
    \GenericError{(gnuplot) \space\space\space\@spaces}{%
      Package color not loaded in conjunction with
      terminal option `colourtext'%
    }{See the gnuplot documentation for explanation.%
    }{Either use 'blacktext' in gnuplot or load the package
      color.sty in LaTeX.}%
    \renewcommand\color[2][]{}%
  }%
  \providecommand\includegraphics[2][]{%
    \GenericError{(gnuplot) \space\space\space\@spaces}{%
      Package graphicx or graphics not loaded%
    }{See the gnuplot documentation for explanation.%
    }{The gnuplot epslatex terminal needs graphicx.sty or graphics.sty.}%
    \renewcommand\includegraphics[2][]{}%
  }%
  \providecommand\rotatebox[2]{#2}%
  \@ifundefined{ifGPcolor}{%
    \newif\ifGPcolor
    \GPcolortrue
  }{}%
  \@ifundefined{ifGPblacktext}{%
    \newif\ifGPblacktext
    \GPblacktextfalse
  }{}%
  \let\gplgaddtomacro\g@addto@macro
  \gdef\gplbacktext{}%
  \gdef\gplfronttext{}%
  \makeatother
  \ifGPblacktext
    \def\colorrgb#1{}%
    \def\colorgray#1{}%
  \else
    \ifGPcolor
      \def\colorrgb#1{\color[rgb]{#1}}%
      \def\colorgray#1{\color[gray]{#1}}%
      \expandafter\def\csname LTw\endcsname{\color{white}}%
      \expandafter\def\csname LTb\endcsname{\color{black}}%
      \expandafter\def\csname LTa\endcsname{\color{black}}%
      \expandafter\def\csname LT0\endcsname{\color[rgb]{1,0,0}}%
      \expandafter\def\csname LT1\endcsname{\color[rgb]{0,1,0}}%
      \expandafter\def\csname LT2\endcsname{\color[rgb]{0,0,1}}%
      \expandafter\def\csname LT3\endcsname{\color[rgb]{1,0,1}}%
      \expandafter\def\csname LT4\endcsname{\color[rgb]{0,1,1}}%
      \expandafter\def\csname LT5\endcsname{\color[rgb]{1,1,0}}%
      \expandafter\def\csname LT6\endcsname{\color[rgb]{0,0,0}}%
      \expandafter\def\csname LT7\endcsname{\color[rgb]{1,0.3,0}}%
      \expandafter\def\csname LT8\endcsname{\color[rgb]{0.5,0.5,0.5}}%
    \else
      \def\colorrgb#1{\color{black}}%
      \def\colorgray#1{\color[gray]{#1}}%
      \expandafter\def\csname LTw\endcsname{\color{white}}%
      \expandafter\def\csname LTb\endcsname{\color{black}}%
      \expandafter\def\csname LTa\endcsname{\color{black}}%
      \expandafter\def\csname LT0\endcsname{\color{black}}%
      \expandafter\def\csname LT1\endcsname{\color{black}}%
      \expandafter\def\csname LT2\endcsname{\color{black}}%
      \expandafter\def\csname LT3\endcsname{\color{black}}%
      \expandafter\def\csname LT4\endcsname{\color{black}}%
      \expandafter\def\csname LT5\endcsname{\color{black}}%
      \expandafter\def\csname LT6\endcsname{\color{black}}%
      \expandafter\def\csname LT7\endcsname{\color{black}}%
      \expandafter\def\csname LT8\endcsname{\color{black}}%
    \fi
  \fi
  \setlength{\unitlength}{0.0500bp}%
  \begin{picture}(8640.00,4462.00)%
    \gplgaddtomacro\gplbacktext{%
      \csname LTb\endcsname%
      \put(462,1119){\makebox(0,0)[r]{\strut{}-4}}%
      \put(462,1785){\makebox(0,0)[r]{\strut{}-2}}%
      \put(462,2450){\makebox(0,0)[r]{\strut{}0}}%
      \put(462,3116){\makebox(0,0)[r]{\strut{}2}}%
      \put(462,3782){\makebox(0,0)[r]{\strut{}4}}%
      \put(927,566){\makebox(0,0){\strut{}-4}}%
      \put(1593,566){\makebox(0,0){\strut{}-2}}%
      \put(2258,566){\makebox(0,0){\strut{}0}}%
      \put(2924,566){\makebox(0,0){\strut{}2}}%
      \put(3590,566){\makebox(0,0){\strut{}4}}%
      \put(-44,2450){\rotatebox{-270}{\makebox(0,0){\strut{}$y/r_{\rm L}$}}}%
      \put(2258,236){\makebox(0,0){\strut{}$x/r_{\rm L}$}}%
      \put(2258,4005){\makebox(0,0){\strut{}}}%
    }%
    \gplgaddtomacro\gplfronttext{%
    }%
    \gplgaddtomacro\gplbacktext{%
      \csname LTb\endcsname%
      \put(4809,1286){\makebox(0,0)[r]{\strut{}-1}}%
      \put(4809,2450){\makebox(0,0)[r]{\strut{}0}}%
      \put(4809,3615){\makebox(0,0)[r]{\strut{}1}}%
      \put(5523,484){\makebox(0,0){\strut{}-1}}%
      \put(6688,484){\makebox(0,0){\strut{}0}}%
      \put(7853,484){\makebox(0,0){\strut{}1}}%
      \put(4303,2450){\rotatebox{-270}{\makebox(0,0){\strut{}$y/r_{\rm L}$}}}%
      \put(6688,154){\makebox(0,0){\strut{}$x/r_{\rm L}$}}%
      \put(6688,4087){\makebox(0,0){\strut{}}}%
    }%
    \gplgaddtomacro\gplfronttext{%
    }%
    \gplbacktext
    \put(0,0){\includegraphics{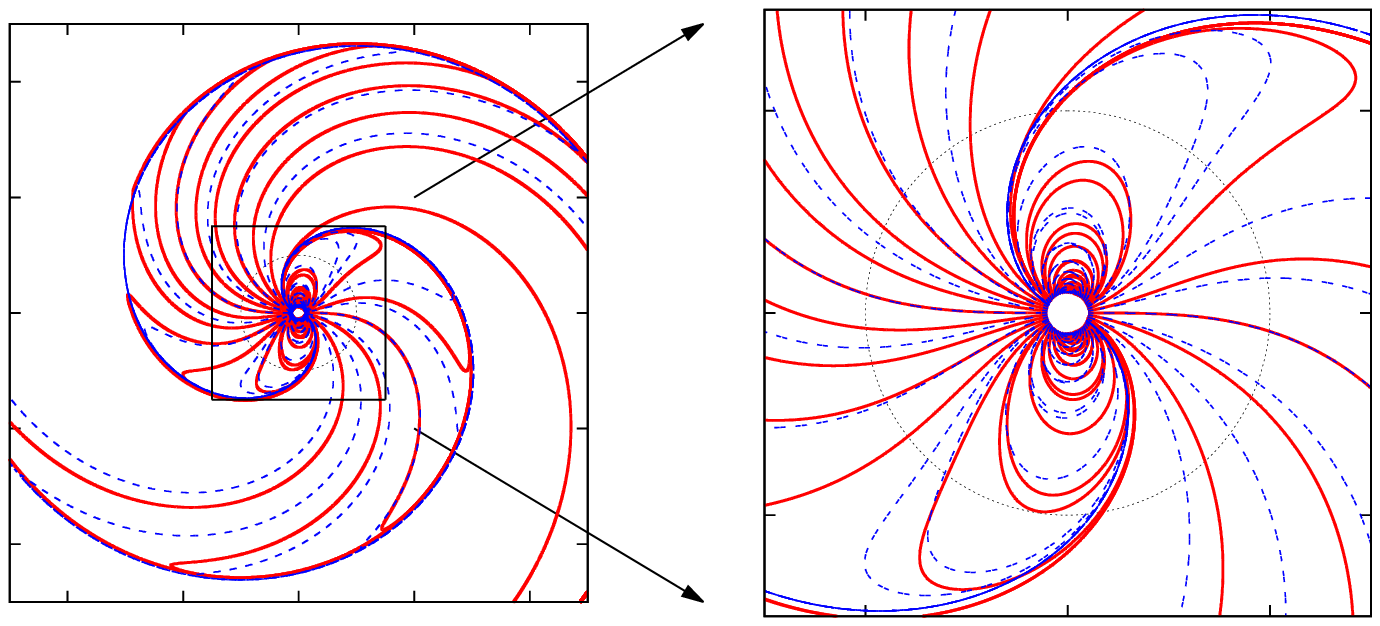}}%
    \gplfronttext
  \end{picture}%
\endgroup

%% file: offset_perp_dipole_zoom_b1.tex
\begingroup
  \makeatletter
  \providecommand\color[2][]{%
    \GenericError{(gnuplot) \space\space\space\@spaces}{%
      Package color not loaded in conjunction with
      terminal option `colourtext'%
    }{See the gnuplot documentation for explanation.%
    }{Either use 'blacktext' in gnuplot or load the package
      color.sty in LaTeX.}%
    \renewcommand\color[2][]{}%
  }%
  \providecommand\includegraphics[2][]{%
    \GenericError{(gnuplot) \space\space\space\@spaces}{%
      Package graphicx or graphics not loaded%
    }{See the gnuplot documentation for explanation.%
    }{The gnuplot epslatex terminal needs graphicx.sty or graphics.sty.}%
    \renewcommand\includegraphics[2][]{}%
  }%
  \providecommand\rotatebox[2]{#2}%
  \@ifundefined{ifGPcolor}{%
    \newif\ifGPcolor
    \GPcolortrue
  }{}%
  \@ifundefined{ifGPblacktext}{%
    \newif\ifGPblacktext
    \GPblacktextfalse
  }{}%
  \let\gplgaddtomacro\g@addto@macro
  \gdef\gplbacktext{}%
  \gdef\gplfronttext{}%
  \makeatother
  \ifGPblacktext
    \def\colorrgb#1{}%
    \def\colorgray#1{}%
  \else
    \ifGPcolor
      \def\colorrgb#1{\color[rgb]{#1}}%
      \def\colorgray#1{\color[gray]{#1}}%
      \expandafter\def\csname LTw\endcsname{\color{white}}%
      \expandafter\def\csname LTb\endcsname{\color{black}}%
      \expandafter\def\csname LTa\endcsname{\color{black}}%
      \expandafter\def\csname LT0\endcsname{\color[rgb]{1,0,0}}%
      \expandafter\def\csname LT1\endcsname{\color[rgb]{0,1,0}}%
      \expandafter\def\csname LT2\endcsname{\color[rgb]{0,0,1}}%
      \expandafter\def\csname LT3\endcsname{\color[rgb]{1,0,1}}%
      \expandafter\def\csname LT4\endcsname{\color[rgb]{0,1,1}}%
      \expandafter\def\csname LT5\endcsname{\color[rgb]{1,1,0}}%
      \expandafter\def\csname LT6\endcsname{\color[rgb]{0,0,0}}%
      \expandafter\def\csname LT7\endcsname{\color[rgb]{1,0.3,0}}%
      \expandafter\def\csname LT8\endcsname{\color[rgb]{0.5,0.5,0.5}}%
    \else
      \def\colorrgb#1{\color{black}}%
      \def\colorgray#1{\color[gray]{#1}}%
      \expandafter\def\csname LTw\endcsname{\color{white}}%
      \expandafter\def\csname LTb\endcsname{\color{black}}%
      \expandafter\def\csname LTa\endcsname{\color{black}}%
      \expandafter\def\csname LT0\endcsname{\color{black}}%
      \expandafter\def\csname LT1\endcsname{\color{black}}%
      \expandafter\def\csname LT2\endcsname{\color{black}}%
      \expandafter\def\csname LT3\endcsname{\color{black}}%
      \expandafter\def\csname LT4\endcsname{\color{black}}%
      \expandafter\def\csname LT5\endcsname{\color{black}}%
      \expandafter\def\csname LT6\endcsname{\color{black}}%
      \expandafter\def\csname LT7\endcsname{\color{black}}%
      \expandafter\def\csname LT8\endcsname{\color{black}}%
    \fi
  \fi
  \setlength{\unitlength}{0.0500bp}%
  \begin{picture}(8640.00,4462.00)%
    \gplgaddtomacro\gplbacktext{%
      \csname LTb\endcsname%
      \put(462,1119){\makebox(0,0)[r]{\strut{}-4}}%
      \put(462,1785){\makebox(0,0)[r]{\strut{}-2}}%
      \put(462,2450){\makebox(0,0)[r]{\strut{}0}}%
      \put(462,3116){\makebox(0,0)[r]{\strut{}2}}%
      \put(462,3782){\makebox(0,0)[r]{\strut{}4}}%
      \put(927,566){\makebox(0,0){\strut{}-4}}%
      \put(1593,566){\makebox(0,0){\strut{}-2}}%
      \put(2258,566){\makebox(0,0){\strut{}0}}%
      \put(2924,566){\makebox(0,0){\strut{}2}}%
      \put(3590,566){\makebox(0,0){\strut{}4}}%
      \put(-44,2450){\rotatebox{-270}{\makebox(0,0){\strut{}$y/r_{\rm L}$}}}%
      \put(2258,236){\makebox(0,0){\strut{}$x/r_{\rm L}$}}%
      \put(2258,4005){\makebox(0,0){\strut{}}}%
    }%
    \gplgaddtomacro\gplfronttext{%
    }%
    \gplgaddtomacro\gplbacktext{%
      \csname LTb\endcsname%
      \put(4809,1286){\makebox(0,0)[r]{\strut{}-1}}%
      \put(4809,2450){\makebox(0,0)[r]{\strut{}0}}%
      \put(4809,3615){\makebox(0,0)[r]{\strut{}1}}%
      \put(5523,484){\makebox(0,0){\strut{}-1}}%
      \put(6688,484){\makebox(0,0){\strut{}0}}%
      \put(7853,484){\makebox(0,0){\strut{}1}}%
      \put(4303,2450){\rotatebox{-270}{\makebox(0,0){\strut{}$y/r_{\rm L}$}}}%
      \put(6688,154){\makebox(0,0){\strut{}$x/r_{\rm L}$}}%
      \put(6688,4087){\makebox(0,0){\strut{}}}%
    }%
    \gplgaddtomacro\gplfronttext{%
    }%
    \gplbacktext
    \put(0,0){\includegraphics{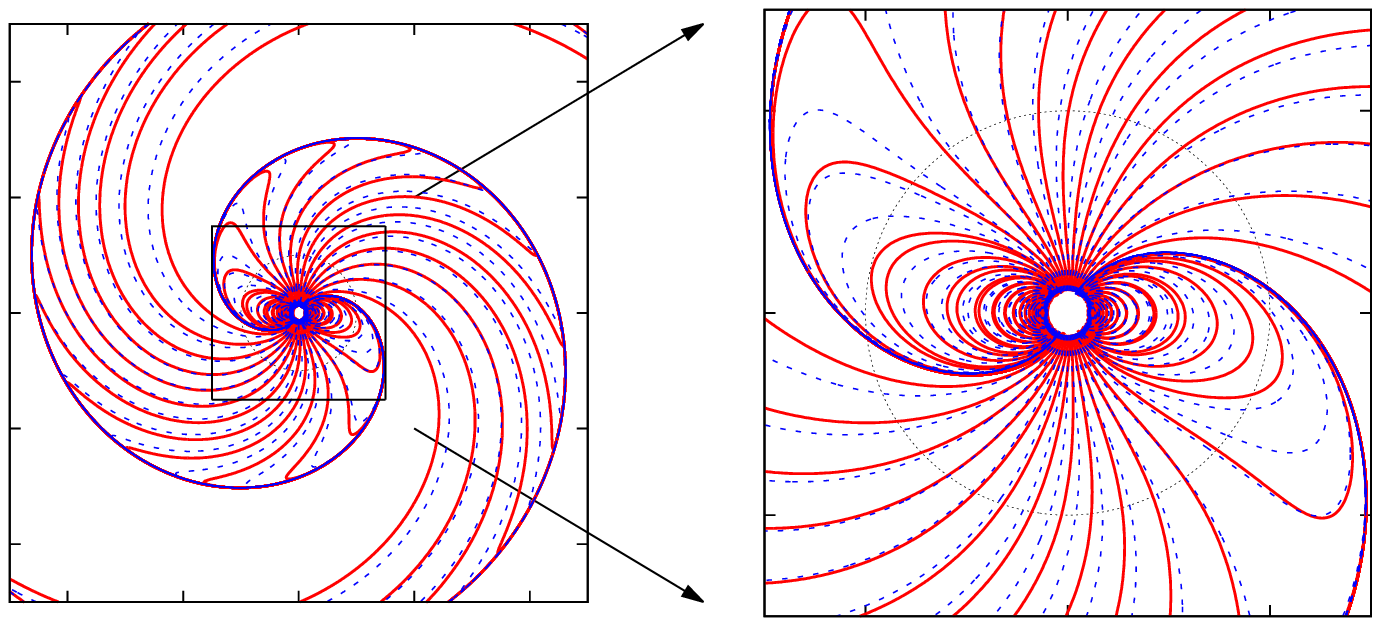}}%
    \gplfronttext
  \end{picture}%
\endgroup

%% file: indice_freinage.tex
\begingroup
  \makeatletter
  \providecommand\color[2][]{%
    \GenericError{(gnuplot) \space\space\space\@spaces}{%
      Package color not loaded in conjunction with
      terminal option `colourtext'%
    }{See the gnuplot documentation for explanation.%
    }{Either use 'blacktext' in gnuplot or load the package
      color.sty in LaTeX.}%
    \renewcommand\color[2][]{}%
  }%
  \providecommand\includegraphics[2][]{%
    \GenericError{(gnuplot) \space\space\space\@spaces}{%
      Package graphicx or graphics not loaded%
    }{See the gnuplot documentation for explanation.%
    }{The gnuplot epslatex terminal needs graphicx.sty or graphics.sty.}%
    \renewcommand\includegraphics[2][]{}%
  }%
  \providecommand\rotatebox[2]{#2}%
  \@ifundefined{ifGPcolor}{%
    \newif\ifGPcolor
    \GPcolortrue
  }{}%
  \@ifundefined{ifGPblacktext}{%
    \newif\ifGPblacktext
    \GPblacktextfalse
  }{}%
  \let\gplgaddtomacro\g@addto@macro
  \gdef\gplbacktext{}%
  \gdef\gplfronttext{}%
  \makeatother
  \ifGPblacktext
    \def\colorrgb#1{}%
    \def\colorgray#1{}%
  \else
    \ifGPcolor
      \def\colorrgb#1{\color[rgb]{#1}}%
      \def\colorgray#1{\color[gray]{#1}}%
      \expandafter\def\csname LTw\endcsname{\color{white}}%
      \expandafter\def\csname LTb\endcsname{\color{black}}%
      \expandafter\def\csname LTa\endcsname{\color{black}}%
      \expandafter\def\csname LT0\endcsname{\color[rgb]{1,0,0}}%
      \expandafter\def\csname LT1\endcsname{\color[rgb]{0,1,0}}%
      \expandafter\def\csname LT2\endcsname{\color[rgb]{0,0,1}}%
      \expandafter\def\csname LT3\endcsname{\color[rgb]{1,0,1}}%
      \expandafter\def\csname LT4\endcsname{\color[rgb]{0,1,1}}%
      \expandafter\def\csname LT5\endcsname{\color[rgb]{1,1,0}}%
      \expandafter\def\csname LT6\endcsname{\color[rgb]{0,0,0}}%
      \expandafter\def\csname LT7\endcsname{\color[rgb]{1,0.3,0}}%
      \expandafter\def\csname LT8\endcsname{\color[rgb]{0.5,0.5,0.5}}%
    \else
      \def\colorrgb#1{\color{black}}%
      \def\colorgray#1{\color[gray]{#1}}%
      \expandafter\def\csname LTw\endcsname{\color{white}}%
      \expandafter\def\csname LTb\endcsname{\color{black}}%
      \expandafter\def\csname LTa\endcsname{\color{black}}%
      \expandafter\def\csname LT0\endcsname{\color{black}}%
      \expandafter\def\csname LT1\endcsname{\color{black}}%
      \expandafter\def\csname LT2\endcsname{\color{black}}%
      \expandafter\def\csname LT3\endcsname{\color{black}}%
      \expandafter\def\csname LT4\endcsname{\color{black}}%
      \expandafter\def\csname LT5\endcsname{\color{black}}%
      \expandafter\def\csname LT6\endcsname{\color{black}}%
      \expandafter\def\csname LT7\endcsname{\color{black}}%
      \expandafter\def\csname LT8\endcsname{\color{black}}%
    \fi
  \fi
  \setlength{\unitlength}{0.0500bp}%
  \begin{picture}(5040.00,3600.00)%
    \gplgaddtomacro\gplbacktext{%
      \csname LTb\endcsname%
      \put(946,806){\makebox(0,0)[r]{\strut{}3.00}}%
      \put(946,1314){\makebox(0,0)[r]{\strut{}3.50}}%
      \put(946,1822){\makebox(0,0)[r]{\strut{}4.00}}%
      \put(946,2329){\makebox(0,0)[r]{\strut{}4.50}}%
      \put(946,2837){\makebox(0,0)[r]{\strut{}5.00}}%
      \put(1078,484){\makebox(0,0){\strut{}1e-05}}%
      \put(1791,484){\makebox(0,0){\strut{}1e-04}}%
      \put(2504,484){\makebox(0,0){\strut{}1e-03}}%
      \put(3217,484){\makebox(0,0){\strut{}1e-02}}%
      \put(3930,484){\makebox(0,0){\strut{}1e-01}}%
      \put(4643,484){\makebox(0,0){\strut{}1e+00}}%
      \put(176,1821){\rotatebox{-270}{\makebox(0,0){\strut{}$n$}}}%
      \put(2860,154){\makebox(0,0){\strut{}$\alpha \textrm{ (in } \degr)$}}%
      \put(2860,3269){\makebox(0,0){\strut{}Braking index}}%
    }%
    \gplgaddtomacro\gplfronttext{%
      \csname LTb\endcsname%
      \put(3819,2766){\makebox(0,0){\strut{}$\epsilon$}}%
      \csname LTb\endcsname%
      \put(3656,2546){\makebox(0,0)[r]{\strut{}0.01}}%
      \csname LTb\endcsname%
      \put(3656,2326){\makebox(0,0)[r]{\strut{}0.02}}%
      \csname LTb\endcsname%
      \put(3656,2106){\makebox(0,0)[r]{\strut{}0.05}}%
      \csname LTb\endcsname%
      \put(3656,1886){\makebox(0,0)[r]{\strut{}0.1}}%
      \csname LTb\endcsname%
      \put(3656,1666){\makebox(0,0)[r]{\strut{}0.2}}%
    }%
    \gplbacktext
    \put(0,0){\includegraphics{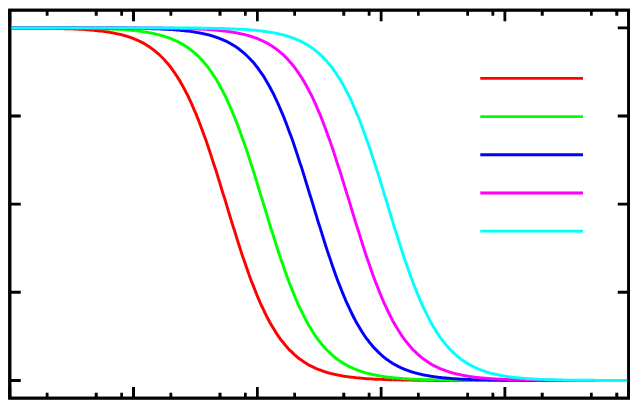}}%
    \gplfronttext
  \end{picture}%
\endgroup

%% file: spirale_in_zoom_b0.tex
\begingroup
  \makeatletter
  \providecommand\color[2][]{%
    \GenericError{(gnuplot) \space\space\space\@spaces}{%
      Package color not loaded in conjunction with
      terminal option `colourtext'%
    }{See the gnuplot documentation for explanation.%
    }{Either use 'blacktext' in gnuplot or load the package
      color.sty in LaTeX.}%
    \renewcommand\color[2][]{}%
  }%
  \providecommand\includegraphics[2][]{%
    \GenericError{(gnuplot) \space\space\space\@spaces}{%
      Package graphicx or graphics not loaded%
    }{See the gnuplot documentation for explanation.%
    }{The gnuplot epslatex terminal needs graphicx.sty or graphics.sty.}%
    \renewcommand\includegraphics[2][]{}%
  }%
  \providecommand\rotatebox[2]{#2}%
  \@ifundefined{ifGPcolor}{%
    \newif\ifGPcolor
    \GPcolortrue
  }{}%
  \@ifundefined{ifGPblacktext}{%
    \newif\ifGPblacktext
    \GPblacktextfalse
  }{}%
  \let\gplgaddtomacro\g@addto@macro
  \gdef\gplbacktext{}%
  \gdef\gplfronttext{}%
  \makeatother
  \ifGPblacktext
    \def\colorrgb#1{}%
    \def\colorgray#1{}%
  \else
    \ifGPcolor
      \def\colorrgb#1{\color[rgb]{#1}}%
      \def\colorgray#1{\color[gray]{#1}}%
      \expandafter\def\csname LTw\endcsname{\color{white}}%
      \expandafter\def\csname LTb\endcsname{\color{black}}%
      \expandafter\def\csname LTa\endcsname{\color{black}}%
      \expandafter\def\csname LT0\endcsname{\color[rgb]{1,0,0}}%
      \expandafter\def\csname LT1\endcsname{\color[rgb]{0,1,0}}%
      \expandafter\def\csname LT2\endcsname{\color[rgb]{0,0,1}}%
      \expandafter\def\csname LT3\endcsname{\color[rgb]{1,0,1}}%
      \expandafter\def\csname LT4\endcsname{\color[rgb]{0,1,1}}%
      \expandafter\def\csname LT5\endcsname{\color[rgb]{1,1,0}}%
      \expandafter\def\csname LT6\endcsname{\color[rgb]{0,0,0}}%
      \expandafter\def\csname LT7\endcsname{\color[rgb]{1,0.3,0}}%
      \expandafter\def\csname LT8\endcsname{\color[rgb]{0.5,0.5,0.5}}%
    \else
      \def\colorrgb#1{\color{black}}%
      \def\colorgray#1{\color[gray]{#1}}%
      \expandafter\def\csname LTw\endcsname{\color{white}}%
      \expandafter\def\csname LTb\endcsname{\color{black}}%
      \expandafter\def\csname LTa\endcsname{\color{black}}%
      \expandafter\def\csname LT0\endcsname{\color{black}}%
      \expandafter\def\csname LT1\endcsname{\color{black}}%
      \expandafter\def\csname LT2\endcsname{\color{black}}%
      \expandafter\def\csname LT3\endcsname{\color{black}}%
      \expandafter\def\csname LT4\endcsname{\color{black}}%
      \expandafter\def\csname LT5\endcsname{\color{black}}%
      \expandafter\def\csname LT6\endcsname{\color{black}}%
      \expandafter\def\csname LT7\endcsname{\color{black}}%
      \expandafter\def\csname LT8\endcsname{\color{black}}%
    \fi
  \fi
  \setlength{\unitlength}{0.0500bp}%
  \begin{picture}(5040.00,3600.00)%
    \gplgaddtomacro\gplbacktext{%
      \csname LTb\endcsname%
      \put(1545,704){\makebox(0,0)[r]{\strut{} 10}}%
      \put(1545,1263){\makebox(0,0)[r]{\strut{} 5}}%
      \put(1545,1822){\makebox(0,0)[r]{\strut{} 0}}%
      \put(1545,2380){\makebox(0,0)[r]{\strut{} 5}}%
      \put(1545,2939){\makebox(0,0)[r]{\strut{} 10}}%
      \put(1677,484){\makebox(0,0){\strut{} 10}}%
      \put(2236,484){\makebox(0,0){\strut{} 5}}%
      \put(2795,484){\makebox(0,0){\strut{} 0}}%
      \put(3353,484){\makebox(0,0){\strut{} 5}}%
      \put(3912,484){\makebox(0,0){\strut{} 10}}%
      \put(907,1821){\rotatebox{-270}{\makebox(0,0){\strut{}$y/r_{\rm L}$}}}%
      \put(2794,154){\makebox(0,0){\strut{}$x/r_{\rm L}$}}%
      \put(2794,3269){\makebox(0,0){\strut{}Spirals}}%
    }%
    \gplgaddtomacro\gplfronttext{%
      \csname LTb\endcsname%
      \put(2925,2766){\makebox(0,0)[r]{\strut{}off-centred}}%
      \csname LTb\endcsname%
      \put(2925,2546){\makebox(0,0)[r]{\strut{}centred}}%
    }%
    \gplgaddtomacro\gplbacktext{%
    }%
    \gplgaddtomacro\gplfronttext{%
    }%
    \gplgaddtomacro\gplbacktext{%
    }%
    \gplgaddtomacro\gplfronttext{%
    }%
    \gplbacktext
    \put(0,0){\includegraphics{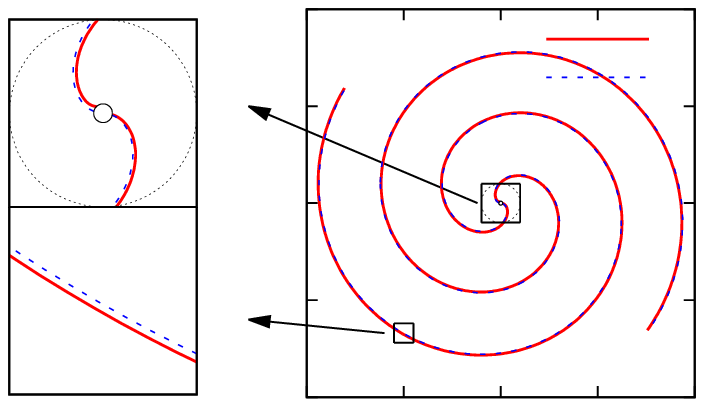}}%
    \gplfronttext
  \end{picture}%
\endgroup

%% file: spirale_in_zoom_b1.tex
\begingroup
  \makeatletter
  \providecommand\color[2][]{%
    \GenericError{(gnuplot) \space\space\space\@spaces}{%
      Package color not loaded in conjunction with
      terminal option `colourtext'%
    }{See the gnuplot documentation for explanation.%
    }{Either use 'blacktext' in gnuplot or load the package
      color.sty in LaTeX.}%
    \renewcommand\color[2][]{}%
  }%
  \providecommand\includegraphics[2][]{%
    \GenericError{(gnuplot) \space\space\space\@spaces}{%
      Package graphicx or graphics not loaded%
    }{See the gnuplot documentation for explanation.%
    }{The gnuplot epslatex terminal needs graphicx.sty or graphics.sty.}%
    \renewcommand\includegraphics[2][]{}%
  }%
  \providecommand\rotatebox[2]{#2}%
  \@ifundefined{ifGPcolor}{%
    \newif\ifGPcolor
    \GPcolortrue
  }{}%
  \@ifundefined{ifGPblacktext}{%
    \newif\ifGPblacktext
    \GPblacktextfalse
  }{}%
  \let\gplgaddtomacro\g@addto@macro
  \gdef\gplbacktext{}%
  \gdef\gplfronttext{}%
  \makeatother
  \ifGPblacktext
    \def\colorrgb#1{}%
    \def\colorgray#1{}%
  \else
    \ifGPcolor
      \def\colorrgb#1{\color[rgb]{#1}}%
      \def\colorgray#1{\color[gray]{#1}}%
      \expandafter\def\csname LTw\endcsname{\color{white}}%
      \expandafter\def\csname LTb\endcsname{\color{black}}%
      \expandafter\def\csname LTa\endcsname{\color{black}}%
      \expandafter\def\csname LT0\endcsname{\color[rgb]{1,0,0}}%
      \expandafter\def\csname LT1\endcsname{\color[rgb]{0,1,0}}%
      \expandafter\def\csname LT2\endcsname{\color[rgb]{0,0,1}}%
      \expandafter\def\csname LT3\endcsname{\color[rgb]{1,0,1}}%
      \expandafter\def\csname LT4\endcsname{\color[rgb]{0,1,1}}%
      \expandafter\def\csname LT5\endcsname{\color[rgb]{1,1,0}}%
      \expandafter\def\csname LT6\endcsname{\color[rgb]{0,0,0}}%
      \expandafter\def\csname LT7\endcsname{\color[rgb]{1,0.3,0}}%
      \expandafter\def\csname LT8\endcsname{\color[rgb]{0.5,0.5,0.5}}%
    \else
      \def\colorrgb#1{\color{black}}%
      \def\colorgray#1{\color[gray]{#1}}%
      \expandafter\def\csname LTw\endcsname{\color{white}}%
      \expandafter\def\csname LTb\endcsname{\color{black}}%
      \expandafter\def\csname LTa\endcsname{\color{black}}%
      \expandafter\def\csname LT0\endcsname{\color{black}}%
      \expandafter\def\csname LT1\endcsname{\color{black}}%
      \expandafter\def\csname LT2\endcsname{\color{black}}%
      \expandafter\def\csname LT3\endcsname{\color{black}}%
      \expandafter\def\csname LT4\endcsname{\color{black}}%
      \expandafter\def\csname LT5\endcsname{\color{black}}%
      \expandafter\def\csname LT6\endcsname{\color{black}}%
      \expandafter\def\csname LT7\endcsname{\color{black}}%
      \expandafter\def\csname LT8\endcsname{\color{black}}%
    \fi
  \fi
  \setlength{\unitlength}{0.0500bp}%
  \begin{picture}(5040.00,3600.00)%
    \gplgaddtomacro\gplbacktext{%
      \csname LTb\endcsname%
      \put(1545,704){\makebox(0,0)[r]{\strut{} 10}}%
      \put(1545,1263){\makebox(0,0)[r]{\strut{} 5}}%
      \put(1545,1822){\makebox(0,0)[r]{\strut{} 0}}%
      \put(1545,2380){\makebox(0,0)[r]{\strut{} 5}}%
      \put(1545,2939){\makebox(0,0)[r]{\strut{} 10}}%
      \put(1677,484){\makebox(0,0){\strut{} 10}}%
      \put(2236,484){\makebox(0,0){\strut{} 5}}%
      \put(2795,484){\makebox(0,0){\strut{} 0}}%
      \put(3353,484){\makebox(0,0){\strut{} 5}}%
      \put(3912,484){\makebox(0,0){\strut{} 10}}%
      \put(907,1821){\rotatebox{-270}{\makebox(0,0){\strut{}$y/r_{\rm L}$}}}%
      \put(2794,154){\makebox(0,0){\strut{}$x/r_{\rm L}$}}%
      \put(2794,3269){\makebox(0,0){\strut{}Spirals}}%
    }%
    \gplgaddtomacro\gplfronttext{%
      \csname LTb\endcsname%
      \put(2925,2766){\makebox(0,0)[r]{\strut{}off-centred}}%
      \csname LTb\endcsname%
      \put(2925,2546){\makebox(0,0)[r]{\strut{}centred}}%
    }%
    \gplgaddtomacro\gplbacktext{%
    }%
    \gplgaddtomacro\gplfronttext{%
    }%
    \gplgaddtomacro\gplbacktext{%
    }%
    \gplgaddtomacro\gplfronttext{%
    }%
    \gplbacktext
    \put(0,0){\includegraphics{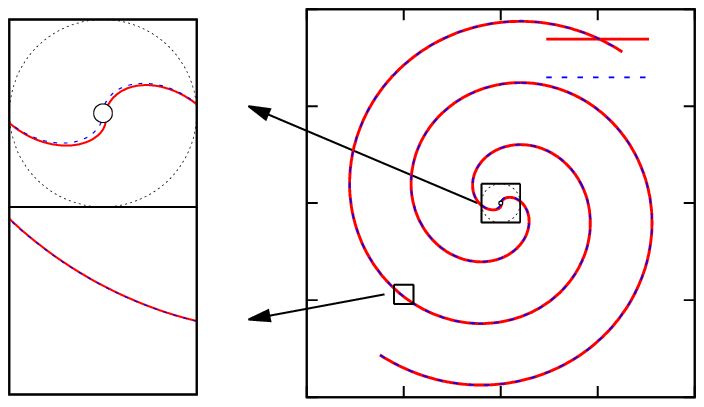}}%
    \gplfronttext
  \end{picture}%
\endgroup

%% file: offset_dipole.bbl
\begin{thebibliography}{50}
\expandafter\ifx\csname natexlab\endcsname\relax\def\natexlab#1{#1}\fi

\bibitem[{{Abdo} {et~al}\mbox{.}(2013){Abdo}, {Ajello}, {Allafort}, {Baldini},
  {Ballet}, {Barbiellini}, {Baring}, {Bastieri}, {Belfiore}, {Bellazzini}, \&
  et~al.}]{2013ApJS..208...17A}
{Abdo} A.~A. {et~al.}, 2013, \apjs, 208, 17

\bibitem[{{Archibald} {et~al}\mbox{.}(2016){Archibald}, {Gotthelf}, {Ferdman},
  {Kaspi}, {Guillot}, {Harrison}, {Keane}, {Pivovaroff}, {Stern}, {Tendulkar},
  \& {Tomsick}}]{2016ApJ...819L..16A}
{Archibald} R.~F. {et~al.}, 2016, \apjl, 819, L16

\bibitem[{{Arfken} \& {Weber}(2005)}]{2005mmp..book.....A}
{Arfken} G.~B., {Weber} H.~J., 2005, {Mathematical methods for physicists 6th
  ed.} Elsevier

\bibitem[{{Arzamasskiy} {et~al}\mbox{.}(2015){Arzamasskiy}, {Philippov}, \&
  {Tchekhovskoy}}]{2015MNRAS.453.3540A}
{Arzamasskiy} L., {Philippov} A., {Tchekhovskoy} A., 2015, \mnras, 453, 3540

\bibitem[{{Barsukov} \& {Tsygan}(2010)}]{2010MNRAS.409.1077B}
{Barsukov} D.~P., {Tsygan} A.~I., 2010, \mnras, 409, 1077

\bibitem[{{Beskin} \& {Zheltoukhov}(2014)}]{2014PhyU...57..799B}
{Beskin} V.~S., {Zheltoukhov} A.~A., 2014, Physics Uspekhi, 57, 799

\bibitem[{{Bonazzola} {et~al}\mbox{.}(2015){Bonazzola}, {Mottez}, \&
  {Heyvaerts}}]{2015A&A...573A..51B}
{Bonazzola} S., {Mottez} F., {Heyvaerts} J., 2015, \aap, 573, A51

\bibitem[{{Borra}(1974)}]{1974ApJ...187..271B}
{Borra} E.~F., 1974, \apj, 187, 271

\bibitem[{{Burnett} \& {Melatos}(2014)}]{2014MNRAS.440.2519B}
{Burnett} C.~R., {Melatos} A., 2014, \mnras, 440, 2519

\bibitem[{{Coelho} {et~al}\mbox{.}(2016){Coelho}, {Pereira}, \& {de
  Araujo}}]{2016ApJ...823...97C}
{Coelho} J.~G., {Pereira} J.~P., {de Araujo} J.~C.~N., 2016, \apj, 823, 97

\bibitem[{Cohen-Tannoudji {et~al}\mbox{.}(1973)Cohen-Tannoudji, Diu, \&
  Lalo{\"e}}]{cohen1973mecanique2}
Cohen-Tannoudji C., Diu B., Lalo{\"e} F., 1973, Mecanique quantique, Collection
  Enseignement des sciences, 16 No. vol.~2. Hermann

\bibitem[{{Davis} \& {Goldstein}(1970)}]{1970ApJ...159L..81D}
{Davis} L., {Goldstein} M., 1970, \apjl, 159, L81

\bibitem[{Devanathan(2006)}]{devanathan2006angular}
Devanathan V., 2006, Angular Momentum Techniques in Quantum Mechanics,
  Fundamental Theories of Physics. Springer Netherlands

\bibitem[{{Ek{\c s}i} {et~al}\mbox{.}(2016){Ek{\c s}i}, {Anda{\c c}}, {{\c
  C}{\i}k{\i}nto{\u g}lu}, {G{\"u}gercino{\u g}lu}, {Vahdat Motlagh}, \&
  {K{\i}z{\i}ltan}}]{2016ApJ...823...34E}
{Ek{\c s}i} K.~Y., {Anda{\c c}} I.~C., {{\c C}{\i}k{\i}nto{\u g}lu} S.,
  {G{\"u}gercino{\u g}lu} E., {Vahdat Motlagh} A., {K{\i}z{\i}ltan} B., 2016,
  \apj, 823, 34

\bibitem[{{Garfinkel}(1964)}]{1964AJ.....69..567G}
{Garfinkel} B., 1964, \aj, 69, 567

\bibitem[{{Geppert} \& {Vigan{\`o}}(2014)}]{2014MNRAS.444.3198G}
{Geppert} U., {Vigan{\`o}} D., 2014, \mnras, 444, 3198

\bibitem[{{Gil} {et~al}\mbox{.}(2003){Gil}, {Melikidze}, \&
  {Geppert}}]{2003A&A...407..315G}
{Gil} J., {Melikidze} G.~I., {Geppert} U., 2003, \aap, 407, 315

\bibitem[{{Goglichidze} {et~al}\mbox{.}(2015){Goglichidze}, {Barsukov}, \&
  {Tsygan}}]{2015MNRAS.451.2564G}
{Goglichidze} O.~A., {Barsukov} D.~P., {Tsygan} A.~I., 2015, \mnras, 451, 2564

\bibitem[{{Goldreich}(1970)}]{1970ApJ...160L..11G}
{Goldreich} P., 1970, \apjl, 160, L11

\bibitem[{{Good} \& {Ng}(1985)}]{1985ApJ...299..706G}
{Good} M.~L., {Ng} K.~K., 1985, \apj, 299, 706

\bibitem[{{Gourgouliatos} \& {Cumming}(2014)}]{2014MNRAS.438.1618G}
{Gourgouliatos} K.~N., {Cumming} A., 2014, \mnras, 438, 1618

\bibitem[{{Gralla} {et~al}\mbox{.}(2016){Gralla}, {Lupsasca}, \&
  {Philippov}}]{2016arXiv160404625G}
{Gralla} S.~E., {Lupsasca} A., {Philippov} A., 2016, ArXiv e-prints

\bibitem[{{Harding} \& {Muslimov}(2011)}]{2011ApJ...726L..10H}
{Harding} A.~K., {Muslimov} A.~G., 2011, \apjl, 726, L10+

\bibitem[{{Harrison} \& {Tademaru}(1975)}]{1975ApJ...201..447H}
{Harrison} E.~R., {Tademaru} E., 1975, \apj, 201, 447

\bibitem[{{Istomin}(2005)}]{2005pnsr.conf...27I}
{Istomin} Y.~N., 2005, in Progress in Neutron Star Research, {Wass} A.~P., ed.,
  p.~27

\bibitem[{{Jackson}(2001)}]{2001elcl.book.....J}
{Jackson} J.~D., 2001, {\'Electrodynamique classique}. Dunod, 2001

\bibitem[{{Komesaroff}(1976)}]{1976PASAu...3...51K}
{Komesaroff} M.~M., 1976, Proceedings of the Astronomical Society of Australia,
  3, 51

\bibitem[{{Landstreet}(1970)}]{1970ApJ...159.1001L}
{Landstreet} J.~D., 1970, \apj, 159, 1001

\bibitem[{{Landstreet}(1980)}]{1980AJ.....85..611L}
{Landstreet} J.~D., 1980, \aj, 85, 611

\bibitem[{MacRobert \& Sneddon(1967)}]{macrobert1967spherical}
MacRobert T., Sneddon I., 1967, Spherical Harmonics: An Elementary Treatise on
  Harmonic Functions, with Applications, International series of monographs in
  pure and applied mathematics. Pergamon Press

\bibitem[{{Martin} \& {Wickramasinghe}(1984)}]{1984MNRAS.206..407M}
{Martin} B., {Wickramasinghe} D.~T., 1984, \mnras, 206, 407

\bibitem[{{Melatos}(1997)}]{1997MNRAS.288.1049M}
{Melatos} A., 1997, \mnras, 288, 1049

\bibitem[{{Melatos}(2000)}]{2000MNRAS.313..217M}
{Melatos} A., 2000, \mnras, 313, 217

\bibitem[{{Michaud} {et~al}\mbox{.}(1981){Michaud}, {Charland}, \&
  {Megessier}}]{1981A&A...103..244M}
{Michaud} G., {Charland} Y., {Megessier} C., 1981, \aap, 103, 244

\bibitem[{{Michel} \& {Goldwire}(1970)}]{1970ApL.....5...21M}
{Michel} F.~C., {Goldwire}, Jr. H.~C., 1970, \aplett, 5, 21

\bibitem[{Morse \& Feshbach(1953{\natexlab{a}})}]{morse1953methodsvol2}
Morse P., Feshbach H., 1953{\natexlab{a}}, Methods of theoretical physics,
  International series in pure and applied physics No. vol.~2. McGraw-Hill

\bibitem[{Morse \& Feshbach(1953{\natexlab{b}})}]{morse1953methodsvol1}
Morse P., Feshbach H., 1953{\natexlab{b}}, Methods of theoretical physics,
  International series in pure and applied physics No. vol.~1. McGraw-Hill

\bibitem[{{Moss}(1986)}]{1986PhR...140....1M}
{Moss} D., 1986, \physrep, 140, 1

\bibitem[{{Ostriker} \& {Gunn}(1969)}]{1969ApJ...157.1395O}
{Ostriker} J.~P., {Gunn} J.~E., 1969, \apj, 157, 1395

\bibitem[{{Palomba}(2000)}]{2000A&A...354..163P}
{Palomba} C., 2000, \aap, 354, 163

\bibitem[{{P{\'e}tri}(2013)}]{2013MNRAS.433..986P}
{P{\'e}tri} J., 2013, \mnras, 433, 986

\bibitem[{{P{\'e}tri}(2015)}]{2015MNRAS.450..714P}
{P{\'e}tri} J., 2015, \mnras, 450, 714

\bibitem[{{Philippov} {et~al}\mbox{.}(2014){Philippov}, {Tchekhovskoy}, \&
  {Li}}]{2014MNRAS.441.1879P}
{Philippov} A., {Tchekhovskoy} A., {Li} J.~G., 2014, \mnras, 441, 1879

\bibitem[{{Putney} \& {Jordan}(1995)}]{1995ApJ...449..863P}
{Putney} A., {Jordan} S., 1995, \apj, 449, 863

\bibitem[{{Roberts}(1979)}]{1979ApJS...41...75R}
{Roberts} W.~J., 1979, \apjs, 41, 75

\bibitem[{Rose(1995)}]{rose1995elementary}
Rose M., 1995, Elementary Theory of Angular Momentum, Dover books on physics
  and chemistry. Dover

\bibitem[{Smirnov(1989)}]{smirnov1989cours}
Smirnov V., 1989, Cours de mathematiques superieures, Cours de
  math{\'e}matiques sup{\'e}rieures No. Tome~3, 2e partie. Mir

\bibitem[{{Stift}(1974)}]{1974MNRAS.169..471S}
{Stift} M.~J., 1974, \mnras, 169, 471

\bibitem[{{Tademaru}(1976)}]{1976ApJ...209..245T}
{Tademaru} E., 1976, \apj, 209, 245

\bibitem[{{Zanazzi} \& {Lai}(2015)}]{2015MNRAS.451..695Z}
{Zanazzi} J.~J., {Lai} D., 2015, \mnras, 451, 695

\end{thebibliography}
